\documentclass[a4paper,11pt]{article}
\usepackage{jinstpub} 
\usepackage{lineno}

\usepackage{graphicx} 
\usepackage{caption}
\usepackage{subcaption}
\usepackage[dvipsnames]{xcolor}
\usepackage{tabularray}
\usepackage{multirow, makecell}
\usepackage{lscape}
\usepackage{booktabs}
\usepackage{units}

\DeclareMathAlphabet{\mathcal}{OMS}{cmsy}{m}{n} 
\usepackage{anyfontsize} 

\title{\boldmath Radiopurity material assays and radiation exposure projections for superconducting qubit measurements at SNOLAB}

\author[a,b]{Y.~Ahmed,}
\author[a]{B.~Binoy,}
\author[a,b]{R.~Bunker,}
\author[a]{D.~Chauhan,}
\author[c]{P.~Delsing,}
\author[e]{R.~Germond,}
\author[a,b]{J.~Hall,}
\author[d]{Z.~Hong,}
\author[a,b]{A.~Iqbal,}
\author[a,b,e]{V.~Iyer,}
\author[c]{A.~Klepikova}
\author[a,d]{A.~Kubik,}
\author[c]{S.~P.~Mantry,}
\author[a,b]{A.~C.~Masuskapoe,}
\author[a]{C.~C.~Monk,}
\author[e]{G.~Peng,}
\author[a]{P.~Qin,}
\author[f,g]{W.~Rau,}
\author[d,e]{T.~Reynolds,}
\author[a,d]{M.~Stukel,}
\author[e]{C.~M.~Wilson,}
\author[a,b,d,1]{B.~Zatschler}
\author[a,b,d,1]{S.~Zatschler,\note{Corresponding authors.}}
\author[d]{A.~Zuniga}
\collaboration{QUTEbits collaboration}

\affiliation[a]{SNOLAB,\\ Creighton Mine \#9, 1039 Regional Road 24, Sudbury, ON P3Y 1N2, Canada}
\affiliation[b]{Laurentian University,\\ School of Natural Sciences, 935 Ramsey Lake Road, Sudbury, ON P3E 2C6, Canada}
\affiliation[c]{Department of Microtechnology and Nanoscience,\\
Chalmers University of Technology, 41296 Gothenburg, Sweden}
\affiliation[d]{Department of Physics,\\ University of Toronto, Toronto, ON M5S 1A7, Canada}
\affiliation[e]{Institute for Quantum Computing and Department of Electrical \& Computer Engineering,\\ University of Waterloo, Waterloo, ON N2L 3G1, Canada}
\affiliation[f]{TRIUMF,\\ Vancouver, BC V6T 2A3, Canada}
\affiliation[g]{Department of Physics,\\ Queen's University, Kingston, ON K7L 3N6, Canada}

\emailAdd{stefan.zatschler@snolab.ca}
\emailAdd{birgit.zatschler@snolab.ca}

\abstract{
Interactions of cosmic rays and other forms of ionizing radiation pose a significant challenge to the reliable operation of state-of-the-art quantum devices and error correction in quantum computing based on superconducting circuits which are typically fabricated on semiconductor substrates.
Shielded by 2\,km of rock overburden, the Cryogenic Underground TEst facility (CUTE) at SNOLAB provides a unique ultra-low radiation environment to probe the performance of quantum technologies with a particular interest in quantum coherence studies.
In this article, we present the findings of an extensive material assaying program in preparation for the first underground operation of superconducting qubits at SNOLAB.
The radioactivity levels identified by the material assays enter a thorough Monte Carlo study based on the \textsc{Geant4} particle physics tracking code.
From these simulations, we estimate the rates of energy deposits from radiogenic sources expected for a quantum-device assembly operated in the CUTE facility.
We further characterize the spectral components of the projected background and identify the dominant particle interaction types.
Finally, we outline how crystal dynamics simulations using the G4CMP solid-state physics extension for \textsc{Geant4} can inform the community-wide efforts to identify effective strategies to mitigate the effects of high-energy particle impacts.
}

\keywords{Interaction of radiation with matter; Superconducting devices and qubits; Detector modeling and simulations}

\begin{document}
\maketitle
\flushbottom

\section{Introduction}
\label{sec:intro}

In recent decades, the emergence of quantum information science (QIS) has intensified research toward using quantum circuits as quantum bits (qubits) \cite{Kjaergaard2020Review, MWL2024QComp}.
The realization that superconducting qubits can be controlled and read out reliably with microwave pulses \cite{Blais2007QIS, GU2017Microwaves} led to the creation of the field of circuit quantum electrodynamics \cite{BlaisCircuitQED2021}.
Although many platforms are still being pursued to implement quantum information processing, superconducting circuit architectures are one of the leading candidates \cite{Blais2004SCcircuits, Blais2007QIS, BlaisCircuitQED2021, GU2017Microwaves}, partly because of their ease of fabrication using well-established semiconductor manufacturing techniques.
One specific realization of a hardware-level implementation of a qubit is the so-called \textit{transmon qubit} \cite{Koch2007Transmons, Roth2023Transmons}.
Transmon qubits are designed to have reduced sensitivity to charge noise and operate as anharmonic LC\footnote{The term ``LC'' refers to an inductance $L$ and capacitance $C$.} oscillators with distinct, non-equidistant energy levels.
These systems utilize the quantum dynamics of electromagnetic fields by embedding a nonlinear inductor, in the form of a Josephson junction, into complex systems of planar microwave circuitry made from superconductors \cite{Oliver2013SCmaterials, Blais2004SCcircuits}.
By engineering Josephson junctions into various topologies, the nonlinearity of the resulting circuit can be used as the basis of a qubit with individually addressable quantum states \cite{Roth2023Transmons}.
The use of superconductors for the surrounding microwave circuitry also minimizes dissipation.
This circumstance is important because it allows the typically fragile qubit states to survive for long enough that the fundamental interactions to create and process quantum information can be leveraged at microwave frequencies \cite{Blais2004SCcircuits, Blais2007QIS, BlaisCircuitQED2021, GU2017Microwaves}.

A key characteristic affecting the computing potential of qubits in real-world applications is the coherence time which indicates, on average, how long a qubit will remain in a prepared quantum state.
Improving the coherence time of superconducting qubits has been a major research focus for the last decade \cite{Kjaergaard2020Review, Roth2023Transmons}.
Experimental studies demonstrated that ionizing radiation can cause decoherence in superconducting qubits \cite{Vepsalainen2020decoherence, Wilen2021chargenoise, McEwen2022CatastrophicErrors, Cardani2023disentangling}, superconducting resonators \cite{Cardani2021quantumUG} and superconducting quantum interference devices (SQUIDs) \cite{SQUIDGAME2025}.
Further studies observed sequences of quantum errors correlated in time which occur across multiple qubits and extend over entire device substrates \cite{Cardani2021quantumUG, Wilen2021chargenoise, Bratrud2025chargenoise, Casagranda2025RadiationQD}.
These correlated quantum error events exhibit signatures consistent with the generation of non-equilibrium quasiparticles induced by ionizing radiation \cite{Serniak2018qps, Larson2025qppoisoning, Celi2026QPdynamics} and have been shown to arise in part from cosmic-ray interactions with the substrate \cite{Harrington2025cosmicrays, Li2025CosmicErrors}.
Importantly, a single high-energy interaction can create cascades of excitations that simultaneously affect multiple qubits, producing strongly correlated error bursts.
Such events violate the core assumptions of most quantum error correction schemes, including the surface code \cite{Fowler2012SurfaceCode}, which rely on sparse and uncorrelated errors in time and space \cite{Shor1995Decoherence, Steane1996ErrorCorrection, Klesse2005ErrorCorrection}.
As a result, radiation-induced errors can be exceptionally difficult to detect and correct with existing techniques, motivating ongoing efforts to engineer more resilient qubit designs \cite{Siddiqi2021scqubits, Martinis2020Decays}.

Identifying the fundamental mechanisms that lead to quantum decoherence in the presence of ionizing radiation requires the ability to test quantum devices while reliably controlling the radiation environment. 
To address this need, new experimental facilities were developed in shallow and deep underground laboratories worldwide \cite{Loer2024PNNL, Bratrud2025chargenoise, Cardani2021quantumUG}.
At this interface, the QUTEbits collaboration\footnote{The QUTEbits collaboration consists of researchers from the Institute for Quantum Computing (IQC) at the University of Waterloo, the University of Toronto, SNOLAB and Laurentian University in Ontario, Canada, as well as Chalmers University of Technology in Sweden.} was formed to investigate the impact of radiation and cosmic rays on quantum technologies.
A key focus of this new collaboration is to study how ionizing radiation affects the coherence of individual qubits and how it causes correlated errors that are problematic for quantum error correction.
The goal of the project is to perform an advanced characterization of superconducting qubits by comparing the results of coherence times measured in surface laboratories to those measured deep underground at SNOLAB \cite{Duncan2010SNOLAB, Smith2012Snolab} with identical quantum devices.

SNOLAB is located at the 6800-foot level in Vale’s Creighton Mine near Sudbury, Ontario, and uses the Canadian Shield to protect experiments from the cosmic rays present at the Earth's surface.
As one of the deepest and cleanest underground science facilities in operation in the world, the rock overburden of 2\,km (equivalent to 6000\,m of water coverage) achieves a suppression of the cosmic ray induced muon flux by a factor of fifty million compared to the surface \cite{Pec2024muonflux}.
One of the underground facilities available to SNOLAB users is CUTE -- the Cryogenic Underground TEst facility -- which is a platform for testing and operating cryogenic devices in an environment with low levels of background radiation \cite{Camus2023CUTE}.
CUTE was constructed originally for testing cryogenic detectors for the SuperCDMS SNOLAB dark matter experiment \cite{Agnese2016SCDMS, Albakry2026CUTE-GeHV, Kennard2026HVeV}.
CUTE is now a SNOLAB user facility and is available for projects based on proposals assessed for their scientific and technological merits.
The facility consists of a 3.5\,m wide tank filled with highly purified water that has a drywell of about 50\,cm diameter in its center.
The drywell is lined with a magnetic shield and 11\,cm of lead for shielding against environmental radiation.
Located within this shielding is a cryogen-free dilution refrigerator mounted on a vibration-isolating damper system. 
Inside the cryostat, directly above the experimental volume ($\sim$20\,L) is a 13\,cm thick layer of lead encased in 1.5\,cm of copper acting as an internal radiation shield.
It covers the direct line of sight above the experimental space and shields from radioactive materials in the brazing of the dilution unit.

The first quantum research project approved to be hosted in SNOLAB's CUTE facility is the QUTEbits project.
In this report, we present a detailed Monte Carlo study for this project, propagating the results of an extensive material-assay program with the \textsc{Geant4} particle physics tracking code \cite{Geant42003, Geant42006, Geant42016}.
The Monte Carlo simulations allow us to estimate the expected rates of energy deposits from ambient radioactive background sources as observed by the proposed quantum-device payload to be operated in CUTE.

This article is organized as follows: in section~\ref{sec:radioactivity_measurements} we report on the radioactivity measurements, their results, and highlight notable findings.
Section~\ref{sec:radiation_modeling} describes the radiation modeling workflow with \textsc{Geant4}, covering the geometry implementation of the experimental apparatus and our approaches to account for different categories of background sources.
A detailed breakdown of the composition of each background category is provided.
In section~\ref{sec:background_projection}, we present the results of the simulation data analysis with a summary of the background composition.
This section also includes a comparison of the total expected background rate to the rate of ionizing radiation hits caused by the presence of radioactive calibration sources in the vicinity of the quantum-device payload.
In section~\ref{sec:G4CMP_studies}, we extend the particle tracking simulations to include crystal dynamics effects by making use of the \textsc{Geant4} Condensed Matter Physics (G4CMP) package \cite{Kelsey2023G4CMP}.
The results obtained from G4CMP, such as the phonon energy collection time, feed back into the processing of the \textsc{Geant4} background and radiation source simulations.
In section~\ref{sec:conclusion}, we close with our conclusions.

\section{Radioactivity measurements}
\label{sec:radioactivity_measurements}

Radioactive isotopes are prevalent in our everyday world and are present to some degree in all materials due to their natural abundance and long half-lives.
Additionally, radioactive contaminants may be introduced at various stages of manufacturing processes.
A common form of contamination arises from the accumulation of radioisotopes on the surfaces of materials through exposure to dust and radon in the air.
Dust can have relatively high concentrations of $^{238}$U, $^{232}$Th, and $^{40}$K~\cite{DiVacri2021Dust}.
Likewise, the decay products from airborne $^{222}$Rn can implant into surfaces, leading to an accumulation of the long-lived radioisotope $^{210}$Pb (and by extension its $^{210}$Bi and $^{210}$Po progeny), producing a nearly constant emission of $\gamma$ rays, $\beta$ electrons, and $\alpha$ particles \cite{Bunker2020Pb210}.
Radioisotopes can also be introduced as byproducts from cosmic-ray interactions -- a process known as cosmogenic activation.
Common examples of radioisotopes produced through cosmogenic activation include tritium, $^{54}$Mn, and several cobalt species \cite{Laubenstein2009Cosmics}.
It should be noted that the activity produced is typically lower than the activity of contaminants introduced by dust or radon \cite{DiVacri2021Dust}.

The level of radioactivity in a material can be probed with different techniques.
In order to measure the level of radioactive contamination from a complex decay chain such as the primordial decay chains of $^{238}$U and $^{232}$Th, not all of the elements need to be separated out.
Instead, typical trace elements can be used to evaluate whether there are any deviations from secular equilibrium in the decay chain.
Because of the presence of long-lived isotopes and the mobile noble gas radon in the $^{238}$U decay chain, activities for this chain are often reported separately for the ``top'' and ``bottom'' part where the break is at $^{226}$Ra.
The same convention will be used in this report.

\subsection{Low-background counting techniques}
\label{sec:counting_techniques}

Prior to upgrading the CUTE cryostat at SNOLAB for microwave measurements of superconducting qubits, the collaboration characterized the radioactivity of common components and materials.
SNOLAB provides a world-class low-background measurement laboratory in its underground facility to screen materials for radioactive impurities \cite{Lawson2020Snolab, Lawson2023Snolab}.
The counting facility’s primary goal is to identify materials which have low concentrations of naturally occurring $^{40}$K and the long-lived decay chains of $^{238}$U and $^{232}$Th, often well below the parts per trillion (ppt) level.
These radioactivity levels are below what is generally accessible by chemical and analytical techniques, therefore assay methods are often performed through radiation counting using high-purity germanium (HPGe) detectors.
HPGe spectroscopy can reach sensitivities of $\mathcal{O}(10)$ to $\mathcal{O}(100)$ $\mu$Bq/kg for isotopes in the primordial decay chains.

The presence of radioactive isotopes in a sample can be identified via their characteristic $\gamma$~rays.
A comparison with background rates together with Monte Carlo simulations assessing the impact of the sample characteristics allows for a quantification of the contaminants, where upper limits are extracted when no significant excess above background is observed.

For the QUTEbits project, radioactivity measurements were performed with several HPGe detectors, each with varying sensitivity to the relevant isotopes of interest and optimized for different sample sizes.
All of the detectors are situated in the low-background counting laboratory at \mbox{SNOLAB} (approximately 2\,km below surface), and protected from environmental $\gamma$ rays with roughly 20\,cm of lead shielding and 5\,cm of copper shielding.
The detectors are constantly flushed with evaporated dry nitrogen at a rate of 2\,L/min to suppress radon which is present in the environmental air \cite{SnolabCountingWeb}.
Characteristics of the five SNOLAB HPGe detectors (PGT, Canberra, Lively, Gopher, VDA) and their isotope-specific sensitivities are presented in table~\ref{tab:sensitivities}.

\begin{table}[ht]
\centering
\caption{Characteristics and sensitivities of the SNOLAB HPGe detectors \cite{SnolabCountingWeb}. The relative efficiencies are reported with respect to a standard NaI(Tl) detector.}
\label{tab:sensitivities}
\begin{tabular}{lcccccc}
\toprule
Detector & PGT & Canberra & Lively & Gopher & VDA\\
\midrule
Detector volume [cm$^3$] & 210 & 300 & 400 & 400 & 400 \\
Relative efficiency & 55\% & 80\% & 107\% & 120\% & 120\%\\ 
Nominal sample size & 1\,L & 8\,mL & 1\,L & 1\,L & 1\,L\\
Energy range [keV] & 90--3000 & 10--900 & 40--3000 & 40--3000 & 40--3000\\
\midrule
$^{238}$U sensitivity [mBq]  & 0.11  & 0.02  & 0.05  & 0.17 & 0.09\\
$^{235}$U sensitivity [mBq]  & 0.16  & 0.01  & 0.02  & 0.08 & 0.06\\
$^{232}$Th sensitivity [mBq] & 0.10  & 0.02  & 0.06  & 0.21 & 0.08\\
$^{40}$K sensitivity [mBq]   & 1.42  & 0.92  & 0.45  & 1.01 & 1.22\\
$^{137}$Cs sensitivity [mBq] & 0.13  & 0.02  & 0.02  & 0.08 & 0.05\\
$^{60}$Co sensitivity [mBq]  & 0.04  & 0.03  & 0.02  & 0.04 & 0.02 \\
$^{54}$Mn sensitivity [mBq]  & 0.043 & 0.033 & 0.021 & 0.044 & 0.034 \\
$^{210}$Pb sensitivity [mBq] & - & 0.55  & 31.53 & 16.49 & 7.71\\
\bottomrule
\end{tabular}
\end{table}

Most of the recorded HPGe detector spectra are evaluated for the same set of radioactive isotopes including the top and bottom parts of the $^{238}$U decay chain, the $^{235}$U decay chain, the $^{232}$Th decay chain and single-isotope decays including $^{40}$K, $^{137}$Cs and $^{60}$Co.
For the bottom part of the $^{238}$U decay chain, only isotopes down to $^{214}$Bi are included in the evaluation, leaving $^{210}$Pb to be determined separately.
For some materials, we also looked for typical cosmogenic activation products such as $^{57}$Co, $^{58}$Co and $^{54}$Mn.

\subsection{Material assay results}
\label{sec:assay_results}

A selection of the collected radioactivity data of the assayed components is presented in table~\ref{tab:AssayResults}.
This selection is limited to the naturally occurring decay chains and $^{40}$K. It lists the assays of the components which we are going to deploy for the QUTEbits measurements and thus are considered in the subsequent background simulation studies as discussed in section~\ref{sec:radiation_modeling}.
A complete list of all assayed components can be found in appendix~\ref{sec:appendix_assay_results}, table~\ref{tab:AssayResultsComplete}.
All assay results (including results of isotopes not listed in this report) have been made publicly available on \href{https://www.radiopurity.org/}{radiopurity.org}\footnote{Search for keyword ``QBITS-CUTE'' or ``QUBITS-CUTE'' on the website.}~\cite{Loach2016radiopurity,RadiopurityWeb}.

\begin{table}[ht!]
\centering
\caption{Assay results of components considered in the background simulation studies. Listed are the component names, the corresponding quantity and the mass of a single component, and the determined radioactivity level of the natural decay chains and $^{40}$K. Some assay results are reported as 90\% C.L. upper limits. In order to account for disequilibrium within the $^{238}$U decay chain, it is split at $^{226}$Ra into the top (t) and bottom (b) of the decay chain. If no quantity value is given in the second column, the mass value in the third column is the total component mass.}
\label{tab:AssayResults}
\resizebox{\textwidth}{!}{
\begin{tabular}{cccccccc} 
\toprule
Component & Quantity & Mass & $^{238}$U & $^{235}$U & $^{232}$Th & $^{210}$Pb & $^{40}$K\\
 & in setup & [g] & [mBq/kg] & [mBq/kg] & [mBq/kg] & [mBq/kg] & [mBq/kg] \\ 
\midrule
\multirow{2}{*}{Si chip} & \multirow{2}{*}{2} &  \multirow{2}{*}{0.057} & t: $<$516.8 & \multirow{2}{*}{$<$22.0} & \multirow{2}{*}{$<$28.5} & \multirow{2}{*}{(29$\pm$22)$\cdot10^{3}$} & \multirow{2}{*}{$<$336.7} \\
 & & & b: $<$12.6 & & & & \vspace{0.3em}\\

\multirow{2}{*}{PCB in OQTO holder} & \multirow{2}{*}{2} & \multirow{2}{*}{3.9} & t: 6132$\pm$1973 &\multirow{2}{*}{106.5$\pm$27.9}&\multirow{2}{*}{2232.0$\pm$138.9}&\multirow{2}{*}{-}&\multirow{2}{*}{1380.1$\pm$567.8} \\
 & & & b: 1499.0$\pm$101.2 & & & & \vspace{0.3em}\\

\multirow{2}{*}{OQTO holder} & \multirow{2}{*}{2} & \multirow{2}{*}{204.6} & t: 1099.0$\pm$195.4 &\multirow{2}{*}{28.6$\pm$2.6}& \multirow{2}{*}{38.4$\pm$4.7}&\multirow{2}{*}{-}& \multirow{2}{*}{60.2$\pm$25.5} \\
 & & & b: 20.4$\pm$3.3 & & & & \vspace{0.3em}\\

\multirow{2}{*}{Cu in QUTEbits stack} & \multirow{2}{*}{-} & \multirow{2}{*}{1688} & t: $<$130.8 &\multirow{2}{*}{1.2$\pm$1.6}&\multirow{2}{*}{$<$4.9}&\multirow{2}{*}{-}& \multirow{2}{*}{$<$65.1} \\
 & & & b: $<$2.4 & & & & \vspace{0.3em}\\

\multirow{2}{*}{Microwave switch} & \multirow{2}{*}{3} & \multirow{2}{*}{95.5} & \multicolumn{1}{c}{t: 7554.0$\pm$667.9} & \multirow{2}{*}{133.5$\pm$8.2}  & \multirow{2}{*}{826.0$\pm$32.7} & \multirow{2}{*}{(41$\pm$15)$\cdot10^{4}$} & \multirow{2}{*}{806.2$\pm$92.6} \\
 & & & \multicolumn{1}{c}{b: 453.0$\pm$19.8} & & & &  \vspace{0.3em}\\

\multirow{2}{*}{Al cavity} & \multirow{2}{*}{6} & \multirow{2}{*}{37.2} & t: $<$85.7 &\multirow{2}{*}{ $<$2.8 }&\multirow{2}{*}{ 9.5$\pm$3.6 }&\multirow{2}{*}{ - }& \multirow{2}{*}{$<$26.9} \\
 & & & b: $<$3.9 & & & & \vspace{0.3em}\\

CM shield at QUTEbits stack & 1 & 1354 & \multicolumn{1}{c}{t: $<$171.4} & \multirow{2}{*}{$<$14.6} & \multirow{2}{*}{134.1$\pm$21.4} & \multirow{2}{*}{$<$3703} & \multirow{2}{*}{$<$126.6}  \\
CM shield at internal lead & 1 & 2005 &  \multicolumn{1}{c}{b: 3.7$\pm$19.8} & & & & \vspace{0.3em}\\

Formable non-magnetic & \multirow{2}{*}{14} & \multirow{2}{*}{10.1} & \multicolumn{1}{c}{t: 787.7$\pm$1024.0} & \multirow{2}{*}{$<$14.2} & \multirow{2}{*}{$<$38.9} & \multirow{2}{*}{-} & \multirow{2}{*}{233.6$\pm$289.6}   \\
cable assembly & & & \multicolumn{1}{c}{b:$<$37.7} &  &  & & \vspace{0.3em}\\

\multirow{2}{*}{SMA connectors} & \multirow{2}{*}{16} & \multirow{2}{*}{3.59} & t: 96.7$\pm$41.8 & \multirow{2}{*}{5.6$\pm$1.4} & \multirow{2}{*}{1.2$\pm$3.9} & \multirow{2}{*}{6922.0$\pm$407.5} & \multirow{2}{*}{91.7$\pm$99.8} \\
& & & b: $<$3.8 \\

\multirowcell{2}{Brass screws} & \multirow{2}{*}{50} & \multirow{2}{*}{0.87} & t: $<$62.2 & \multirow{2}{*}{$<$6.6} & \multirow{2}{*}{ $<$11.8} & \multirow{2}{*}{$<$603.7} & \multirow{2}{*}{391.5$\pm$525.7} \\
& & & b: $<$15.0 \\

\multirow{2}{*}{BF-6 glue at Si chip}& \multirow{2}{*}{1} & \multirow{2}{*}{0.012}  & t: $<$44.7  &\multirow{2}{*}{$<$1.6}&\multirow{2}{*}{26.6$\pm$4.5}&\multirow{2}{*}{$<$130}& \multirow{2}{*}{42.9$\pm$24.4} \\
 & & & b: $<$1.3 & & & & \vspace{0.3em}\\

\multirow{2}{*}{CliQ10 IR filter} &\multirow{2}{*}{4} & \multirow{2}{*}{25.3} & t: 39.6$\pm$23.3 &\multirow{2}{*}{2.4$\pm$1.2}&\multirow{2}{*}{46.4$\pm$6.1}&\multirow{2}{*}{$<$146.4}& \multirow{2}{*}{720.3$\pm$246.5} \\
 & & & b: 32.5$\pm$6.1 & & &  & \\

\bottomrule
\end{tabular}
}
\end{table}

The proximity of a component with respect to the quantum-device substrate (silicon in this case) is related to its impact in terms of additional radiation; closer components may contribute more, whereas components located farther away are likely to be shielded and may contribute less.
In the simulations we denote the ``QUTEbits stack'' as the assembly of all components inside the cryogenic magnetic (CM) shield which completely encloses the OQTO holder \cite{OQTOWeb} containing the QUTEbits Si chip, up to six 3D Al resonator cavities, and several electronic components (see section~\ref{sec:geometry_overview} for a detailed description).
The components listed in table~\ref{tab:AssayResults} are all within this CM shield, except for a second CM shield surrounding the internal lead shield of CUTE and the microwave switches located in between the internal shield and the QUTEbits stack.
The OQTO holder, PCB and the BF-6 glue are the only components which are in direct line of sight with the Si chips.
However, the Al cavities are in direct line of sight to several cables, screws, the CliQ10 IR filters and various Cu parts in the QUTEbits stack.

\begin{figure}[ht!]
\begin{center}
\includegraphics[width=\textwidth]{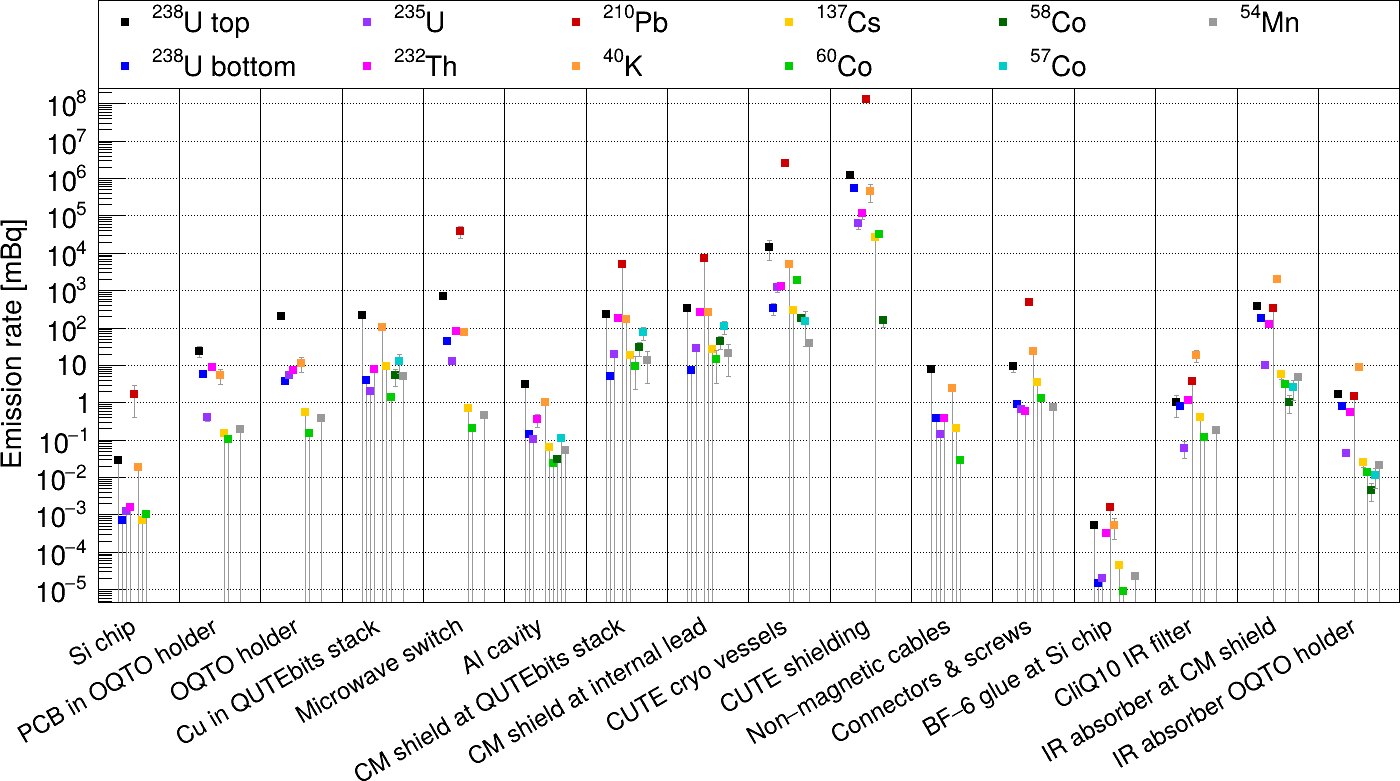}
\end{center}
\caption{Emission rate of each component for each measured isotope or decay chain. The emission rate is calculated per piece, not taking into account the actual quantity of components to be used in the QUTEbits setup. A line extending from the marker to the bottom of the plot indicates an upper limit of the corresponding assay result at 90\% C.L., while all other values are measurements with symmetric error bars, which appear asymmetric on logarithmic scale.}
\label{fig:bulk_assays}
\end{figure}

The expected emission rate of a component can be calculated from the specific activity reported in table~\ref{tab:AssayResults} and the component mass.
Figure~\ref{fig:bulk_assays} displays a summary of the emission rates for the components and isotopes of interest.

In addition to the assayed components for the QUTEbits project, figure~\ref{fig:bulk_assays} also includes a combined representation of the assays of the CUTE cryogenic vessels as well as the CUTE shielding layers, which includes the internal lead shield above the QUTEbits stack, the different layers of thermal shielding within the cryostat, several stainless steel components such as the outer vacuum chamber, the outer radiation lead shield and the ultra pure water in the CUTE water tank.
The CUTE material assay results are available online at~\cite{SnolabCountingWeb}.

\subsection{Notable assay findings}
\label{sec:assay_findings}

While a comprehensive simulation study (see section~\ref{sec:radiation_modeling}) provides an estimate of the impact of individual components, the emission rate of each component can already be a helpful indicator for material selection criteria and design choices.
Considering the assembly geometry, the most significant background contributions are from components with a direct line of sight to the Si chips and from components that are part of the Si chip assembly.

Single-crystal Si is a common substrate material for quantum devices.
The screening results from several Si wafers show similar contamination levels compared to each other (see table~\ref{tab:AssayResultsComplete}). However, most of the results are upper limits and those not quoted as upper limits have large uncertainties and could arguably be considered to be upper limits. Thus, it is expected that the Si samples are actually much cleaner than indicated by the assay results which is consistent with other findings \cite{Arevalo2021Damic}.
In any case, because the Al circuit layer is directly deposited onto the Si substrate, the radiopurity of the silicon is a central contribution to the overall background budget.
By considering these assays in our background simulation campaign, we adopt a very conservative approach leading to the background rate being dominated by the Si assay results.

For several of the assayed components, the reported $^{210}$Pb activities and uncertainties appear unrealistically high.
These measurements are likely imprecise because the HPGe detector sensitivity is poor for $^{210}$Pb (see table~\ref{tab:sensitivities}). 
The main emitted $\gamma$ ray has a relatively low energy (46.5\,keV) and is readily absorbed within the sample or the HPGe detector enclosure such that it cannot be detected efficiently.
We also note a potential systematic uncertainty related to the distribution of $^{210}$Pb in the sample.
The assumption that $^{210}$Pb is homogeneously distributed in the sample may be incorrect, because $^{210}$Pb can accumulate on surfaces when a material is exposed to air which contains radon.
We explore this possibility in section~\ref{sec:surface_contamination}.

Several raw aluminum samples used for the deposition of the thin superconducting Al films on top of the Si substrates were screened, and their rates were generally found to be quite low with a few exceptions of rather high levels of $^{238}$U, $^{232}$Th and $^{40}$K (see table~\ref{tab:AssayResultsComplete}).
However, because of the tiny mass of the films, we can safely neglect their contribution.

Printed circuit boards (PCBs) from multiple vendors were assayed and their respective activities were found to be relatively high.
While this is not surprising as the many steps in the manufacturing process can lead to the introduction of additional radioisotopes that can be held in the materials, this does point to an important avenue for reducing backgrounds as this component would be in direct contact with the Si chip.
Notably, some of the measured PCBs were much more radioactive than others (see table~\ref{tab:AssayResultsComplete}), indicating which products are suitable for the anticipated low-background assembly to be operated in CUTE.

For the OQTO holder made from copper, the assay results turned out to be low in radioactivity.
The OQTO holder was assayed with its brass screws and SMA connectors in place, which were also measured separately.
The radioactivity of the brass screws was below the detector's sensitivity and the SMA connectors were relatively clean except for $^{210}$Pb.
The most radioactive part of the SMA connectors are the SMA pins, which were assayed separately as well (see table~\ref{tab:AssayResultsComplete}).

Components that would be mounted below the internal lead shield of CUTE, but not in direct line of sight of the Si chips, include up to six 3D Al resonator cavities and two cylindrical CM shields (used for shielding against environmental magnetic fields).
Some of the assayed Amumetal samples showed high levels of $^{238}$U and $^{210}$Pb.
For the CM shield material, we requested samples from different vendors, and the sample provided by Ad-Vance Magnetics Inc (referred to as ``CP-EXP-1184'') met our radiopurity requirements.

For the formable non-magnetic cable assembly (EZ-FLEX.86-CU from EZ Form cable) inside the QUTEbits stack that connects the qubit holder to the feedthrough plate mounted on the magnetic shield, the measured activities were close to the sensitivity of the HPGe detector. 
The readout components above the level of the CUTE internal lead shield include additional microwave cables, filters, and circulators.
Most of the cables that would be above the internal lead shield did not show high levels of radioactivity, nor did the circulators.
Since the internal lead shield absorbs radiation very well, none of the components above the internal shield were considered in the simulation studies discussed in section~\ref{sec:radiation_modeling}.

We considered two potential locations for dedicated IR absorbers: one directly at the inside of the CM shield, and another one inside the OQTO holder on surfaces with direct line of sight to the Si chip.
For both cases, the components of a hypothetical recipe consisting of a Cu sheet, Stycast 1266, glass beads and carbon black (also known as Berkeley Black \cite{Persky1999}) were assayed separately.
Considering the constituents' mass contribution, carbon black would dominate the emission rate, and we decided to replace it with a copper powder in our simulation studies (see assay results in table~\ref{tab:AssayResultsComplete}).
While this reduced the expected radioactivity of the IR absorber assembly, more work is needed to identify vendors for specific components such as the glass beads with sufficiently low levels of radioactivity.
Consequently, the initial QUTEbits measurements will be performed without such IR absorbers and the total background rates reported in this article do not include their contribution.
We do, however, discuss their potential impact in appendix~\ref{sec:potential_components}.

Finally, it shall be noted that the BF-6 glue provided by Ukrvet Biopharm, Ukraine, turned out to be quite clean and can be recommended for use in low-background environments.

\subsection{Cosmogenic activation products}
\label{sec:activation_products}

Secondary cosmic-ray particles such as neutrons, protons and muons can trigger nuclear reactions and in consequence lead to the activation of materials.
As the particle flux is strongly suppressed underground, this effect is negligible for materials stored at locations like SNOLAB.
Due to the rather high energy of these particles, spallation reactions can produce various radionuclides, e.g.\ $^{60}$Co, $^{58}$Co, $^{57}$Co and $^{54}$Mn which may contribute noticeably to the overall background budget, as indicated in figure~\ref{fig:bulk_assays}.
For copper, the most important activation product is $^{60}$Co because of its high production rate, relatively long half-life and high-energy gamma emission~\cite{Laubenstein2009Cosmics}.

A complete tracking of the location of materials allows the activity of each radionuclide to be estimated using production rates provided in the literature, such as \cite{Laubenstein2009Cosmics}.
However, the emission rates were deemed to be not significant enough compared to the overall radiopurity of the considered materials to justify this effort.
Thus, for this project the assay results of $^{60}$Co, $^{58}$Co, $^{57}$Co and $^{54}$Mn were used instead, even though the assayed component may have a different exposure history regarding cosmogenic activation than the actual component used in the final QUTEbits setup.

A cosmogenic activation product commonly found in silicon is $^{32}$Si, a $\beta$ emitter with a half-life of 153~years.
It is primarily produced in the atmosphere and can be introduced into silicon in different ways and at different stages of the production process~\cite{Orrell2018Si}, leading to significant sample-to-sample variations in terms of specific activity~\cite{Caldwell1990Si, Arevalo2015Damic, Arevalo2021Damic}.
The DAMIC collaboration measured the $^{32}$Si content of different batches of their silicon CCDs \cite{Arevalo2015Damic, Arevalo2021Damic}, with the latest measurement indicating $140 \pm 30\,\unit{\mu Bq/kg}$ of $^{32}$Si in the CCD substrates.
Compared to other sources of radioactivity in the QUTEbits setup, this level of $^{32}$Si activity is negligible for us.

\section{Radiation exposure modeling with \textsc{Geant4}}
\label{sec:radiation_modeling}

The \textsc{Geant4} toolkit is a Monte Carlo simulation code which models the passage of particles through matter~\cite{Geant42003, Geant42006, Geant42016}.
Of particular interest for this work is the radioactive decay of various isotopes and the interaction of the emitted particles with the surrounding matter, most importantly the energy deposits in the Si substrates and Al cavities.
A comprehensive description of radioactive decay physics and how these are modeled in \textsc{Geant4}'s radioactive decay module can be found in~\cite{Hauf2013G4decay, Hauf2013G4decayvalidation}.

The general workflow to model the radiation exposure in the Si chips and Al cavities is as follows:
at first \textsc{Geant4} volumes are designed as geometrical representations of all relevant components and material properties are assigned accordingly.
The assays presented in section~\ref{sec:assay_results} are mapped to the corresponding volumes, i.e.\ each volume is homogeneously contaminated with the radioisotopes it contains according to the assay results.
The decay of these contaminants is modeled by \textsc{Geant4} as a primary event and the emission products are tracked through the geometry where they interact with matter, create secondary particles and are eventually absorbed in a material.
For these simulations, the \textit{Shielding} physics lists, including NeutronHP, together with the electromagnetic option 4 (EMZ) were used with \textsc{Geant4}-10.7.4.
In the case of decay chains, all subsequent decays are modeled, if not specified otherwise.
For the top part of the $^{238}$U decay chain, the decays are stopped at $^{226}$Ra, while for the bottom part they start with the decay of $^{226}$Ra, which allows modeling of the disequilibrium indicated in some of the assay results.

Finally, all energy deposits of the particles reaching the Si chips or any of the 3D Al cavities are recorded.
In the analysis procedure described in section~\ref{sec:hit_processing}, these energy deposits are grouped into physical events according to their timestamps.
Together with the total number of simulated primary decays and the isotope emission rates from the assay results (see also figure~\ref{fig:bulk_assays}), the event rates inside the Si chips and Al cavities can be calculated.

In addition to the bulk material radioactive contaminants identified from the assay results, we simulated contamination accumulated on a material's surface due to radon exposure and ionizing radiation from the SNOLAB rock cavern; these are discussed in the following sections.
Note that we neglect any radioactive contribution from dust accumulation in our studies, because SNOLAB is a controlled class 2000 cleanroom environment.
In addition, all components were wiped upon entering the CUTE cleanroom, which achieves class 200 cleanroom conditions on average \cite{Camus2023CUTE}.

\subsection{Geometry overview}
\label{sec:geometry_overview}

\begin{figure}[ht!]
\begin{center}
\includegraphics[width=\textwidth]{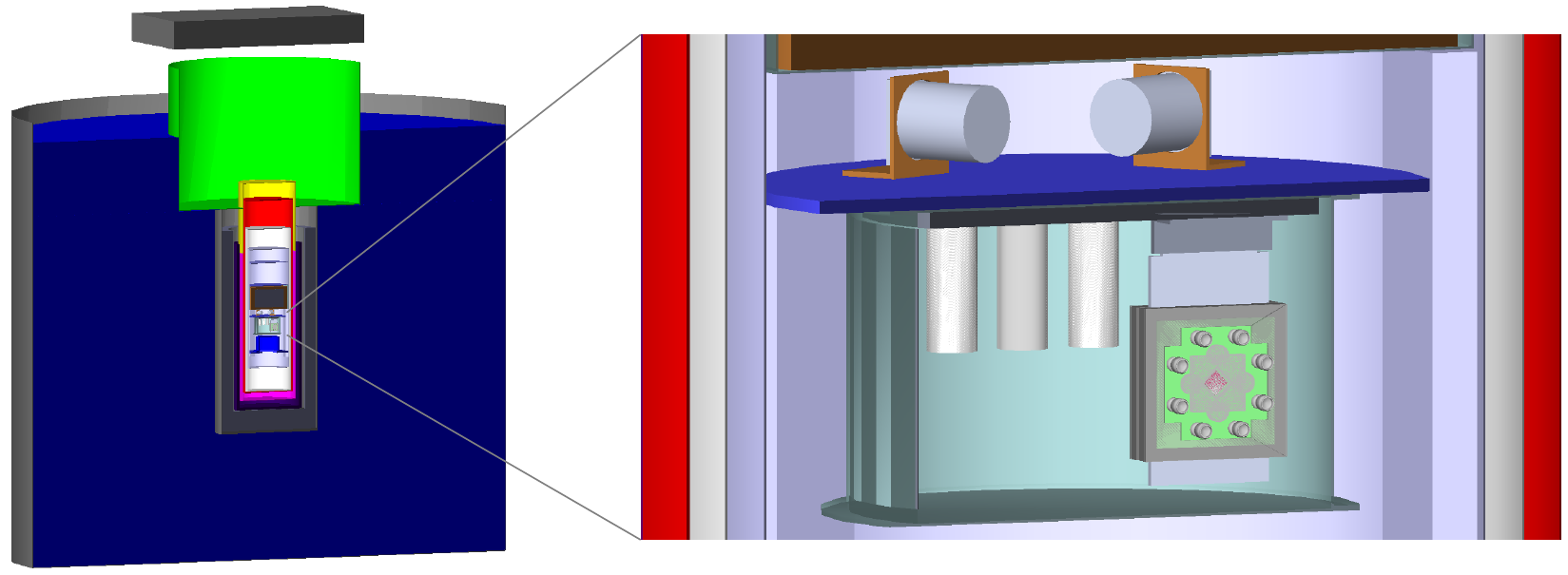}
\end{center}
\caption{Visualization of the CUTE geometry in \textsc{Geant4}. Left: components from outside to inside include: PE block on top (gray), water tank (blue), stainless steel cans (green, yellow, gray), two layers of lead (lilac, dark purple),  mu-metal magnetic shield (dark gray), OVC (pink) and cryogenic stages (50\,K -- red, 4\,K -- white, 1\,K -- light blue, 100\,mK -- mid blue, MC -- blue). The internal lead shield (dark gray) is encased in copper (brown) and a CM shield (mint green). Right: above the QUTEbits stack, there are an additional MC plate (blue) and microwave switches (gray, brown). The QUTEbits stack is enclosed by a CM shield (mint green) and hosts Al cavities (light gray) mounted to the top. The OQTO holder (dark gray) contains the PCB (green) and the Si chip (red), which is partly visible through a transparent lid.}
\label{fig:simulation_geometry}
\end{figure}

The QUTEbits setup is deployed into CUTE at SNOLAB \cite{Camus2023CUTE}.
A visualization of the CUTE shielding and its cryogenic stages is shown in figure~\ref{fig:simulation_geometry}.
Our \textsc{Geant4} geometry model also contains a rough representation of the SNOLAB rock cavern around the CUTE facility.

The CUTE shielding consists of a water tank and a polyethylene (PE) block on top of it to moderate and absorb neutrons emitted by the cavern walls.
Stainless steel cans are submerged into the water tank forming a drywell that hosts the inner shielding as well as mechanical and electronics parts.
Two layers of lead with a total thickness of about 11\,cm absorb high-energy $\gamma$ rays from the cavern walls and other external materials.
A mu-metal shield is present between them to protect sensitive electronics from environmental magnetic fields.

The CUTE cryogenic stages have an outer vacuum chamber (OVC) made of stainless steel at the outside, and several cryogenic stages made of copper: 50\,K, 4\,K, 1\,K, 100\,mK and 10\,mK.
The coldest stage is denoted as the mixing chamber (MC).
Inside the cryostat, there is an internal lead shield above the experimental payload which is thermally anchored to the 1\,K stage.
It serves as an absorber for $\gamma$ rays from electronics located at warmer stages and the cryostat itself.
As lead typically contains several radioactive contaminants, it is encased in radiopure copper. 

The QUTEbits sample space at the MC is fully enclosed in a CM shield.
It is designed to hold up to two OQTO sample holders \cite{OQTOWeb} mounted back-to-back on a copper plate (see figure~\ref{fig:simulation_geometry}).
Each OQTO holder hosts one Si chip with a dimension of $7\times7\times0.5$\,mm$^3$ (mass of 57.1\,mg) and a 300\,nm thick Al circuit layer.
These Si chips are patterned with different quantum circuit designs and are provided by Chalmers University of Technology and the Institute for Quantum Computing (IQC) at the University of Waterloo.
The circuit mask is not relevant for the \textsc{Geant4} radiation transport simulations, but it matters when studying the propagation of charge carriers and lattice vibrations (phonons) with G4CMP (see section~\ref{sec:G4CMP_studies}).

The QUTEbits setup can also host up to six Al cavities acting as 3D resonators.
Inside the CM shield, there are several copper parts supporting the Al cavities and OQTO holders, ensuring thermal conductivity to the MC stage.
The internal lead shield above the payload can be surrounded by an additional CM shield to isolate any trapped magnetic flux from the lead block when it transitions to its superconducting state at a critical temperature of $T_c = 7.2$\,K.

Additionally, there are several small components which do not have a physical geometry in the simulation, because their attenuation of radiation is negligible compared to other volumes.
These include microwave connectors, cables and adapters as well as screws and washers, but also the IR filters at the OQTO holder and glue spots on the Si chip.
To take into account their radioactive contributions, these contaminants are mapped to a representative volume and if applicable limited to a certain region of the volume.

In each QUTEbits run, a different configuration of the internal assembly might be used.
In particular, there could be runs with two OQTO holders and no Al cavities, or with just one OQTO holder and up to six Al cavities.
These modest changes in the setup can be taken into account by switching off components in the overall background budget calculation. 
For the following sections, the full setup described above and shown in figure~\ref{fig:simulation_geometry} is assumed.

\subsection{\textsc{Geant4} detector hit processing}
\label{sec:hit_processing}

This section describes the output format and event processing applied in our simulation pipeline.
Our \textsc{Geant4} application saves energy deposits in the sensitive elements (Si chips, Al cavities) using a particle hit collection in a ROOT TTree format \cite{ROOT_NIMA_1997}.
Along with the particle type that created the hit and the amount of deposited energy, we store additional G4Track information such as the hit time with respect to the time when the primary decay took place (defined as start time $t = 0 \equiv t_0$ in our framework). 
Saving the simulation output per particle hit allows for an offline processing without having to rerun large parts of the particle tracking simulation to extract additional G4Track information of interest.

The hit processing output consists of binned energy histograms for each sensitive detector.
The standard hit processing involves the following steps.
First, we sum all energy deposits that occur in the same geometry element within a time period of $\Delta t = 15\,\mu$s in each \textsc{Geant4} event.
This step virtually splits complex \textsc{Geant4} events that may span large time periods, e.g.\ for decay chains with tens of seconds to years in between consecutive decays, into sensible readout time windows that would be recorded in a real-world device as a single topological event.
The split time window $\Delta t$ is informed by crystal dynamics simulations using G4CMP \cite{Kelsey2023G4CMP} in response to radiation hits as discussed in section~\ref{sec:G4CMP_collection}.
By evaluating the particle type that resulted in an energy deposit, the energy spectra observed by the detectors can be split according to the underlying radiation type. 
In our case, we distinguish between $\alpha$ particle hits, $\beta/\gamma$ hits and nuclear recoils induced by neutron interactions or nuclei (see section~\ref{sec:spectral_analysis}).

For the calibration source simulations using $^{252}$Cf discussed in section~\ref{sec:calibration_sources}, there is an additional step in the hit processing which includes some but not all hits arising from the decay chain seeded by $^{252}$Cf. 
By default, \textsc{Geant4} models the full $^{252}$Cf decay chain which includes several long-lived isotopes such as $^{248}$Cm ($T_{1/2}=3.5\cdot 10^5$\,yr) and $^{244}$Pu ($T_{1/2}=8.1\cdot 10^7$\,yr).
Only a very small fraction of these long-lived decay products are likely to contribute to the radiation exposure over the course of the experiment.
To ensure fair consideration of subsequent decays while limiting the radiation exposure to the reasonably observable fraction, we reject hits that occur more than 6 years after the primary $^{252}$Cf decay.
This choice is also intended to mimic the effects of aging.
Over time, the aging of the source alters its emitted spectrum, evolving from an almost pure $^{252}$Cf spectrum (when the source was manufactured) to a more diverse spectrum with significant contributions from decay products with half-lives longer than $T_{1/2}(^{252}\text{Cf}) = 2.6$\,yr.
The 6-yr cutoff corresponds to the approximate age of the source at the time of the planned QUTEbits measurements.
In practice, this additional hit selection does not have a significant impact on our reported rates.

\subsection{Bulk contamination}
\label{sec:bulk_contamination}

Contaminants in the bulk of a material are simulated by uniformly distributing the identified isotopes throughout the volume.
The characteristic decays are then generated by \textsc{Geant4}, the emitted particles are propagated through the geometry and the energy deposits are recorded and analyzed as described in section~\ref{sec:hit_processing}.

If a component is not represented by a dedicated geometry in the simulation, a representative volume has been identified and contaminated.
For instance, for connectors punching through the CM shield, the corresponding part of the CM shield has been contaminated according to the connectors' assay measurements, while for the BF-6 glue at the Si chip, a small surface area at each of the four corners of the Si chip was contaminated.
On the other hand, for the formable non-magnetic cables inside the CM shield, a virtual cylindrical volume with smaller dimensions than the CM shield was defined to homogeneously sample corresponding decay positions.

\begin{figure}[ht!]
\begin{center}
\includegraphics[width=\textwidth]{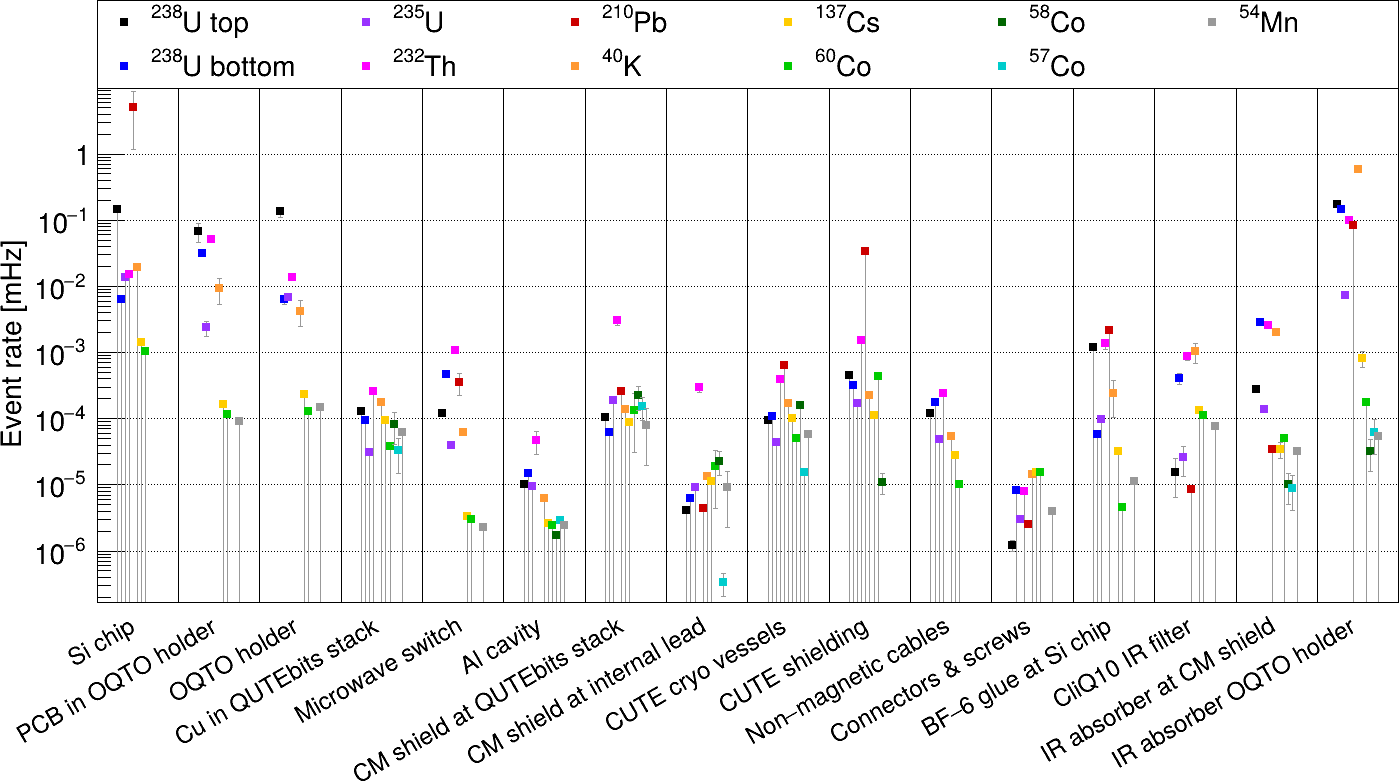}
\end{center}
\caption{Simulated event rates in one of the Si chips according to the assay results and corresponding emission rates for various components. The Si chip itself, as well as components in direct line of sight, are the dominating backgrounds. A line extending from a marker to the bottom of the plot indicates a 90\% C.L.\ upper limit.}
\label{fig:bulk_rates_Si}
\end{figure}

\begin{figure}[ht!]
\begin{center}
\includegraphics[width=\textwidth]{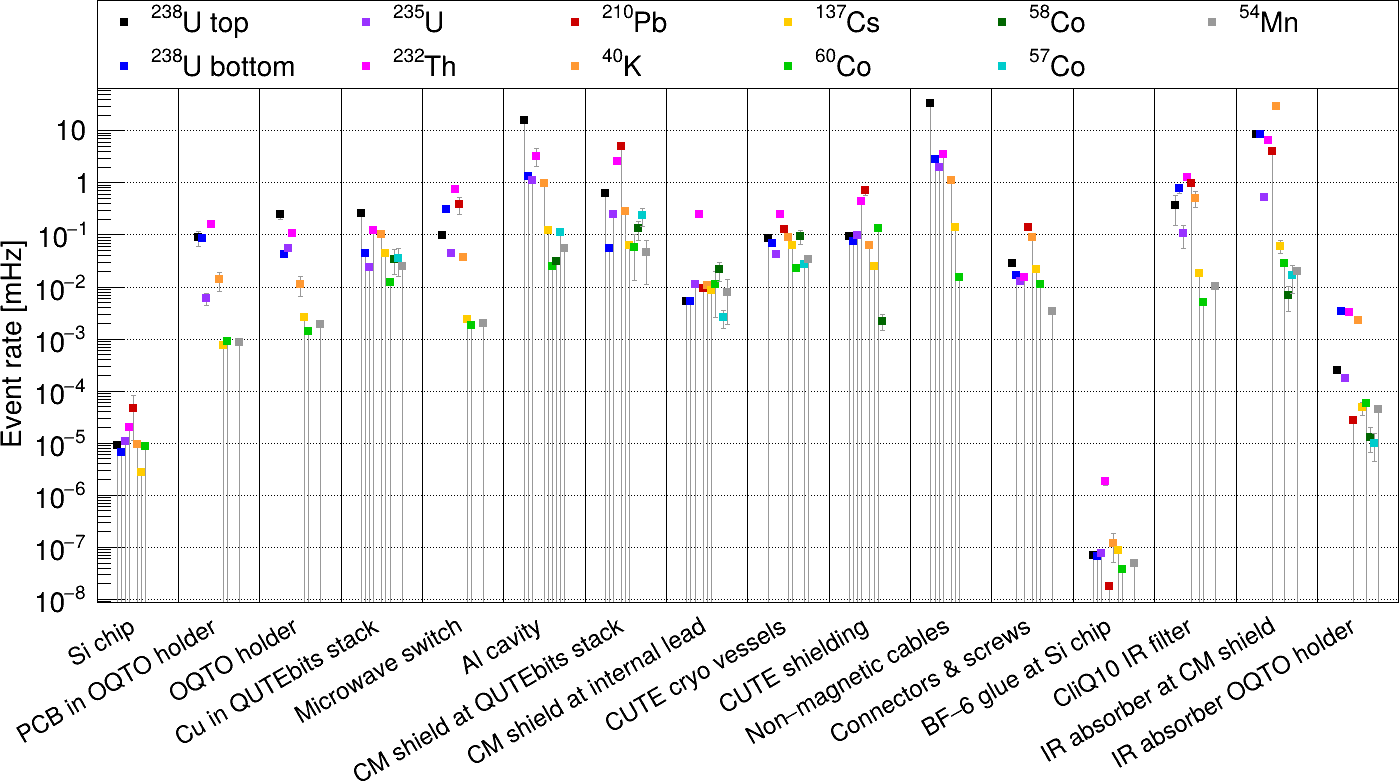}
\end{center}
\caption{Simulated event rates in one of the Al cavities according to the assay results and corresponding emission rates for various components. The dominant background originates from the Al cavity itself and components which are in direct line of sight. A line extending from a marker to the bottom of the plot indicates a 90\% C.L.\ upper limit.}
\label{fig:bulk_rates_AlCav}
\end{figure}

The individual contributions to the overall background rate of each of the 482 combinations of components and corresponding isotopes (see appendix~\ref{sec:computing_resources} for more details) were grouped into component categories.
The resulting event rates for these categories are shown in figure~\ref{fig:bulk_rates_Si} for the example of one of the two Si chips and in figure~\ref{fig:bulk_rates_AlCav} for one of the six Al cavities.
The event rates are very similar for both Si chips and among the six Al cavities, respectively.

The event rate calculations take into account whether an assay result was reported as an upper limit or as a measurement with uncertainties.
The statistical uncertainties from the simulations are generally negligible with the exception of volumes far away from the payload or for isotopes which emit only low-energy $\gamma$ rays.
If no hits were registered in a simulation, a 90\% C.L.\ upper limit is calculated using the Feldman-Cousins method for Poisson signals \cite{Feldman_Cousins_1998}.
Consequently, the event rate of a component which originally had an assay measurement with an uncertainty as shown in figure~\ref{fig:bulk_assays}, may appear as an upper limit in figure~\ref{fig:bulk_rates_Si} or figure~\ref{fig:bulk_rates_AlCav}.

Figure~\ref{fig:bulk_rates_Si} indicates that the dominating background contribution for the Si chip originates from the substrate itself as well as the PCB (see appendix~\ref{sec:potential_components} for a comparison of different PCBs) and the OQTO holder, i.e.\ the components which are in direct contact with it.
While the BF-6 glue is also in direct contact with the Si chip, due to its radiopurity and tiny mass, its contribution is two orders of magnitude lower.

The event rate in the Si chip induced by the connectors and screws is several orders of magnitude below the contribution of the Si chip itself, mainly because none of the screws or connectors are in direct line of sight, but also because they are relatively low in radioactivity.

We also simulated two potential IR absorbers -- one at the CM shield, and one inside the OQTO holder -- but do not include them in the total background projections, because their emission rate was deemed to be too high.
Their background estimates are shown in figures~\ref{fig:bulk_rates_Si} and \ref{fig:bulk_rates_AlCav} and are discussed in appendix~\ref{sec:potential_components} considering different recipes.

The event rates in one of the Al cavities, displayed in figure~\ref{fig:bulk_rates_AlCav}, show similar results.
The largest contributions are from the Al cavities themselves, as well as components in direct line of sight.
For both, the Si chips and the Al cavities, we observe that even though the CUTE shielding layers have by far the highest emission rate (see figure~\ref{fig:bulk_assays}) among all components, their contribution to the background event rate in the devices is at least one order of magnitude lower compared to the contributions from the internal device components.
This validates the design of the CUTE shielding, e.g.\ the high emission rate of the outer lead shield is compensated by its distance to the payload and the presence of the inner shielding layers absorbing most of the emitted $\gamma$ rays.

As pointed out in section~\ref{sec:assay_findings}, it is challenging to make a precise assessment of the $^{210}$Pb contamination in the bulk of a material by using HPGe spectroscopy.
Hence, we report the simulated event rates in table~\ref{tab:bulk_background_rates} for one of the Si chips and Al cavities with and without taking into account the reported $^{210}$Pb bulk assays.
Table~\ref{tab:bulk_background_rates} also shows the composition of the total event rate broken down by particle type.
For the Si chips, the total background rate is one order of magnitude higher when taking into account the $^{210}$Pb bulk assay results, while for the Al cavity the effect is subdominant because the assay of aluminum does not include results for $^{210}$Pb itself.

\begin{table}[ht!]
\centering
\caption{Simulated background rates in one Si chip and one Al cavity from bulk contaminants of all considered components. The rates are reported with and without the contribution of the  $^{210}$Pb assays. In addition, the total rates are broken down by particle type.}
\label{tab:bulk_background_rates}
\begin{tabular}{c|cc|cc}
\toprule
Detector & \multicolumn{2}{c|}{Si chip} & \multicolumn{2}{c}{Al cavity} \\
         & with $^{210}$Pb & without $^{210}$Pb & with $^{210}$Pb & without $^{210}$Pb \\
\midrule
Total rate [mHz]           & 5.57 & $5.51\cdot 10^{-1}$ & $8.50\cdot 10^{1}$ & $7.78\cdot 10^{1}$\\
\midrule
$\alpha$ rate [mHz]        & 1.82 & $1.28\cdot 10^{-1}$ & $2.41\cdot 10^{1}$ & $2.38\cdot 10^{1}$ \\
$\beta/\gamma$ rate [mHz]  & 3.81 & $4.52\cdot 10^{-1}$ & $5.68\cdot 10^{1}$ & $5.01\cdot 10^{1}$\\
NR rate [mHz]              & 5.22 & $2.08\cdot 10^{-1}$ & $4.32\cdot 10^{1}$ & $4.27\cdot 10^{1}$\\
\bottomrule
\end{tabular}
\end{table}

The breakdown per particle type indicates that the total background rate is roughly equally shared by the energy deposits of $\alpha$ particles, electromagnetic interactions from $\beta$ electrons, positrons or $\gamma$ rays and nuclear recoils (NR) induced by neutrons interacting with a nucleus or as consequence to the radioactive decay of a nucleus.
When excluding $^{210}$Pb, this ratio is mostly retained.
In all cases, the $\alpha$ particle induced rate is found to be the lowest.
However, an $\alpha$ decay close to the detectors may deposit significantly more energy compared to the other cases (see section~\ref{sec:spectral_analysis}).
Note that the sum of the rates broken down per particle type is not necessarily equal to the total rate because one simulation event can produce hits of different particle types.

\subsection{Interstitial air}
\label{sec:interstitial_air}

The cavern rock surrounding SNOLAB contains traces of the naturally occurring long-lived isotope $^{238}$U.
While the metals of the decay chain usually stay inside the rock, the gaseous and radioactive radon isotope $^{222}$Rn emanates into the mine air.
As a result, the radon level underground is generally higher than in an above ground laboratory.
SNOLAB has a dedicated ventilation system that provides fresh air to the underground laboratory which reduces the constantly monitored $^{222}$Rn concentration to about $123.1 \pm 6.2\,\unit{Bq/m^3}$ inside the laboratory \cite{Lawson2023Snolab}.

This same level of $^{222}$Rn diffuses into the interstitial air between the CUTE shielding layers. However, the air between the OVC and the inner of the two lead shield layers is flushed with low-radon air ($< 10\,\unit{Bq/m^3}$) \cite{Camus2023CUTE}.
In the simulation, the air inside and outside of the CUTE lead shield has been contaminated with $^{222}$Rn to determine the contribution to the overall background rate.
Table~\ref{tab:interstitial_air_background_rates} reports the resulting rates.

\begin{table}[ht!]
\centering
\caption{Simulated background rate from $^{222}$Rn decays in the interstitial air inside and outside of the low-radioactivity lead shield which surrounds the CUTE cryostat.}
\label{tab:interstitial_air_background_rates}
\begin{tabular}{ccc}
\toprule
Detector                    &  Si chip            & Al cavity \\
\midrule
Total rate [mHz]            & $1.39\cdot 10^{-3}$ & $4.55\cdot 10^{-1}$ \\
\midrule
Air inside lead shield [mHz]  & $<4.29\cdot 10^{-4}$ & $<1.95\cdot 10^{-1}$\\
Air outside lead shield [mHz] & $9.57\cdot 10^{-4}$ & $2.60\cdot 10^{-1}$\\
\bottomrule
\end{tabular}
\end{table}

The contribution of the interstitial air to the total background rate is about two orders of magnitude lower than from the bulk contaminants; it does not introduce a significant background rate that needs to be mitigated.
Broken down by particle type, there were no recorded $\alpha$ particle hits, no NR hits in the Si chips, and only a few NR hits in the Al cavities.
The latter contribution is three orders of magnitude below the $\beta/\gamma$ rate and is therefore negligible.

\subsection{Surface contamination}
\label{sec:surface_contamination}

In this section, we perform a dedicated estimate of the $^{210}$Pb activity due to radon exposure under different scenarios.
As noted before in section~\ref{sec:interstitial_air}, the average $^{222}$Rn level underground at SNOLAB is $123.1 \pm 6.2\,\unit{Bq/m^3}$ \cite{Lawson2023Snolab}.
On Earth's surface, the $^{222}$Rn level can widely vary depending on location, season, and ventilation.
In the SNOLAB surface building, it has been measured to be $5.55 \pm 4.44\,\unit{Bq/m^3}$ \cite{SNOLAB_Tech_Ref_Man}, which we use as the reference value for our estimates.

When $^{222}$Rn decays in air, its progeny can deposit (or ``plate out'') onto exposed material surfaces.
The subsequent $\alpha$ decays of $^{218}$Po and $^{214}$Po impart significant momentum to their nuclear decay products such that they can be implanted into the material.
Further down the decay chain is the long-lived $^{210}$Pb (half-life of 22.2\,yr).
Exposure to radon in air therefore results in the accumulation of $^{210}$Pb surface contamination through these plate-out and implantation processes.

With a half-life of 3.8 days, $^{222}$Rn can easily diffuse in air due to Brownian motion and ventilation; the volume distribution is well-approximated as uniform.
Depending on the ventilation rate and the size of the room, a so-called plate-out height\footnote{Height of air above a surface from which it is expected that all radon progeny adhere to the surface below.} can be determined and used to estimate the $^{210}$Pb surface contamination rate \cite{Nero_2008}.
For the area at SNOLAB where CUTE is located, the plate-out height for copper was measured to be $36.3 \pm 0.8\,\unit{cm}$ \cite{Stein_2018}.
Because the CUTE cleanroom is supplied with low-radon air with $< 10\,\unit{Bq/m^3}$ of $^{222}$Rn \cite{Camus2023CUTE}, the $^{210}$Pb accumulation inside the cleanroom can be neglected, and only the $^{222}$Rn exposure in the underground laboratory and on Earth's surface is taken into account. 

The $^{210}$Pb surface contamination $A_\text{Pb}$ can be estimated as follows:
\begin{equation}
A_\text{Pb}(t) = A_\text{Rn} \cdot h \cdot s \cdot \left( 1 - \mathrm e^{-\lambda_\text{Pb}\cdot t} \right) \quad \text{with} \quad \lambda_\text{Pb} = \frac{\ln 2}{T_{1/2}\left(^{210}\text{Pb}\right)}.
\end{equation}
where $A_\text{Rn}$ is the radon concentration in air and is assumed to be constant, $h$ is the plate-out height taken from \cite{Stein_2018}, and $s$ is the exposed surface area of a component.
Table~\ref{tab:Rn_exposure_scenarious} summarizes the total emission rates from $^{210}$Pb surface contamination for three radon exposure scenarios, summed over all considered components.
Because of the long half-life of $^{210}$Pb, the emission rates and corresponding simulated background rates depend approximately linearly on the exposure time.

\begin{table}[ht!]
\centering
\caption{Radon exposure scenarios assuming different time spans for materials being stored on Earth's surface and inside SNOLAB. The total emission rates of $^{210}$Pb accumulated at the components' surfaces were calculated for each of the exposure scenarios. These emission rates were used to scale the simulation results of a $^{210}$Pb surface contamination to obtain an estimate of the expected surface background rates.}
\label{tab:Rn_exposure_scenarious}
\begin{tabular}{cccc}
\toprule
Radon exposure scenario                           & short & medium & long \\
\midrule
Exposure on Earth's surface [years]               & 1     & 2     & 3 \\
Exposure underground at SNOLAB [days]             & 7     & 14    & 21 \\
\midrule
Emission rate from Earth's surface exposure [mBq] & 90.4  & 178.2 & 263.2 \\
Emission rate from SNOLAB exposure [mBq]          & 39.1  & 78.1  & 117.1 \\
Total emission rate [mBq]                         & 129.5 & 256.3 & 380.3\\
\midrule
Surface background rate in Si chip [mHz] & $6.94\cdot 10^{-2}$ & $1.37\cdot 10^{-1}$ & $2.04\cdot 10^{-1}$ \\
Surface background rate in Al cavity [mHz] & 5.17 & $1.02\cdot 10^{1}$ & $1.52\cdot 10^{1}$ \\
\bottomrule
\end{tabular}
\end{table}

Compared to the simulated background rates determined from radiocontaminants in the bulk of the materials, the simulated rate from $^{210}$Pb on component surfaces is about an order of magnitude lower, which is more in line with the other bulk contaminants when neglecting the assay results for $^{210}$Pb.
In the following, the radon exposure scenario with the medium time span is used.

The contribution from $^{210}$Pb on a material's surface is composed of two different categories: $^{210}$Pb adsorbed on the surface, and $^{210}$Pb implanted in the surface.
For the corresponding simulations, we apply the Jacobi model \cite{Jacobi_1972,Nero_2008} to determine the ratio of adsorbed vs.\ implanted $^{210}$Pb.
In combination with \textsc{Geant4} simulations, we found that this ratio is independent of the material, and about $55.8\%$ of the $^{210}$Pb is implanted, while the remaining fraction is adsorbed on the surface (see appendix~\ref{sec:uncertainties} for a discussion of the systematic uncertainties).
For these simulations, as well as for determining the implantation depth profile of $^{210}$Pb, dedicated \textsc{Geant4} simulations were run using the Screened Nuclear Recoil Physics List \cite{Mendenhall_2005}.
This physics list has been added to our \textsc{Geant4} application \cite{Redl_2014} and validated against results obtained with SRIM \cite{Ziegler_2010} for individual ions.
While SRIM is limited to simulating the transport of individual ion species with a fixed momentum direction, \textsc{Geant4} allows simulation of complex decay chains, effectively modeling subsequent implantation of isotopes with randomized nuclear recoil direction.
This eventually leads to an implantation depth profile of $^{210}$Pb which depends on the material properties.
For our study, we determined $^{210}$Pb implantation depth profiles for Si, Cu, Al and mu-metal.

\begin{table}[ht!]
\centering
\caption{Simulated background rates for $^{210}$Pb adsorbed on surfaces and implanted in surfaces using the medium exposure scenario presented in table~\ref{tab:Rn_exposure_scenarious}. The rates were scaled with the determined ratio between adsorbed and implanted $^{210}$Pb.}
\label{tab:surface_background_rates}
\begin{tabular}{c|cc|cc}
\toprule
Detector                       &  \multicolumn{2}{c|}{Si chip}   & \multicolumn{2}{c}{Al cavity} \\
Contamination mode             & adsorbed            & implanted & adsorbed & implanted \\
\midrule
Total rate [mHz]               & $6.69\cdot 10^{-2}$ & $7.05\cdot 10^{-2}$ & 5.74 & 4.50 \\
\midrule
Earth's surface exposure [mHz] & $4.65\cdot 10^{-2}$ & $4.90\cdot 10^{-2}$ & 3.99 & 3.13 \\
SNOLAB exposure [mHz]          & $2.04\cdot 10^{-2}$ & $2.15\cdot 10^{-2}$ & 1.75 & 1.37 \\
\midrule
$\alpha$ rate [mHz]            & $1.05\cdot 10^{-2}$ & $1.63\cdot 10^{-2}$ & 0.63 & 0.91 \\
$\beta/\gamma$ rate [mHz]      & $4.72\cdot 10^{-2}$ & $5.19\cdot 10^{-2}$ & 4.18 & 3.46 \\
NR rate [mHz]                  & $2.25\cdot 10^{-2}$ & $3.31\cdot 10^{-2}$ & 1.84 & 1.92 \\
\bottomrule
\end{tabular}
\end{table}

Table~\ref{tab:surface_background_rates} summarizes the simulated background rates obtained for $^{210}$Pb adsorbed on the surface and when using the implantation depth profiles for implanted $^{210}$Pb.
The total simulated $^{210}$Pb activity is a sum of the contributions due to the radon exposure on Earth's surface and inside SNOLAB.
The breakdown per particle type may result in a higher sum because an event can be composed of multiple particle interaction types.

The contribution of adsorbed and implanted $^{210}$Pb is almost equal, while the exposure on Earth's surface has a larger impact due to the longer exposure time, which compensates the lower radon activity as indicated by the emission rates shown in table~\ref{tab:Rn_exposure_scenarious}.
The breakdown per particle type shows that the rate of $\beta$ particles and $\gamma$ rays is the largest, but the contribution of $\alpha$ particles and NRs is relevant as well.

\subsection{External radiation sources}
\label{sec:external_sources}

The muon flux is strongly suppressed 2\,km underground at SNOLAB and has been measured to be roughly two muons per square meter per week \cite{SNO_2009}.
Taking into account the tiny size of our Si chip and its vertical orientation which reduces the effective area exposed to vertically incoming particles, the background introduced by muons is completely negligible. 

The cavern rock surrounding SNOLAB contains traces of $^{40}$K and the natural decay chains of $^{238}$U and $^{232}$Th, which all emit $\gamma$ rays.
The natural decay chains also emit $\alpha$ particles which can produce neutrons via $(\alpha,n)$ reactions in the rock itself.
Additionally, for some isotopes in the decay chains there is a small probability that neutrons are produced by spontaneous fission.
The thermal neutron flux at SNOLAB was measured as $4144.9 \pm 49.8(\text{stat.}) \pm 105.3(\text{syst.})\,\unit{n/m^2/day}$ and the fast neutron flux was estimated to be roughly $4000\,\unit{n/m^2/day}$ \cite{SNOLAB_Tech_Ref_Man}.
The cavern $\gamma$ ray flux has been measured in SNOLAB's J-drift to be about $4.25 \cdot 10^4 \,\unit{\gamma/m^2/s}$ \cite{Robinson_2015}.

To determine the background rate introduced by cavern $\gamma$ rays and neutrons in our \textsc{Geant4} simulation framework, we draw primary particle configurations from the known energy distributions of these particle species and start particle tracks from inside the cavern wall around the CUTE facility.
We simulated $3\cdot10^{12}$ primary $\gamma$ rays and $10^9$ primary neutrons.
Because of the CUTE facility's extensive shielding, no hits were observed in any of the Si chips or Al cavities.
This null result yields a 90\% C.L.\ upper limit on the expected background rate of $<8.16\cdot 10^{-3}\,\unit{mHz}$ in a single Si chip or Al cavity, which breaks down into $<8.14\cdot 10^{-3}\,\unit{mHz}$ from $\gamma$ rays and $<1.81\cdot 10^{-5}\,\unit{mHz}$ from neutrons.
Since these background rates are well below the rate from bulk contaminants, they are deemed negligible.

\section{Background projection results}
\label{sec:background_projection}

This section summarizes the outcome of our background and calibration source simulation studies.
Table~\ref{tab:BG_rates_summary} reports the results of the previously discussed background categories and the composition of the estimated total background rate for one of the Si chips and one of the Al cavities.
The bulk material contaminants inferred from the assay results dominate, while the contribution of the interstitial air in the CUTE shield is negligible.
The assumptions made for the radon exposure leading to an excess of $^{210}$Pb on the materials' surfaces indicate that the accumulated $^{210}$Pb activity may cause a rate on the same order of magnitude as the bulk contaminants for very thin volumes such as the Si chips, and an order of magnitude below the bulk contaminants for larger volumes like the Al cavities.
Because the emitted particles of the $^{210}$Pb decay chain have rather low energies, their penetration power is poor, thus the $^{210}$Pb contribution is crucial for thin volumes, but subdominant for thicker volumes.
In all cases, the expected rate introduced by $\gamma$ rays or neutrons emitted from the cavern rock is much lower than the intrinsic bulk contamination of the materials proving the efficiency of CUTE's multilayered radiation shield.

Note that we omit the simulated bulk $^{210}$Pb rates corresponding to the $^{210}$Pb-specific HPGe assay results.
The relatively low sensitivity of the $^{210}$Pb-specific assays provides low confidence in the measurements.
Instead, we infer the bulk $^{210}$Pb contribution from the HPGe assays of the bottom part of the $^{238}$U decay chain assuming secular equilibrium.
We also include the $^{210}$Pb surface contamination from radon exposure as discussed in section~\ref{sec:surface_contamination}.

\begin{table}[ht!]
\centering
\caption{Summary of simulated background contributions. The total rate is composed of the four considered categories: bulk contaminants according to the assay results with $^{210}$Pb in secular equilibrium with $^{238}$U, interstitial air in the CUTE shield, $^{210}$Pb accumulated on material surfaces due to radon exposure and $\gamma$ rays and neutrons emitted from the SNOLAB cavern rock.}
\label{tab:BG_rates_summary}
\begin{tabular}{ccc}
\toprule
Detector                   &  Si chip             & Al cavity \\
\midrule
Total rate [mHz]           & $6.98\cdot 10^{-1}$  & $8.85\cdot 10^{1}$ \\
\midrule
Bulk [mHz]  & $5.51\cdot 10^{-1}$  & $7.78\cdot 10^{1}$ \\
Interstitial air [mHz]     & $1.39\cdot 10^{-3}$  & $4.55\cdot 10^{-1}$ \\
Surface $^{210}$Pb [mHz]   & $1.37\cdot 10^{-1}$  & $1.02\cdot 10^{1}$ \\
Cavern rock [mHz]          & $<8.16\cdot 10^{-3}$ & $<8.16\cdot 10^{-3}$ \\
\midrule
$\alpha$ rate [mHz]        & $1.55\cdot 10^{-1}$  & $2.53\cdot 10^{1}$ \\
$\beta/\gamma$ rate [mHz]  & $5.52\cdot 10^{-1}$  & $5.82\cdot 10^{1}$ \\
NR rate [mHz]              & $2.64\cdot 10^{-1}$  & $4.65\cdot 10^{1}$ \\
\bottomrule
\end{tabular}
\end{table}

The breakdown per particle type shows that while the contribution of $\beta$ particles and $\gamma$ rays is higher than from $\alpha$ particles or NRs, all particle types are relevant.
The sum of the individual particle rates is larger than the total background rate, which indicates that some events are composed of more than one particle interaction type.

A discussion of the uncertainties of the background rates shown in table~\ref{tab:BG_rates_summary} and the fraction of upper limits contributing to the total rate can be found in appendix~\ref{sec:uncertainties}.

\subsection{Spectral component analysis}
\label{sec:spectral_analysis}

\begin{figure}[ht]
    \centering
    \includegraphics[width=1.0\linewidth]{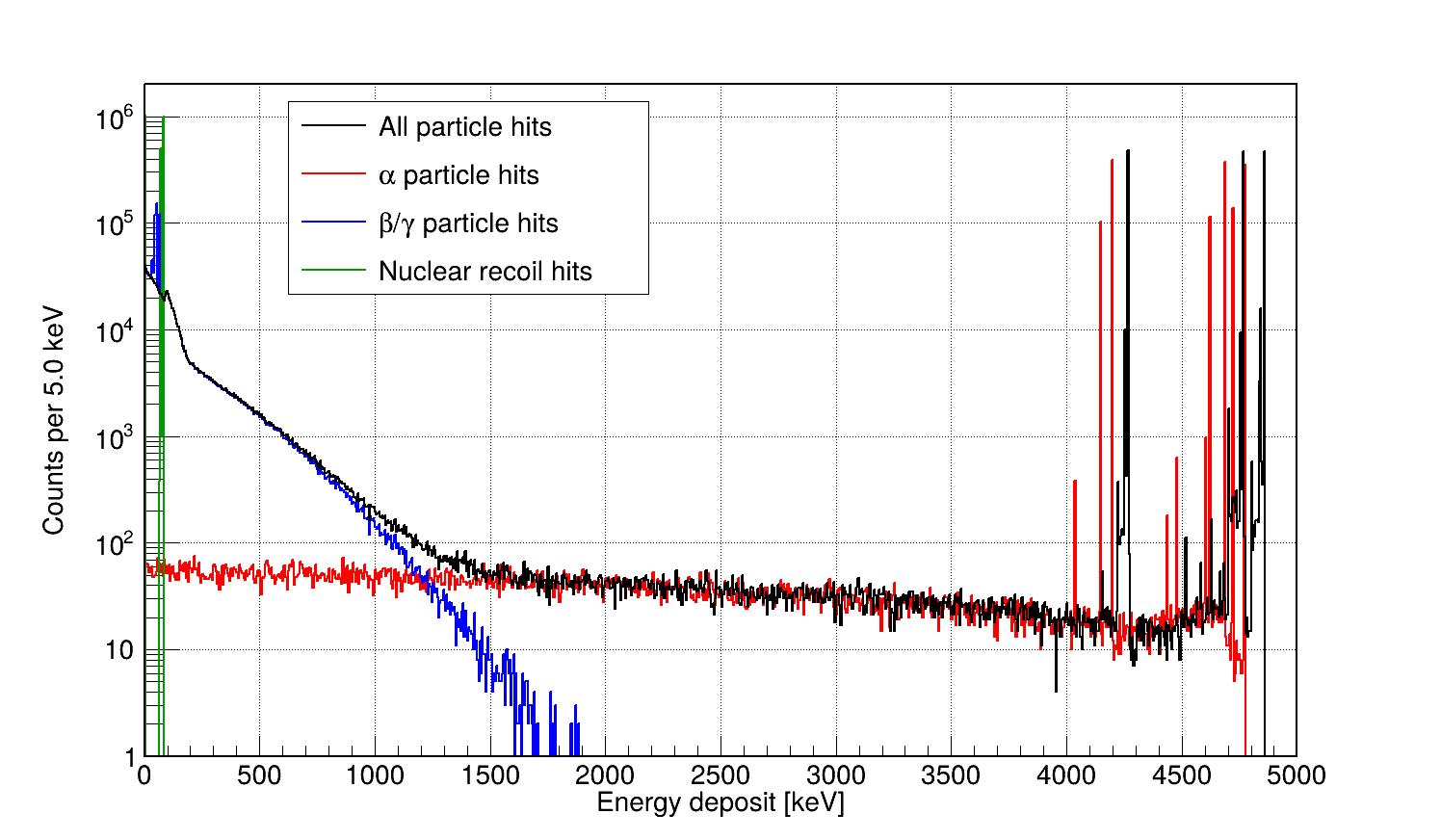}
    \caption{Recorded spectrum of the top part of the $^{238}$U decay chain simulated as bulk contamination in one of the Si chips. The combined spectrum of all particle hits (black) is broken down into different particle interaction types: $\alpha$ particles (red), $\beta/\gamma$ particles (blue), nuclear recoils (green). The displacement between the full-energy $\alpha$ peaks in the combined spectrum compared to the \mbox{$\alpha$ particle} spectrum is due to the separation of the nuclear recoil contribution.}
    \label{fig:U238_Si_chip}
\end{figure}

In addition to characterizing the rate of ionizing radiation hits in the simulation, we are also interested in the energy spectrum of the expected background radiation field and its composition.
Figure~\ref{fig:U238_Si_chip} illustrates the spectrum observed by one of the Si chips for the simulated decays of the top part of the $^{238}$U decay chain in the bulk of the Si substrate.

The spectrum of all particles hits, grouped into topological events as discussed in section~\ref{sec:hit_processing}, is dominated by the full-energy absorption of $\alpha$ particles emitted by the decays of $^{238}$U ($Q_\alpha = 4270$\,keV), $^{234}$U ($Q_\alpha = 4858$\,keV) and $^{230}$Th ($Q_\alpha = 4770$\,keV) following the decay sequence:

\begin{equation}
    {^{238}\text{U}} \overset{\alpha}{\longrightarrow} {^{234}\text{Th}} \overset{\beta^-}{\longrightarrow} {^{234}\text{Pa}} \overset{\beta^-}{\longrightarrow} {^{234}\text{U}} \overset{\alpha}{\longrightarrow} {^{230}\text{Th}} \overset{\alpha}{\longrightarrow} {^{226}\text{Ra}}.
\end{equation}

The highest-energy lines in the spectrum correspond to the $\alpha$ decays transitioning into the ground state of the progeny nuclides.
The line features are caused by the full-energy absorption of $\alpha$ particles carrying the kinetic energy $E_\alpha = Q_\alpha - E_\text{NR}$, with $Q_\alpha$ being the energy released in the nuclear reaction (Q-value) and $E_\text{NR}$ the kinetic energy transferred to the decay nucleus.
Additional lines at lower energies than $Q_\alpha$ are visible for transitions into excited states which involve the emission of discrete $\gamma$ rays and typically a lower decay probability than the ground state transition.

The endpoint of the $\beta/\gamma$ component of the spectrum agrees with the Q-value of the $\beta^-$ decay of $^{234}$Pa ($Q_\beta = 2194$\,keV).
Because of the small size of the Si chip, there is a moderate chance for high-energy $\beta$ electrons to escape the Si substrate without depositing their full kinetic energy, which is why the recorded spectrum does not extend all the way to the Q-value of $^{234}$Pa with the simulated statistics.
At lower energies, the $\beta/\gamma$ spectrum is a superposition of the two $\beta^-$ decays present in the chain from $^{234}$Pa and $^{234}$Th ($Q_\beta = 274$\,keV).

The energy range in between the $\alpha$ peaks and the endpoint of the combined $\beta/\gamma$ spectrum is populated by $\alpha$ particle hits that deposit only a fraction of $Q_\alpha$.
For $\alpha$ decays that are simulated close to the surface of the Si chip, the $\alpha$ particle can escape the volume without depositing its full kinetic energy.
This partial energy deposit occurs for about 2.1\% of the simulated events in this particular configuration.

In the energy range from 30\,keV to 80\,keV, characteristic X-ray and $\gamma$ lines are visible in the $\beta/\gamma$ spectrum.
Similarly, the energy spectrum of nuclear recoils shows distinct lines for the recoiling nuclei from the present $\alpha$ decays of $^{238}$U ($E_\text{NR} = 70$\,keV), $^{234}$U ($E_\text{NR} = 81$\,keV) and $^{230}$Th ($E_\text{NR} = 81$\,keV).
These discrete lines are not visible in the total spectrum of all particle hits because there are usually multiple hits of different particle types that get grouped into the same topological event characterized by the sum of all energy deposits within $\Delta t = 15\,\mu$s (see section~\ref{sec:hit_processing}).
Because of the separation of particle type interactions, the $\alpha$ peaks in the $\alpha$ spectrum appear at the kinetic energy of the emitted $\alpha$ particles $E_\alpha$ while they appear at $Q_\alpha = E_\alpha + E_\text{NR}$ in the total spectrum of all particle hits.

Based on the total rate estimates reported in table~\ref{tab:BG_rates_summary}, we do not expect significant contributions from external background radiation sources such as the SNOLAB rock cavern or from radon decays in the interstitial air inside the CUTE lead shield.
These contributions will henceforth be neglected.

\begin{figure}[ht!]
    \centering
    \includegraphics[width=1.0\linewidth]{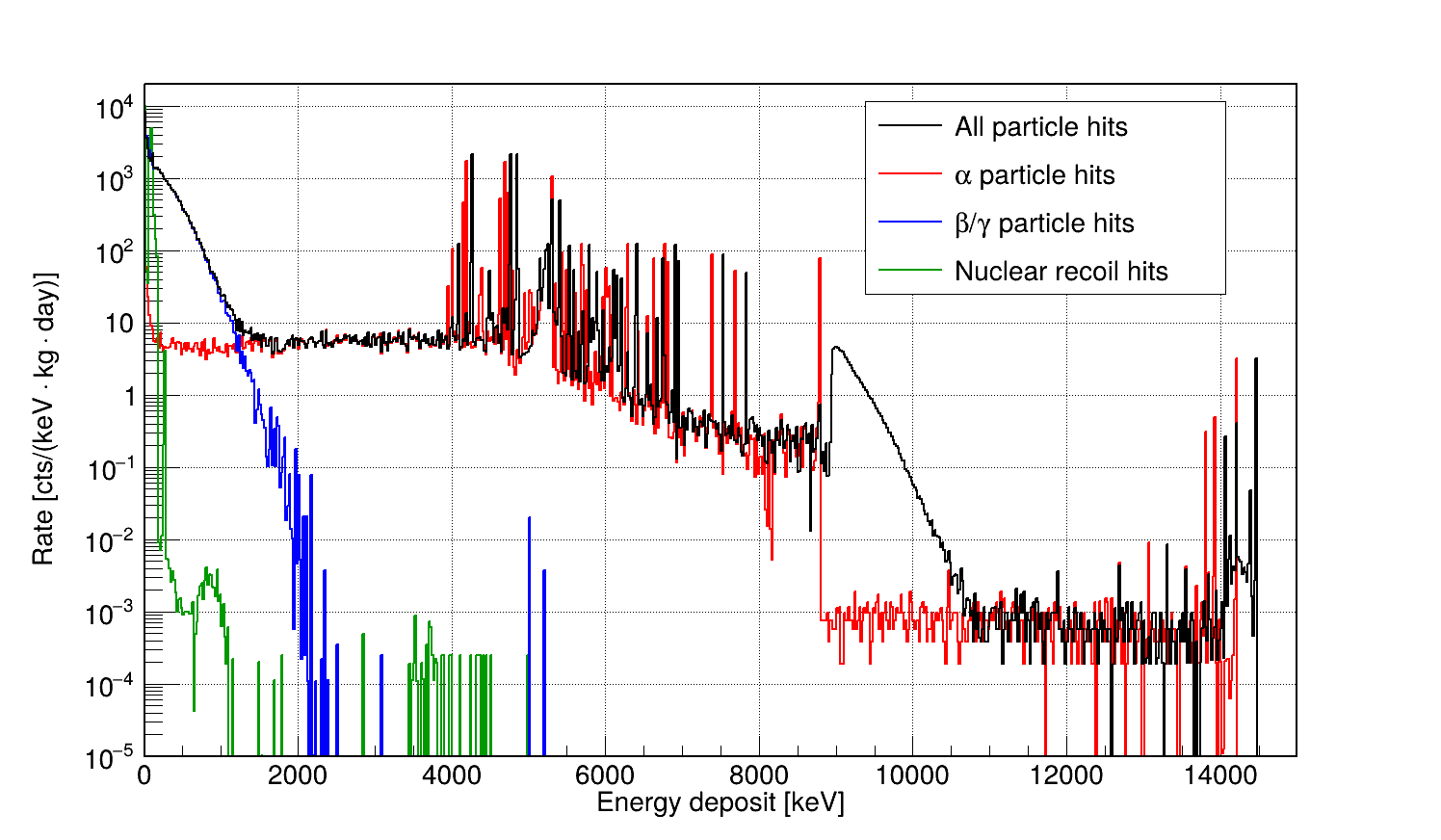}
    \caption{Projected total background spectrum of all simulated radioisotopes from bulk and surface contaminants for one of the Si chips. The total spectrum (black) is composed of all particle hit types whereas the other histograms show the contributions of specific particle type interactions: $\alpha$ particles (red), $\beta/\gamma$ particles (blue), nuclear recoils (green). The spectral features are described in the text. Each spectrum has been normalized to decay rate units.}
    \label{fig:BG_total_particles}
\end{figure}

Figure~\ref{fig:BG_total_particles} depicts the projected total background spectrum in one of the Si chips for all the simulated radioisotopes arising from bulk and surface contaminants.
The total spectrum is broken down into particle type interactions following the single bulk contaminant example presented in figure~\ref{fig:U238_Si_chip}.  
Each spectrum has been normalized to standardized decay rate units to report the simulated event rate in terms of counts per energy bin, device mass and live time.
The integral over the normalized spectra multiplied by the energy bin width of 20\,keV, the device mass of 57\,mg and conversion from days to seconds yields the background rates reported in table~\ref{tab:BG_rates_summary}.

\begin{figure}[ht!]
    \centering
    \includegraphics[width=1.0\linewidth]{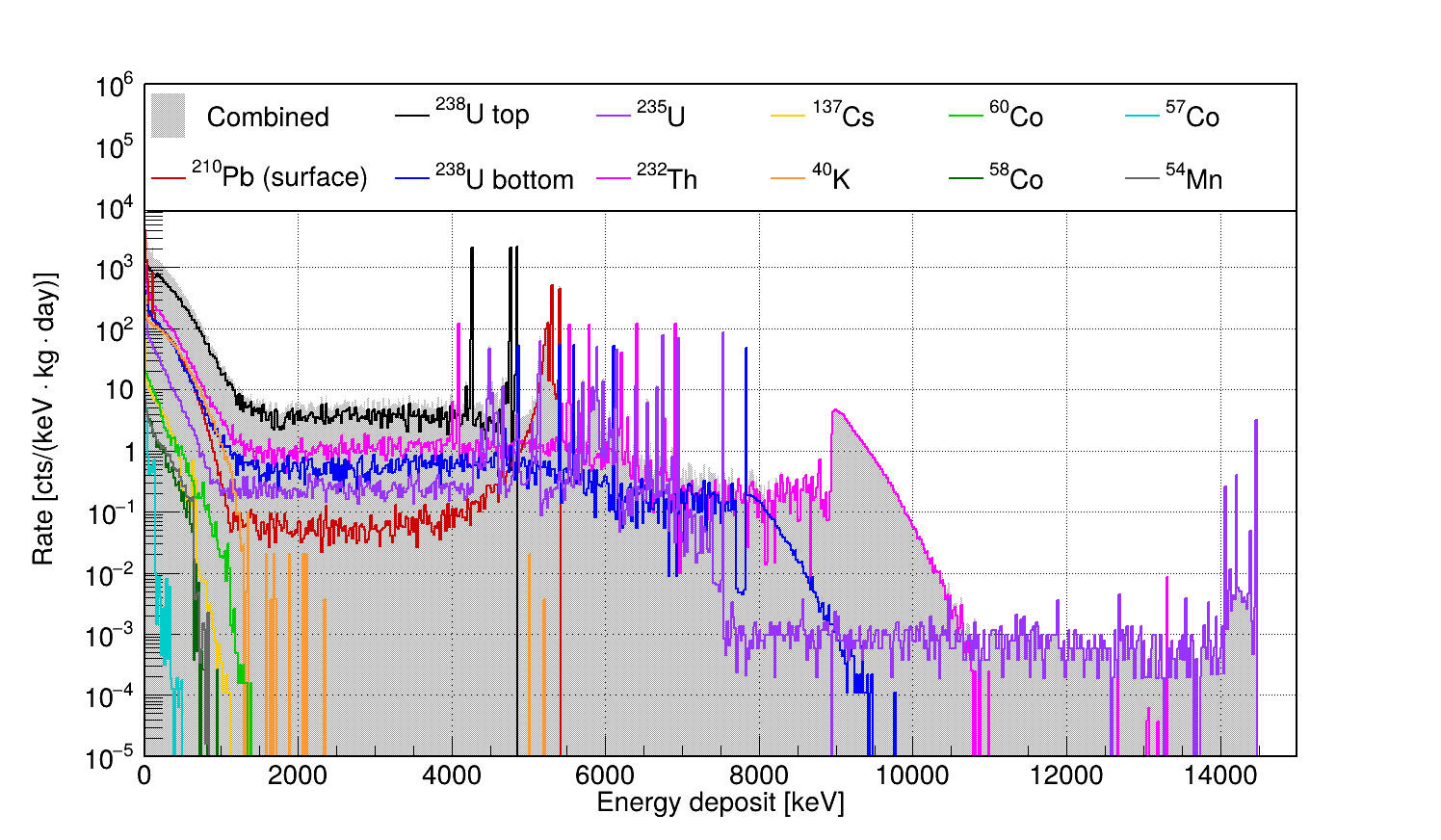}
    \caption{Isotopic composition of the total background projection for one of the Si chips. The total spectrum (gray filling) is the sum of the individual radioisotope spectra (colored histograms). Each spectrum has been normalized to decay rate units.}
    \label{fig:BG_total_isotopes}
\end{figure}

The same characteristic features as discussed for the example shown in figure~\ref{fig:U238_Si_chip} are present in the total background projection in figure~\ref{fig:BG_total_particles}.
From the highest energies observed in the simulation down to about 1.5\,MeV, $\alpha$ decays dominate the spectrum both in intensity and rate.
The highest-energy deposits are caused by a summation of two subsequent $\alpha$ decays within the $^{235}$U decay chain (see also figure~\ref{fig:BG_total_isotopes}): $^{219}$Rn with the most probable emission of $E_\alpha = 6819$\,keV and $^{215}$Po with $E_\alpha = 7386$\,keV yield a summed energy of $E_\alpha = 14205$\,keV.
This summation effect is a consequence of the split time window $\Delta t = 15\,\mu$s applied in our processing of the \textsc{Geant4} events discussed in section~\ref{sec:hit_processing}.
Because the half-life of $^{215}$Po is only 1.78\,ms, there is a small likelihood for this decay to occur within $\Delta t$ with respect to the previous decay in the chain from $^{219}$Rn.
The lower $\alpha$ peaks in that high-energy region above 10\,MeV are summation peaks involving less probable transitions into excited states of the progeny nuclei.
The flat region below the summation peak at $E_\alpha = 14205$\,keV and the highest single $\alpha$ decay line at $E_\alpha = 8954$\,keV from $^{212}$Po is caused by coincident $\alpha$ decays in which one of the particles does not deposit its full energy in the substrate.

Another summation effect is visible in the total background spectrum with the sawtooth-like feature around 9\,MeV.
This feature is attributed to the delayed coincidence decay of $^{212}$Bi ($Q_\beta = 2252$\,keV) into $^{212}$Po ($Q_\alpha = 8954$\,keV) as part of the $^{232}$Th decay chain.
The half-life of $^{212}$Po is $0.3\,\mu$s and much shorter than $\Delta t = 15\,\mu$s resulting in a high chance of summing up the energy deposits from the discrete $\alpha$ particle energy from the polonium decay and the continuous $\beta$ spectrum of the bismuth isotope.
A similar decay sequence is found in the bottom part of the $^{238}$U decay chain starting with $^{214}$Bi ($Q_\beta = 3269$\,keV) which decays into $^{214}$Po ($Q_\alpha = 7833.54$\,keV).
The latter isotope $^{214}$Po has a half-life of $164\,\mu$s which is ten times longer than $\Delta t$, but still results in a summation feature visible in the $^{238}$U bottom part spectrum in figure~\ref{fig:BG_total_isotopes}.
However, in the total background spectrum the $^{214}$Bi-$^{214}$Po coincidence is subdominant.
The $^{214}$Bi-$^{214}$Po coincidence would be more prominent in the total spectrum for a much larger value of the split time $\Delta t$.

Below the lowest visible $\alpha$ decay peak from $^{232}$Th ($E_\alpha = 3947$\,keV), degraded energy deposits by $\alpha$ particles make up about 13\% of the total $\alpha$ particle rate.
The $\beta/\gamma$ rate is dominated by the $\beta^-$-decay of $^{234}$Pa  with $Q_\beta = 2194$\,keV from the top part of the $^{238}$U decay chain.
Below $Q_\beta$ of $^{234}$Pa, the integrated $\beta/\gamma$ rate is about forty times higher than the integrated rate over the same energy range in the $\alpha$ particle spectrum.

Another breakdown of the total background spectrum is shown in figure~\ref{fig:BG_total_isotopes} where the contributions of the individual radioisotopes and their decay chains are highlighted.
The main contributions arise from the bulk contamination with the primordial decay chain elements $^{238}$U (lower energy range dominated by top part, mid energy range by bottom part), $^{232}$Th (mid to high energy range) and $^{235}$U (high energy range).
The spectrum of the bottom part of the $^{238}$U chain clearly shows the $^{214}$Bi-$^{214}$Po summation feature around 8\,MeV.
Another significant contribution arises from the $^{210}$Pb surface contamination.
The remaining radioisotopes contribute roughly equally to the total spectrum at low energies.

\subsection{Comparison to radioactive calibration sources}
\label{sec:calibration_sources}

In addition to characterizing the ambient background radiation level for the QUTEbits project, we use the same simulation framework to study the prospects of using available calibration sources at CUTE to expose the quantum devices to a controlled level of ionizing radiation above the expected background level.
Similar studies using radioactive calibration sources were reported in Ref.~\cite{Vepsalainen2020decoherence, Larson2025qppoisoning} using a strong $^{60}$Co source, in Ref.~\cite{Cardani2021quantumUG} using $^{232}$Th and Ref.~\cite{Celi2026QPdynamics} using $^{137}$Cs.

At CUTE there are calibration sources for $\gamma$ ray calibration using a $^{133}$Ba source and neutron exposure using a $^{252}$Cf source.
Both sources have a nominal activity of $37.0\pm 5.6$\,kBq, which has reduced to about 22.5\,kBq for $^{133}$Ba and to about 7.6\,kBq for $^{252}$Cf taking into account the half-lives and reference manufacturing dates of the sources for a projected date of use in April 2026.
More information on the CUTE calibration systems can be found in \cite{Camus2023CUTE}.

The estimate of the calibration source event rates follows analogously to the steps described in section~\ref{sec:radiation_modeling}.
For the sources themselves, a source container assembly is placed at different positions in the geometry resembling CUTE's calibration systems.
The inner volume of the source container is a sphere of 2\,mm diameter which is contaminated with the radioisotope of interest to simulate the decay physics and particle tracking with \textsc{Geant4}.
The simulation output is then processed as described in section~\ref{sec:hit_processing} with the distinction that we split the $^{252}$Cf detector hits into electron recoils (ERs) and nuclear recoils (NRs) in order to separately assess the neutron induced event rate.

\begin{figure}[ht]
    \centering
    \includegraphics[width=0.3\linewidth]{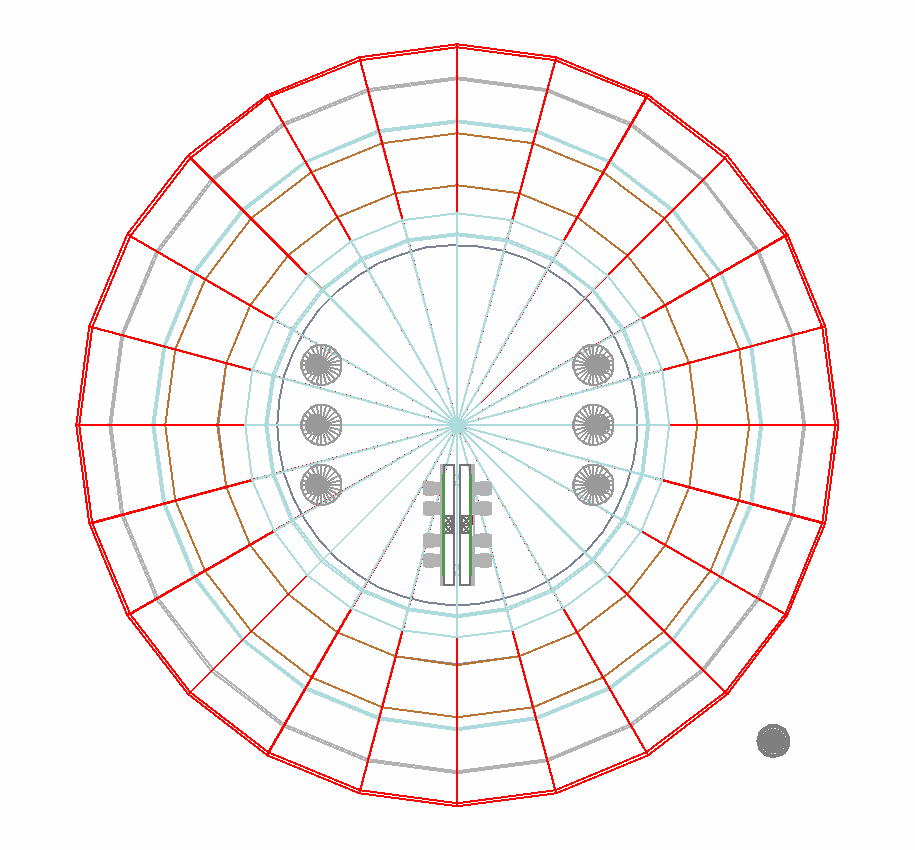} \hspace{0.2cm}
    \includegraphics[width=0.3\linewidth]{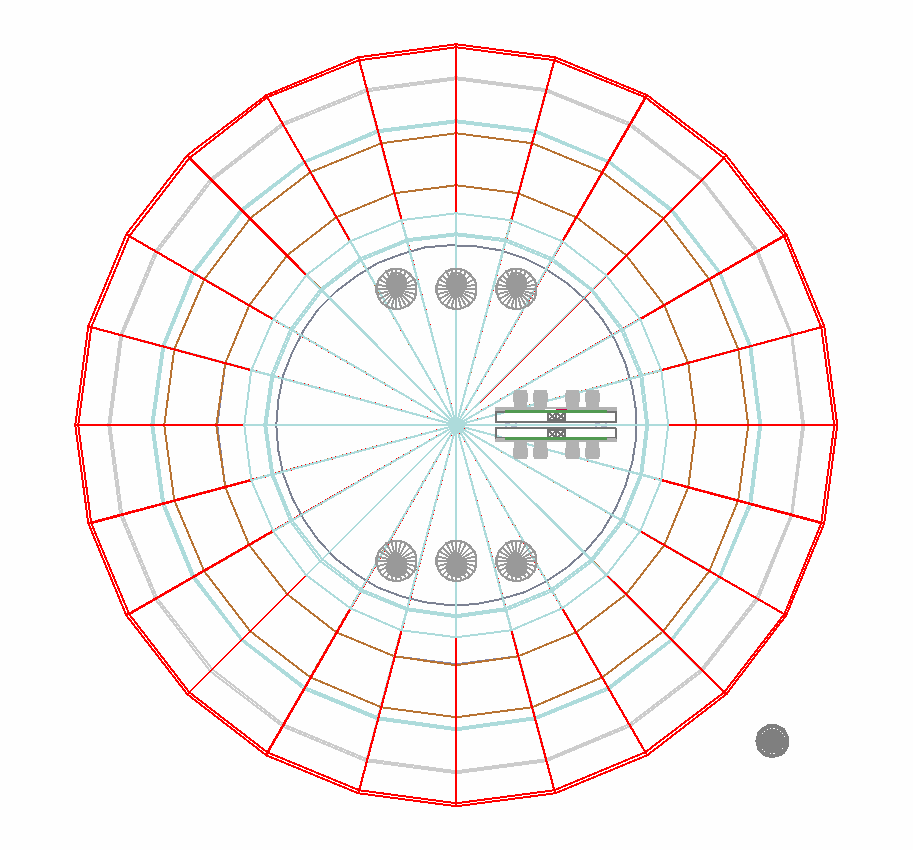}  \hspace{0.2cm}
    \includegraphics[width=0.3\linewidth]{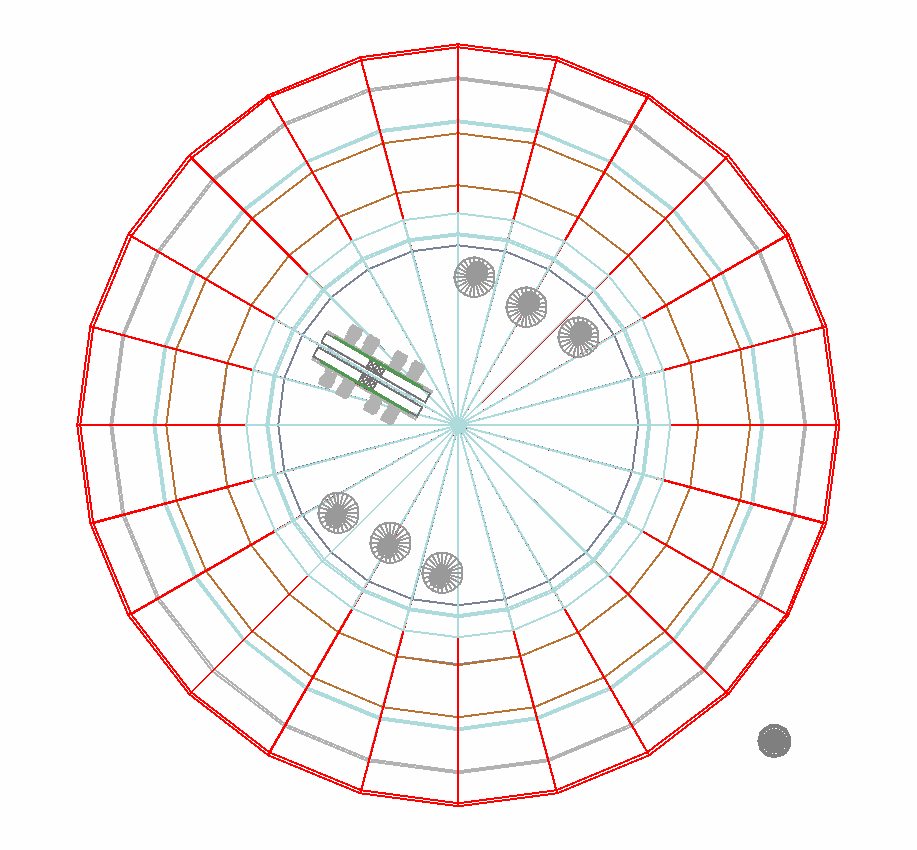} 
    \caption{\textsc{Geant4} visualization of different rotations of the QUTEbits payload with respect to the $^{133}$Ba source position (lower right). From left to right the rotations are referred to as $\alpha_\text{rot} = 90^\circ$, $180^\circ$ and $330^\circ$. The visualization only displays some of the geometry elements for better visibility.}
    \label{fig:Ba_rotation}
\end{figure}

\begin{figure}[ht]
    \centering
    \includegraphics[width=1.0\linewidth]{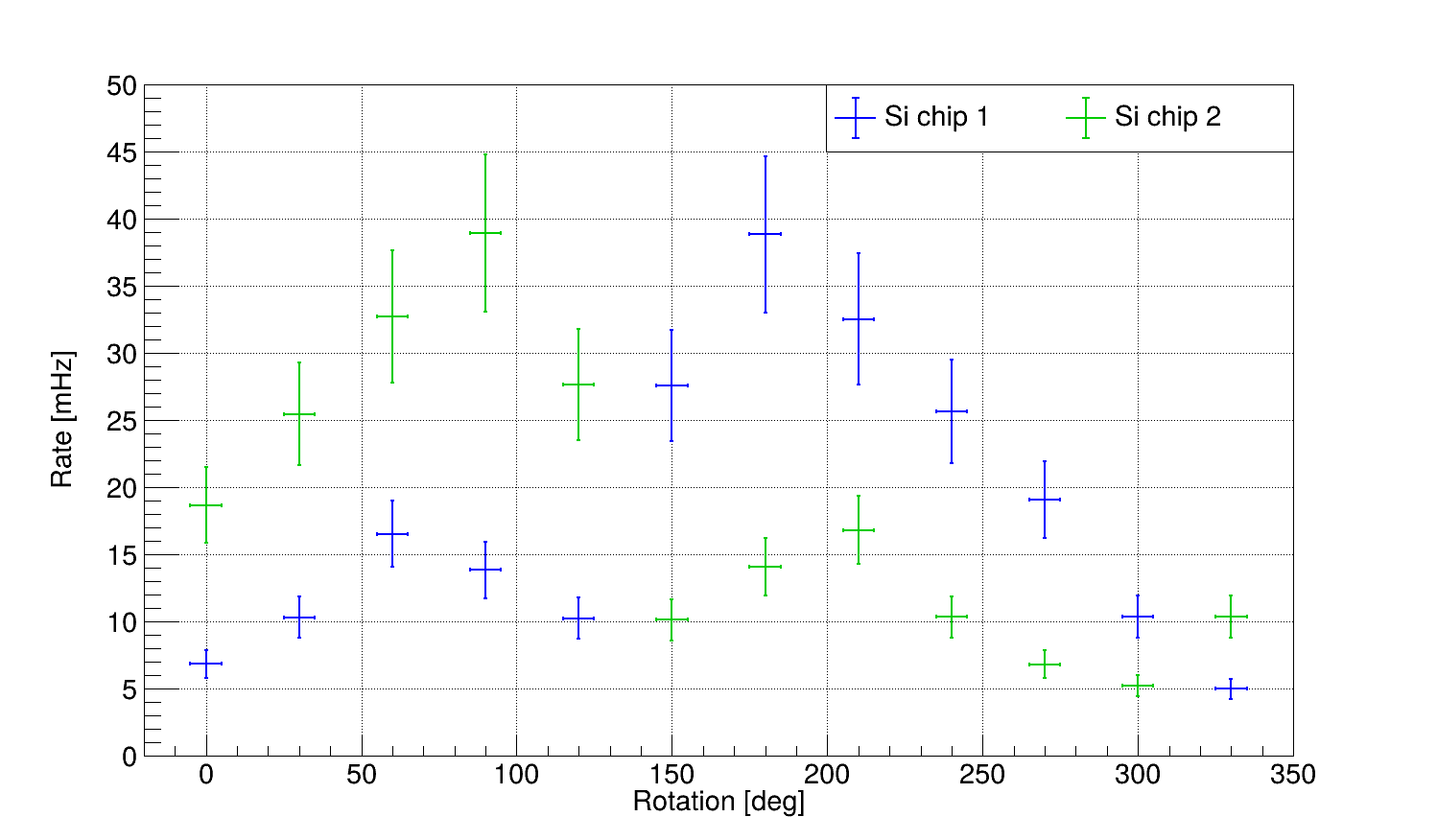}
    \caption{Simulated $^{133}$Ba rates in both Si chips for different payload rotations. The highest event rates are achieved for $\alpha_\text{rot} = 90^\circ$ and $\alpha_\text{rot} = 180^\circ$. The error bars take into account a source position uncertainty of $\Delta \alpha_\text{rot} = \pm 5^\circ$ and the combined statistical and systematical uncertainty on the rate extracted from the \textsc{Geant4} simulations and their normalization.}
    \label{fig:Ba_rates}
\end{figure}

The CM shield surrounding the QUTEbits payload was designed in a way that allows rotation of the payload in steps of $30^\circ$ with respect to a fixed reference as illustrated in figure~\ref{fig:Ba_rotation} for the $^{133}$Ba source position.
In order to optimize the $^{133}$Ba event rate in the Si chips, simulations were performed with varying rotation angles $\alpha_\text{rot}$ while the source was aligned with the center of the Si chips in the same horizontal plane.
The resulting rates versus rotation angle are shown in figure~\ref{fig:Ba_rates}.
The uncertainty on the estimated event rates is dominated by the systematic uncertainty of the source activity which is reported as 15\% of the nominal activity by the manufacturer.

The highest event rates are obtained for $\alpha_\text{rot} = 90^\circ$ (Si chip 2) and $\alpha_\text{rot} = 180^\circ$ (Si chip 1), which is when the Si chips provide the largest effective solid angle coverage.
Likewise, rotation angles $\alpha_\text{rot} \geq 300^\circ$ expose the smallest effective cross-section resulting in the lowest $^{133}$Ba event rate.
The difference in rate between the two Si chips is caused by the additional amount of material from the OQTO holders and their mounting in between the source and the Si chip facing away from the source.
The presence of the material between the Si chips and their small size also result in a very low coincident hit rate of about 0.03\% of the recorded single detector hits for the highest $^{133}$Ba rate positions.

In comparison to the expected $^{133}$Ba rates, the simulated $^{252}$Cf rates are much lower for the most favorable payload rotation.
Because the $^{252}$Cf source is being deployed inside the water tank of CUTE, which is further out from the $^{133}$Ba source position within the outer lead shield, there is a higher position uncertainty associated with its deployment.
A comparison of the expected calibration source event rates is presented in table~\ref{tab:calibration_rates}.

While the combined $^{252}$Cf event rate, composed of roughly the same amount of NRs and ERs, turns out to be comparable to the total background projection (see table~\ref{tab:BG_rates_summary}), the $^{133}$Ba event rates are expected to exceed the background level in all deployment scenarios.
For the highest $^{133}$Ba rate deployment ($\alpha_\text{rot} = 90^\circ, 180^\circ$), the expected excess over background is about fifty times the projected background rate.
For the lowest rate deployment ($\alpha_\text{rot} = 300^\circ, 330^\circ$), the excess is expected to be about a factor of seven.

\begin{table}[ht!]
\centering
\caption{Comparison of simulated calibration source event rates for $^{133}$Ba and $^{252}$Cf in Si chip 1 for different payload rotations. The statistical uncertainties reflect the statistics of the hits achieved in the \textsc{Geant4} simulation. The systematic uncertainties are dominated by the uncertainty of the source activities. The results for the second Si chip are statistically equivalent.}
\label{tab:calibration_rates}
\begin{tabular}{cc}
\toprule
Source configuration &  Event rate in Si chip [mHz]\\
\midrule
$^{133}$Ba with $\alpha_\text{rot} = 90^\circ$  & $13.8 \pm 0.2 (\text{stat.}) \pm 2.5(\text{syst.})$ \\
$^{133}$Ba with $\alpha_\text{rot} = 180^\circ$ & $38.8 \pm 0.3 (\text{stat.}) \pm 5.8(\text{syst.})$ \\
$^{133}$Ba with $\alpha_\text{rot} = 330^\circ$ & $5.0  \pm 0.1 (\text{stat.}) \pm 0.7(\text{syst.})$ \\
\midrule
$^{252}$Cf ERs with $\alpha_\text{rot} = 90^\circ$ & $0.5 \pm 0.1 (\text{stat.}) \pm 0.2(\text{syst.})$ \\
$^{252}$Cf NRs with $\alpha_\text{rot} = 90^\circ$ & $0.4 \pm 0.1 (\text{stat.}) \pm 0.2(\text{syst.})$ \\
$^{252}$Cf combined & $0.9 \pm 0.1 (\text{stat.}) \pm 0.3(\text{syst.})$ \\
\bottomrule
\end{tabular}
\end{table}

\section{Crystal dynamics simulations with G4CMP}
\label{sec:G4CMP_studies}

The \textsc{Geant4} Condensed Matter Physics (G4CMP) package \cite{Kelsey2023G4CMP} is a publicly available addition to the \textsc{Geant4} toolkit \cite{Geant42003, Geant42006, Geant42016}.
It provides the relevant solid-state physics to simulate the response of crystalline substrates to interactions of high-energy particles at cryogenic temperatures ($T \ll 1$\,K).

If sufficiently energetic particles interact in a semiconductor, they produce lattice vibrations (phonons) and electron-hole pairs (e$^-$h$^+$ pairs).
Because of energy conservation, we can assume that the energy expended to generate the charge carriers is eventually deposited into the phonon energy system, in particular when the charge carriers recombine either in the bulk of the semiconductor or thin-film metal electrodes attached on the surface.
Following this assumption, the total phonon energy collected by the sensitive metal films, in our case superconducting Al, is a direct measure of the original particle's energy deposit when interacting with the Si substrate.

G4CMP models the e$^-$h$^+$ pair production from particle impacts in a variety of substrate materials including Si, the subsequent charge carrier transport, their recombination and induced production of acoustic phonons\footnote{G4CMP does not model optical phonons explicitly, because in cryogenic devices held at milikelvin temperatures, optical phonons immediately downconvert to lower-energy (acoustic) phonon modes \cite{Kelsey2023G4CMP}.}, as well as the resulting phonon dynamics.
Additionally, G4CMP comprises a set of generalized physics processes related to the production of Bogoliubov quasiparticles which result from the breaking of Cooper pairs in superconducting films (see Ref.~\cite{Kelsey2023G4CMP} for more details).

In this section, we apply G4CMP to study the phonon and charge carrier dynamics in one of our QUTEbits chip designs with two main objectives.
The first is to inform the event processing of the \textsc{Geant4}-based background simulations by estimating the typical phonon energy collection time (see section~\ref{sec:G4CMP_collection}).
The second goal of this study is to make a connection between the composition of the total background spectrum and the potential of different particle types to cause individual qubit errors or a cluster of correlated errors across multiple qubits (see section~\ref{sec:G4CMP_multiplicity}).

\begin{figure}[ht!]
    \centering
    \includegraphics[width=0.48\linewidth]{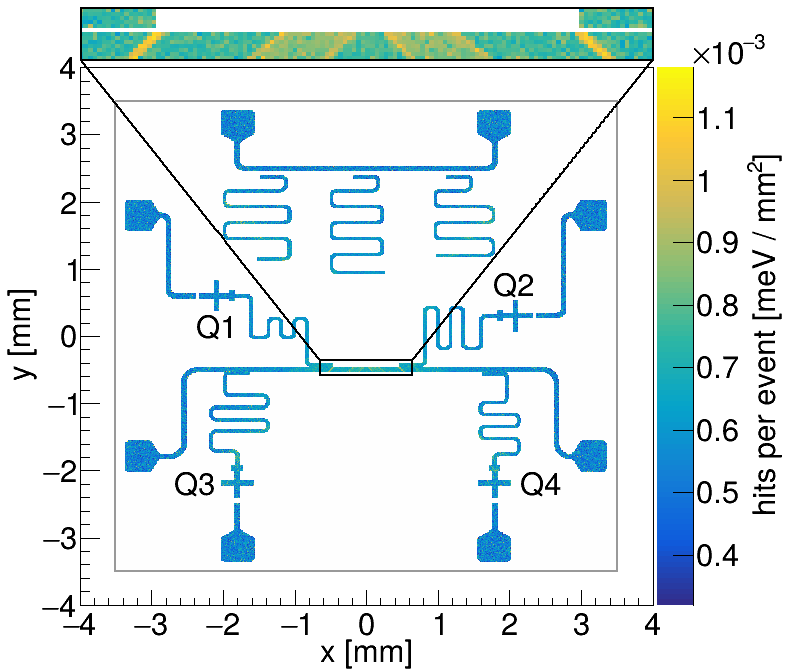} \hspace{0.2cm}
    \includegraphics[width=0.48\linewidth]{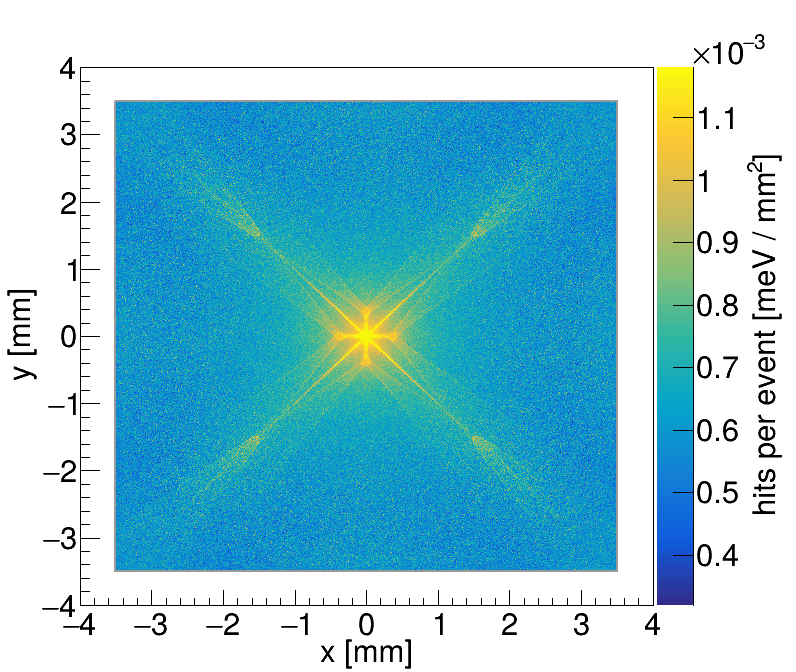}
    \caption{Visualization of a G4CMP simulation with a phonon-only point source emitting single phonons of $E_\text{ph} = 4\,$meV at the center of a chip design with four transmon qubits (Q1 -- Q4). Left: phonon hits collected by the Al layout on top of the Si chip resembling the quantum circuitry with a zoom-in on the central transmission line. Right: bottom view of the spatial hit distribution recorded by the planar backside of the chip. Both figures show the imprints of the characteristic phonon caustics pattern for (100) Si with a $45^\circ$ lattice rotation (see e.g.\ \cite{Kelsey2023G4CMP} for more details). Each hit coordinate has been weighted with the energy deposited by the phonon. In contrast, the response to a homogeneously distributed phonon contamination in the bulk would be completely uniform.}
    \label{fig:G4CMP_hits}
\end{figure}

Both of these objectives can be targeted with a simplified approach that does not require an elaborate implementation of the complex quantum circuit layout as dedicated \textsc{Geant4} geometry elements.
Instead, we split the one-sided chip layout provided by Chalmers University in the form of a GDS file into two layers: one representing the center lines of the quantum circuit layout covering about 8.2\% of the surface and one representing the ground plane (see figure~\ref{fig:G4CMP_hits}).
Furthermore, we attach the two layers to the opposite faces of the Si chip in the simulation, with the top layer resembling the layout of the quantum circuitry and the bottom layer being a solid Al plane covering the entire substrate surface of $7\times7$\,mm$^2$.
With this approach, the total Al coverage of the chip is slightly increased compared to the original single-side layout.
However, because of the Monte Carlo nature of the particle tracking simulations, defining the Al mask in this way does not make a statistically significant difference for studying phenomena such as the localized phonon absorption probability and temporal characteristics of the phonon dynamics.

The planar circuit layout extracted from the GDS file gets converted into a 3D object through voxelization.
The voxel size is informed by the thickness of the Al film of 300\,nm and the dimensions of the circuitry elements such as the transmission lines, 2D resonators and the width of the stripes forming the transmon crosses.
Different voxel sizes were explored and we found that cuboids of \mbox{10$\times$10$\times$0.3\,$\mu$m$^3$} provide a sufficient level of granularity to achieve the targeted fidelity of the geometry description for this study.
The advantage of this simplified approach over a more sophisticated geometry is its flexibility to allow reading in a variety of designs with the same framework and minimal coding work.
An apparent drawback, on the other hand, is the limited fidelity which may not satisfy the needs for more advanced characterization studies such as presented in Ref.~\cite{Yelton2024qpmodeling, Celi2026QPdynamics}.
We hope that our approach may spark new developments in the context of G4CMP and the QIS community with a shared long-term vision of exploring effective strategies for mitigating phonon-mediated quasiparticle poisoning in quantum systems.

The configuration of the material properties and interfaces for this study as well as viable downsample options available in G4CMP are described in appendix~\ref{sec:G4CMP_config}.
All simulations discussed in the following were performed with GCPMP version V09-09-02 and \textsc{Geant4}-10.7.4.

Similar to the previously discussed radioactive source simulations, we record particle hits caused by the specific G4CMP particle types -- acoustic phonons of different polarizations, drift charges representing e$^-$h$^+$ pairs -- incident on the sensitive Al planes.
For the top surface and side walls, the bare Si substrate acts as a diffuse mirror for phonons.
Charge carrier tracks that reach a surface are terminated and we do not model any other form of charge trapping.
When charge carriers are created, we allow for Fano fluctuations of the number of created e$^-$h$^+$ pairs.
For a complete overview of the capabilities of G4CMP and configuration options see Ref.~\cite{Kelsey2023G4CMP}.
While the background studies discussed in section~\ref{sec:radiation_modeling} do not take into account an energy threshold when calculating the projected background rates, in practice the Si bandgap of $\varepsilon_g = 1.17$\,eV (see table~\ref{tab:G4CMP_parameters} in appendix~\ref{sec:G4CMP_config}) acts as a lower bound for ionization to appear.
In the context of radioactive backgrounds, $\varepsilon_g$ is a good approximation for a ``no threshold'' scenario.
For the anticipated G4CMP simulations, there is an additional energy threshold of $2\Delta_\text{Al}$ to be considered which is the minimum phonon energy required to break a Cooper pair in the superconducting Al film.
In our case, we only track phonons that carry $E_\text{ph} \geq 2\Delta_\text{Al}$ with $\Delta_\text{Al} = 0.174\,$meV being our adopted value of the superconducting energy gap of bulk Al at 0\,K (see table~\ref{tab:G4CMP_parameters}).
It should be noted that for thin Al films, there is typically an increase of $\Delta_\text{Al}$ compared to bulk Al \cite{Court2008AlGap}.
However, our film thickness of 300\,nm is still in the regime that can be described by bulk Al properties (see also appendix~\ref{sec:G4CMP_config}).

A summary of the performed G4CMP simulations is given in table~\ref{tab:G4CMP_simulations} with details described in the following sections.
We investigated phonon-only simulations, starting with individual phonons as primary particles of varying energy, electron and nuclear recoils with varying number of charge carrier pairs, and finally high-energy $\beta$ electrons and $\alpha$ particles up to multiple MeV of kinetic energy.
Considering all investigated energy scales, the crystal response to particle impacts was probed over nine orders of magnitude from $\mathcal{O}(1\,\text{meV})$ to $\mathcal{O}(1\,\text{MeV})$.

An example of a phonon-only simulation with $E_\text{ph} = 4$\,meV is shown in figure~\ref{fig:G4CMP_hits}.
All particle types of interest were simulated with two different source configurations: 1) a point source at the center of the Si chip, and 2) a Si bulk contamination source.
The point source option was mainly used to optimize the G4CMP runtime (see appendix~\ref{sec:G4CMP_config}) and perform general consistency checks such as observing the expected phonon caustics pattern for (100) Si as shown in figure~\ref{fig:G4CMP_hits}. 
The phonon caustics would not be visible for higher phonon energies because the pattern gets washed out due to the stronger impact of phonon scattering and phonon downconversion processes such as the anharmonic decay of a high-energy phonon into two lower energy phonons in the crystal bulk.

With the bulk contamination source, points are sampled homogeneously in the Si substrate to start primary particles of the specified kinetic energy.
This approach is representative for the majority of the simulated background events, in particular the radioactive contaminants intrinsic to the Si chip undergoing $\beta$  and $\alpha$ decays.

\subsection{Phonon energy collection time}
\label{sec:G4CMP_collection}

The phonon energy collection time informs the \textsc{Geant4} particle hit processing discussed in section~\ref{sec:hit_processing}.
In this work, we define it as the time it takes to collect all of the athermal phonons created in a particle interaction that are able to break Cooper pairs in the superconducting Al films.
In order to study the effect of different kinds of particle interactions and energy scales, we perform separate G4CMP simulations with a phonon-only bulk source, ER and NR-like energy deposits as well as full particle tracking simulations for $\beta$ electrons and $\alpha$ particles.
Energy deposits by incident particles such as $\gamma$ rays or high-energy $\beta$ electrons will create a population of charge carriers without depositing energy via lattice vibrations (or \textit{prompt phonons}).
These types of interactions are referred to as electron recoils (ERs).

The initial energetic charge carrier pairs ionize subsequent e$^-$h$^+$ pairs in an energy cascade for which the total number of e$^-$h$^+$ pairs pairs created is $N_\text{eh} = E_r /\varepsilon_\text{eh} (E_r)$, where $E_r$ is the recoil energy deposited and $\varepsilon_\text{eh}$ is the average energy required to create one e$^-$h$^+$ pair.
In G4CMP we assume $\varepsilon_\text{eh} = 3.81$\,eV in Si to be constant but allow for Fano fluctuations of the number of charge carrier pairs.
Electrons drifting through the crystal can produce secondary phonons via intervalley scattering.
Both charge carrier species can also emit secondary phonons when they recombine at interface boundaries.
When recombination occurs, half of the Si bandgap energy ($\varepsilon_g = 1.17$\,eV) is emitted via phonons at the Debye frequency of $\omega_D = 15$\,THz in Si (see table~\ref{tab:G4CMP_parameters}).
All of these secondary phonons propagate diffusively and quickly transition to ballistic transport as the phonons downconvert in energy (see Ref.~\cite{Kelsey2023G4CMP} for more details).

If an incident high-energy particle interacts primarily with the nucleus of an atom, such as neutrons causing nuclear recoils (NRs), the energy is split between e$^-$h$^+$ pairs and prompt phonons.
The energy sharing between charge carriers and prompt phonons is determined by the so-called ionization yield $Y$.
The ionization yield is a function of the recoil energy and can be expressed as the ratio of the ionization energy over the recoil energy.
It is generally assumed to be unity for ER-like interactions and smaller than unity for NR-like interactions following a widely accepted model developed by Lindhard \textit{et al.} \cite{Lindhard1963, Sarkis2020}.
By default, we let G4CMP calculate the ionization yield according to the Lindhard theory based on the particle type and target material.
For our low-energy NR-like simulations, we fix it to $Y=0.1$ taking into consideration a recent measurement of the ionization yield in Si at recoil energies as low as 100\,eV \cite{SCDMSImpact2023}.
For $\alpha$ particles, the nature of the energy deposit depends on their kinetic energy and the mass of the target nuclei.
For low $\alpha$ particle energies of $\mathcal{O}$(1\,keV) in Si, the interactions are NR-like with about 80\% of the energy transferred into prompt phonons ($Y=0.2$).
At about 100\,keV, the deposited energy is shared roughly equally between e$^-$h$^+$ pairs and prompt phonons.
For typical $\alpha$ decay energies in the range of 4 -- 10\,MeV, the interactions are ER-like with the ionization yield approaching close to unity (see table~\ref{tab:G4CMP_simulations}).

For our G4CMP-based simulations, we evaluate the hit-time distribution of phonons that are absorbed by the Al sensor planes causing a G4CMP particle hit to find the mean and maximum phonon collection times, $\overline{\tau}$ and $\tau_\text{max}$, reported in table~\ref{tab:G4CMP_simulations}.
The maximum collection time accounts for 99.9\% of the total collected phonon energy per event.

\begin{table}[ht!]
    \centering
    \caption{Overview of G4CMP simulations for different particle interaction types and energy ranges. The Lindhard ionization yield, $Y$, is fixed to the reported values for the respective energy ranges except for $\alpha$ particles for which G4CMP calculates a value for each energy. The uncertainties on the mean ($\overline{\tau}$) and maximum ($\tau_\text{max}$) of the phonon energy collection time represent the standard deviation of the simulations performed for different energies of the primary particles.}
    \label{tab:G4CMP_simulations}
    \begin{tabular}{cccccc}
    \toprule
    Primary particle & Energy range & Lindhard $Y$ & $\overline{\tau}$ [$\mu$s] & $\tau_\text{max}$ [$\mu$s]\\
    \midrule
    Phonon            & 2\,meV -- 2\,eV & - & $1.2 \pm 0.2$ & $14.7 \pm 1.1$ \\
    ER   & 2\,eV -- 192\,eV & 1.0 & $1.2 \pm 0.1$ & $15.9 \pm 1.6$ \\ 
    NR    & 2\,eV -- 192\,eV & 0.1 & $1.3 \pm 0.1$ & $16.1 \pm 0.8$ \\ 
    $\beta$ electron  & 1\,keV -- 3000\,keV & 1.0 & $1.2 \pm 0.1$ & $16.7 \pm 2.6$ \\  
    $\alpha$ particle & 1\,keV -- 8000\,keV & 0.2 -- 1.0 & $1.3 \pm 0.1$ & $16.0 \pm 2.4$ \\ 
    \bottomrule
    \end{tabular}
\end{table}

We find that the results for $\overline{\tau}$ and $\tau_\text{max}$ are consistent across all particle interaction types and investigated energy ranges.
The first three cases (phonon-only, ER and NR) allow us to separate the contributions from the phonon and charge carrier propagation.
Using the same evaluation approach to determine the charge collection time reveals that the overall energy collection is dominated by the phonon propagation with the charge collection being about a factor of three faster.
On a similar note, the full particle tracking simulations with $\beta$ electrons and $\alpha$ particles reveal that the primary particle energy is transferred into e$^-$h$^+$ pairs via ionization and prompt phonons on timescales on the order of $\mathcal{O}$(10\,ps), which is orders of magnitude faster than the typical phonon propagation.

The energy ranges for the different particle type interactions were chosen deliberately to cover the typical energies encountered in different background scenarios (see section~\ref{sec:spectral_analysis}) with some overlap between the categories.
For the $\beta/\alpha$ simulations, the energy deposits were downsampled to an appropriate energy scale that was inferred from the spatial phonon hit distribution analysis presented in the next section (see also appendix~\ref{sec:G4CMP_config}).

Finally, it should be noted that the determined collection times strongly depend on the simulated chip geometry, its aspect ratio and assigned surface properties as well as the substrate material and its lattice orientation.
Moreover, we do not model any quasiparticle dynamics in the Al film beyond the version of Kaplan's model of phonon-quasiparticle behavior implemented in G4CMP \cite{Kelsey2023G4CMP}.
Because of that, the collection times do not include a contribution from quasiparticle diffusion in the superconducting films.
Future versions of G4CMP will include modeling of energy delocalization via quasiparticle diffusion, quasiparticle recombination and concomitant phonon emission\footnote{An extension adding the ability to track quasiparticles in superconducting films was implemented in G4CMP V10.}.

For the processing of the \textsc{Geant4} hits discussed in section~\ref{sec:hit_processing}, we decided to use a split time of $\Delta t = 15\,\mu$s informed by the determined values of $\tau_\text{max}$ for the phonon-only simulations.
A simple interpretation of the determined phonon energy collection time can be made by relating the timescale to the lateral chip dimension and thickness using the speed of sound in Si of $v_s = 9.0\,\mu$m/ns (see table~\ref{tab:G4CMP_parameters}) to express the determined values of $\overline{\tau}$ and $\tau_\text{max}$ in terms of lateral (from side-to-side) or face-to-face phonon reflections.
The mean value of $\overline{\tau} = 1.2\,\mu$s ($\tau_\text{max} = 14.7\,\mu$s) corresponds to 21.6 (264) face-to-face or 1.5 (18.9) lateral reflections prior to absorption.

\subsection{Qubit hit multiplicity}
\label{sec:G4CMP_multiplicity}

The same set of G4CMP simulations evaluated in the previous section with a focus on the temporal crystal dynamics can be used to infer spatial characteristics such as the probability to disturb one or multiple qubits on the same substrate.
To first order, we assume that any phonon hit with $E_\text{ph} \geq 2\Delta_\text{Al}$ depositing energy in the qubit islands represented by the transmon crosses indicated in figure~\ref{fig:G4CMP_hits} would result in a loss of qubit coherence.
If the same occurs for multiple qubits on the same substrate simultaneously, it would be indicative of the rate of correlated qubit errors.

\begin{figure}[ht]
    \centering
    \begin{subfigure}[b]{0.97\linewidth}
        \centering
        \includegraphics[width=\textwidth]{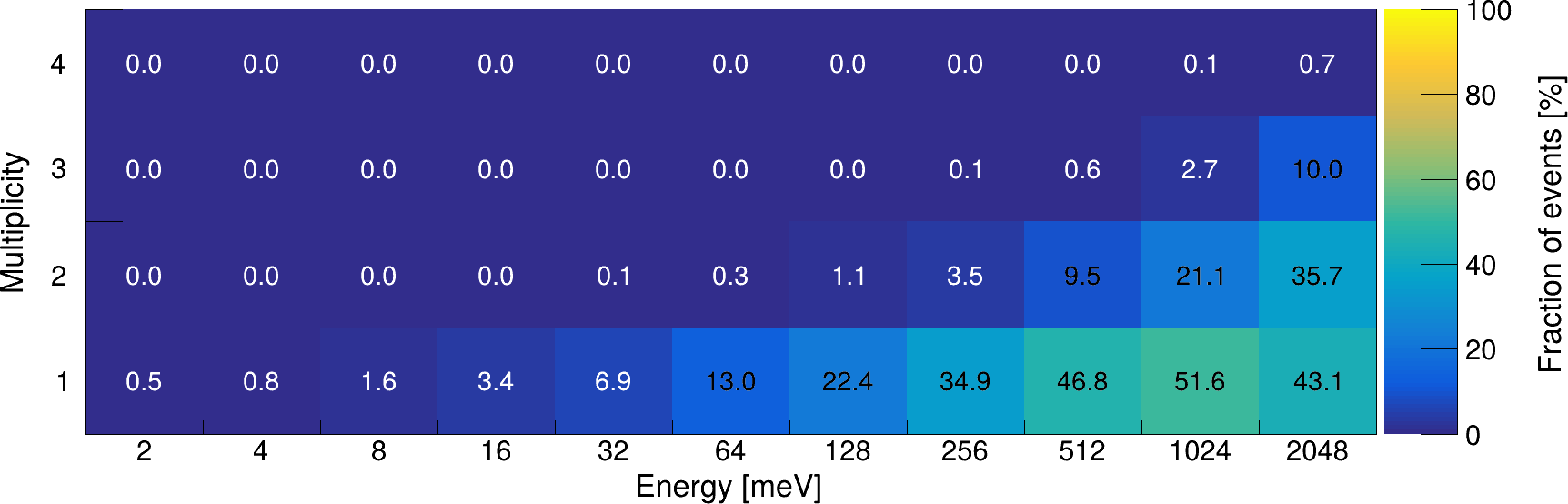}
        \caption{Phonon-only simulation.}
        \label{fig:G4CMP_multiplicity_phonon}
    \end{subfigure}\\
    \begin{subfigure}[b]{0.45\linewidth}
        \centering
        \includegraphics[width=\textwidth]{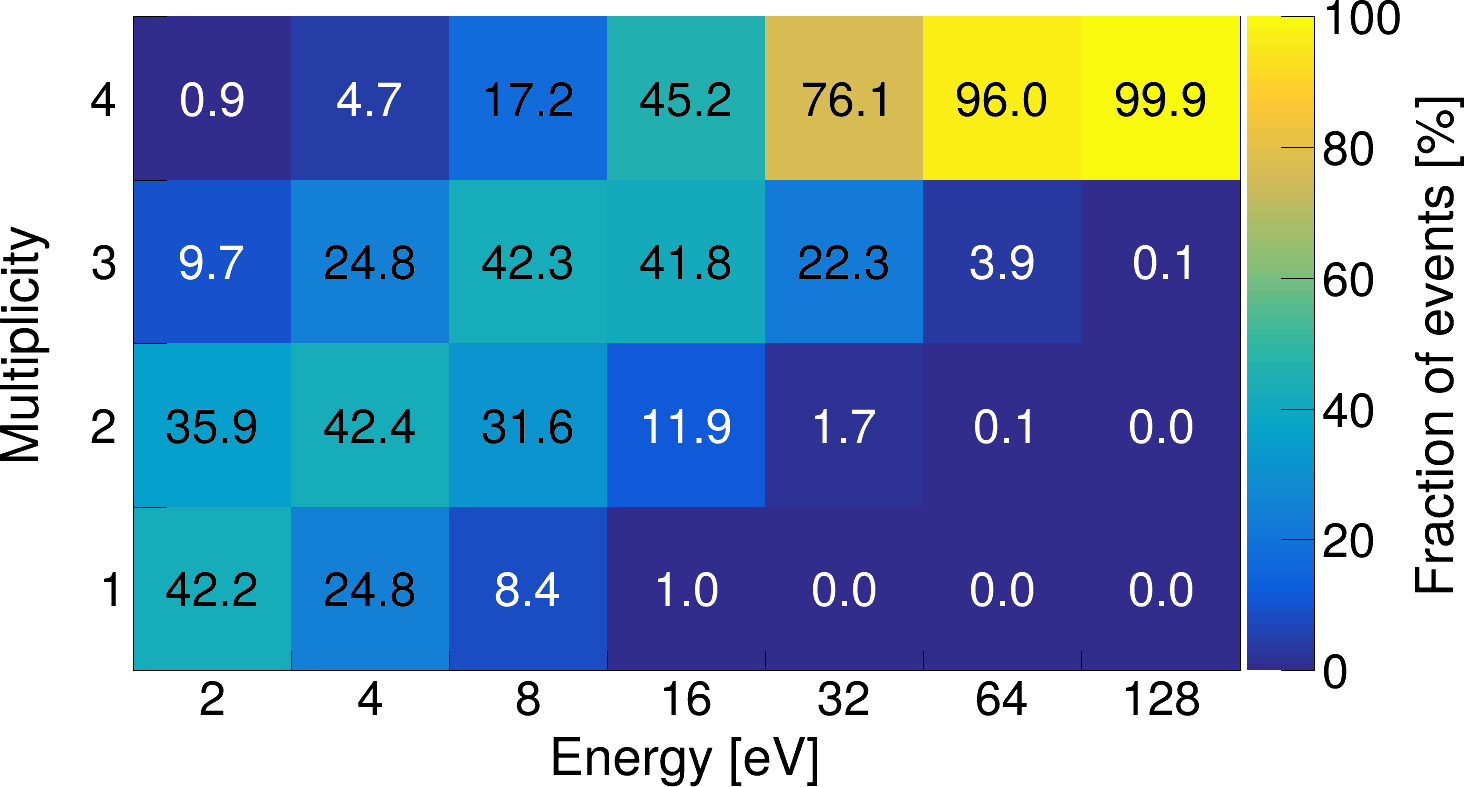}
        \caption{Electron-recoil (ER) simulation.}
        \label{fig:G4CMP_multiplicity_ER}
    \end{subfigure}
    \hspace{0.2cm}
    \begin{subfigure}[b]{0.45\linewidth}
        \centering
        \includegraphics[width=\textwidth]{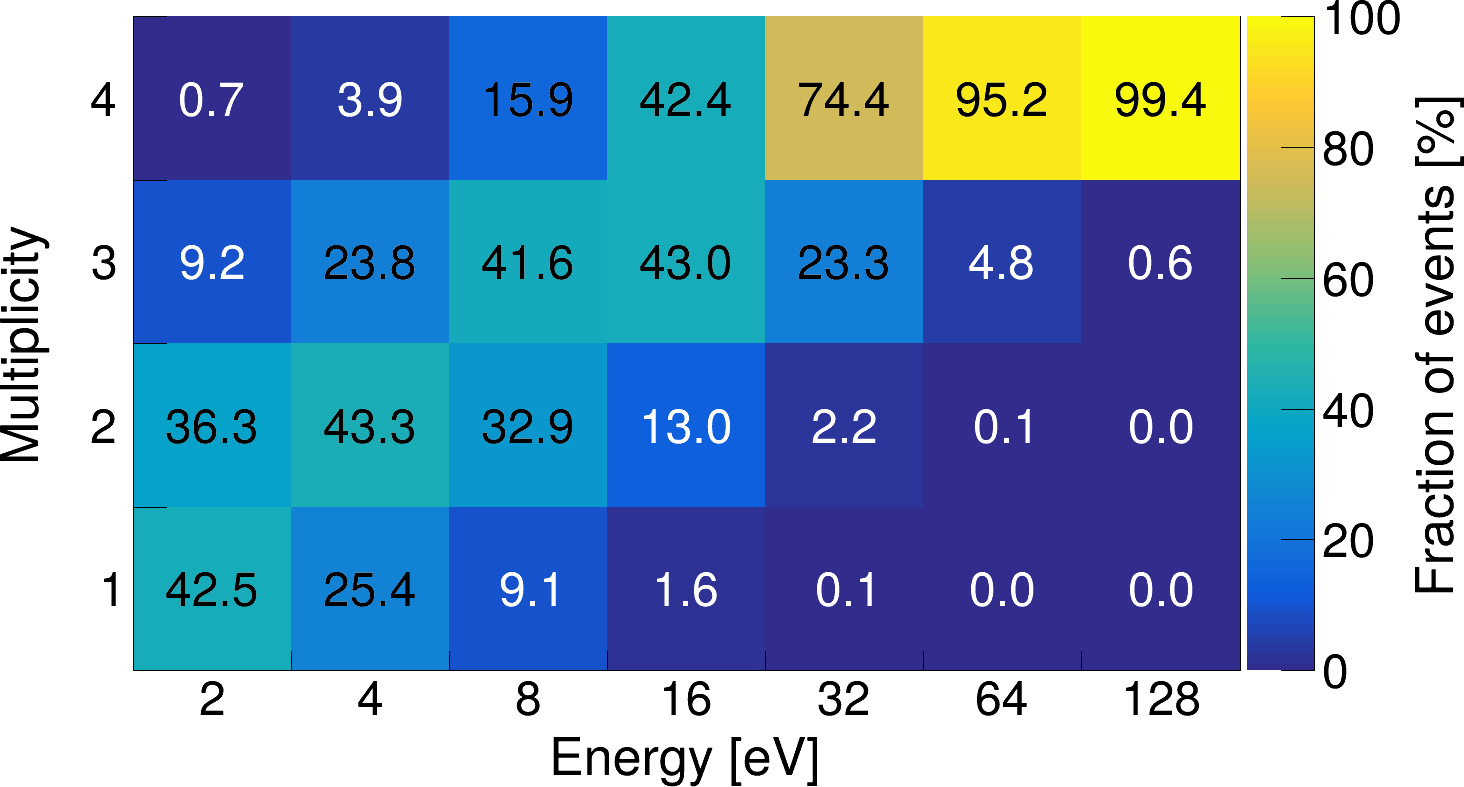}
        \caption{Nuclear-recoil (NR) simulation.}
        \label{fig:G4CMP_multiplicity_NR}
    \end{subfigure}
    \caption{Simulated qubit hit multiplicities for different particle interaction types (\subref{fig:G4CMP_multiplicity_phonon} -- \subref{fig:G4CMP_multiplicity_NR}) modeled as uniform bulk contaminant sources. The color coding and values are indicative for the probability of a specific multiplicity to be observed by the 4-qubit chip design presented in figure~\ref{fig:G4CMP_hits}. For the phonon-only simulations with energy deposits higher than the Debye energy of $\omega_D = 62\,$meV in Si, multiple primary phonons are generated per event. The ER simulations were performed with an ionization yield of $Y = 1$ while the NR simulations were done with $Y = 0.1$.}
    \label{fig:G4CMP_multiplicity}
\end{figure}

In this section, we present the results of the spatial hit evaluation of the simulations summarized in table~\ref{tab:G4CMP_simulations} focusing on the qubit hit multiplicity.
We define the multiplicity as $M \in [0,1,2,3,4]$ with $M = 1$ meaning that one of the four transmon qubits recorded at least one phonon hit and $M = 4$ that all qubits were hit in the same simulation event.
A simulation event is characterized by the maximum phonon collection time reported in table~\ref{tab:G4CMP_simulations}.
To that extent, we analyze the phonon hit coordinates for all simulations and extract the qubit multiplicity per event.
Figure~\ref{fig:G4CMP_multiplicity} illustrates the qubit hit multiplicities for the idealized G4CMP simulations with a bulk contamination source to create phonon-only, ER and NR events of increasing total energy deposits.

The phonon-only simulations starting with a single phonon show a low likelihood to reach $M>1$, whereas the $M=1$ probability increases roughly linearly with the phonon energy.
This linearity is a consequence of the increase in number of secondary phonons from the rapid anharmonic downconversion of the initial phonon which directly depends on the deposited recoil energy.
Above the Debye energy in Si ($\omega_D = 62$\,meV), multiple phonons are created initially which increases the chance for higher multiplicities to occur.

Above a total energy deposit of 2\,eV, we enter the regime in which real particle interactions would produce e$^-$h$^+$ pairs.
We simulate this case separately for ER-like ($Y=1$) and NR-like interactions ($Y=0.1$) with very similar results in terms of qubit multiplicities.
The same level of agreement is apparent for the highest simulated phonon-only energy deposit and the lowest ER and NR energies, respectively.
The observed agreement between the simulations is a strong supportive argument for our original assumption that the total phonon energy collected by the sensor planes is representative of the original particle's energy deposit.
This statement even holds when considering differences in the ionization yield because the energy deposited into the crystal gets efficiently transferred into the phonon system eventually.
Moreover, we conclude that the rapid downconversion of high-energy phonons quickly erases the origin of the energy deposit leading to very similar temporal, spatial and energy distributions of the phonons incident on the sensor planes.

Another conclusion that can be drawn from the charge-carrier simulations is that above an energy scale of $\mathcal{O}$(100\,eV), almost all events result in the maximum qubit multiplicity of $M=4 \equiv M4$.
Above this threshold, the energy deposit is large enough that the resulting secondary particles spread out over the entire volume of the simulated Si chip size regardless of the position of the primary energy deposit.
From this observation, we conclude that any point-like energy deposit above $\sim 200$\,eV would potentially lead to catastrophic qubit errors regardless of the spatial separation of the qubit islands on the $7\times7$ mm$^2$ chip design.
Already at lower energy deposits of $\mathcal{O}$(10\,eV), the likelihood of observing multiplicities $M>1$ dominates with $M\geq 2$ events showing up in more than 90\% of the energy deposits above 8\,eV in our simulations.

As a final step, we evaluate our high-energy $\beta$ electron and $\alpha$ particle simulations to determine representative average multiplicity fractions $\overline{M}_\beta$ and $\overline{M}_\alpha$, respectively.
Based on the results obtained with the ER and NR simulations, the energy deposits of the $\beta$ electrons and $\alpha$ particles were downsampled to an equivalent energy deposit of $E_r = 200\,$eV (see appendix~\ref{sec:G4CMP_config}).
We find that the multiplicity fractions determined for the energy ranges listed in table~\ref{tab:G4CMP_simulations} are largely independent of the energy scale and higher energy simulations do not provide additional information.
However, we observe a small systematic difference between $\beta$ electrons and $\alpha$ particles such that \mbox{$\overline{M}_\alpha (M4) \approx 100\%$} while we find $\overline{M}_\beta (M4) = (96.3 \pm 0.4)\%$ and $\overline{M}_\beta (M3) = (3.8 \pm 0.3)\%$ for $\beta$ electrons.
For the latter values, we take the standard deviation of the average over the simulated energies as a measure of the uncertainty.
We attribute the difference to the softer phonon spectrum created by ER-like interactions and the larger initial volume over which e$^-$h$^+$ pairs get produced, while the initial $\beta$ electron deposits its kinetic energy via ionization.
Another contribution could potentially be arising from the fact that we make use of a G4CMP hit-merging algorithm that groups together energy deposits that are closer together than a specified step size (see also appendix~\ref{sec:G4CMP_config}).

Our findings demonstrate how Monte Carlo tools like G4CMP can be applied for simplified crystal dynamics studies to address questions relevant to the phenomenon of phonon-mediated quasiparticle poisoning in superconducting qubit devices.

\section{Conclusions}
\label{sec:conclusion}

We have performed an extensive material screening campaign which informed design choices for the QUTEbits project.
The assay results serve as an input for sophisticated background projections based on radiation transport simulations with \textsc{Geant4} and crystal dynamics studies with G4CMP.
Because of their associated high uncertainties, we replaced the $^{210}$Pb bulk material assays with the outcome of dedicated studies of the $^{210}$Pb excess accumulated on material surfaces due to radon exposure, which turned out to be on the same order of magnitude as the contribution of bulk radioactivity to the overall background rate.

Our simulation studies predict a total background rate of less than 1\,mHz per Si chip in CUTE and that we can control the radiation environment by switching from the low background rate of the ambient components to a fifty times higher rate by inserting a $^{133}$Ba $\gamma$ ray calibration source.
This feature will be of particular interest when measuring the coherence times of superconducting qubits in CUTE.
In fact, the same type of measurements will be repeated above ground at the University of Waterloo to provide an additional comparison between the impact of the radiation level on Earth's surface and the ultra-low background environment at SNOLAB.

By performing dedicated G4CMP simulations, we determined the phonon energy collection time of our chip design to inform the \textsc{Geant4} particle hit processing.
The phonon collection time depends on the substrate dimensions, the material and its physical properties. 
Because of that, simplified G4CMP studies could serve as a guide to identify quantum circuit layouts which might be better suited to suppress correlated errors, e.g.\ by mechanically isolating the qubit islands from the bulk of the substrate, placing and optimizing suitable phonon absorbers and exploring gap-engineered backside coatings to prevent phonon reflections and increase absorption.

Finally, we investigated how multiple qubits on the same substrate respond to different kinds of particle interaction types.
We interpret the occurrence of events with increasing qubit hit multiplicity as a measure of the projected rate of correlated errors between transmon qubits on the same chip.
It turned out that the probability to observe the maximal qubit multiplicity is fairly independent from the type of particle interaction.
Moreover, we found that energy deposits of $\mathcal{O}$(10\,eV) have a high chance to cause correlated qubit errors even with mm-spaced qubit islands.
Above an energy threshold of $\mathcal{O}$(100\,eV), we expect the entire 
substrate volume to be affected by phonon hits regardless of the origin of the radiation hit impact.

By combining these findings and assuming linear scaling laws, we can make rough projections for quantum computing with fault-tolerant qubits when operated deep underground with a radiation shield comparable to CUTE.
For a benchmark quantum computer processor\footnote{Informed by the approximate dimensions of Google's Willow quantum chip \cite{Google2024Willow}.} based on a Si chip of 20$\times$20$\times$0.5\,mm$^3$ (assuming the same substrate thickness as in our study), we can project a combined background rate of $\mathcal{O}$(10\,mHz).
If we further assume that this is the rate of catastrophic errors which could not be mitigated by means of quantum error correction, it would limit the continuous algorithm execution time to $\mathcal{O}$(100\,s).

We can compare this scenario to a typical surface laboratory environment with an unshielded dilution refrigerator at sea level (see e.g.\ table 6 in Ref.~\cite{Loer2024PNNL}).
The present background rates are dominated by cosmic-ray induced muons and ambient $\gamma$ rays from radioactive contaminants in the environment.
For the projected Si chip size of 20$\times$20$\times$0.5\,mm$^3$ (0.47\,g), the simulated interaction rates reported in Ref.~\cite{Loer2024PNNL} would correspond to 90--140\,mHz from muon interactions and about 200\,mHz from ambient $\gamma$ rays.
These background levels would limit the undisturbed quantum algorithm execution to a few seconds compared to the previously reported $\mathcal{O}$(100\,s) for an operation in a well-shielded deep underground facility.

In the near future, the QUTEbits project will perform an advanced characterization of superconducting qubits deep underground at SNOLAB, focusing on measurements of coherence times, for which the performed assay and simulation studies provide an important foundation.

\section*{Acknowledgments}
The authors gratefully acknowledge SNOLAB and its staff for providing access to the CUTE facility and continued logistical and technical support of the QUTEbits project.
Moreover, the authors thank SNOLAB for performing extensive material screening measurements throughout the course of the project.
The authors would also like to thank the SuperCDMS collaboration for jointly collaborating on their \textsc{Geant4} simulation framework.

This research was sponsored by the U.S.\ Army Research Office (ARO) and the
Laboratory for Physical Sciences (LPS).
It was accomplished under Award Number: W911NF-23-1-0338.
The views and conclusions contained in this document are those of the authors and should not be interpreted as representing the official policies, either expressed or implied, of the Army Research Office or the U.S.\ Government.
The U.S.\ Government is authorized to reproduce and distribute reprints for Government purposes notwithstanding any copyright notation herein.

This research was enabled in part by support provided by Compute Ontario (computeontario.ca) and the Digital Research Alliance of Canada (alliancecan.ca).

\bibliographystyle{JHEP}
\bibliography{biblio}
\appendix

\section{Detailed assay results}
\label{sec:appendix_assay_results}

This section provides the full radioactivity screening results of all assayed components and materials.
Table~\ref{tab:AssayResultsComplete} includes the excerpt presented in table~\ref{tab:AssayResults} of this article but contains more isotopes.
The complete assay results including all isotopes probed were published on \href{https://www.radiopurity.org/}{radiopurity.org} \cite{RadiopurityWeb}. 
In several cases, the products of different vendors were screened to select components which are low in radioactivity.
For several of the listed components and raw materials such as aluminum or copper, samples were assayed which are not necessarily of the exact same batch as was used for producing certain parts, but they were usually provided by the same vendor.

Previously used\footnote{Used PCBs may have residual solder contamination and wire bond footprints from prior assembly.} and also new PCBs of different vendors have been assayed.
In particular, CERcuits from Belgium provided several batches of the individual components used to manufacture PCBs Al$_2$O$_3$ and AlN, which were separately assayed to better identify which specific components contribute most significantly to the radioactivity of PCBs.

\begin{table}
\centering
\caption{Results of the assay and screening measurements showing the contamination levels of different radioisotopes and decay chains in the various components. For $^{238}$U, the top and bottom parts of the chains are reported separately. Entries with a dash (``-'') mean that the detector used for the assay measurement had no sensitivity to the particular radioisotope of interest, or it was not reported. In contrary to the selection presented in table~\ref{tab:AssayResults}, here the actual mass of the assayed sample and if applicable the number of units are provided.}
\label{tab:AssayResultsComplete}
\resizebox{\textwidth}{!}{
\begin{tabular}{ccccccccc} 
\toprule
Component & Mass & $^{238}$U & $^{235}$U & $^{232}$Th & $^{210}$Pb & $^{137}$Cs  & $^{60}$Co & $^{40}$K \\
& [g] & [mBq/kg] & [mBq/kg] & [mBq/kg] & [mBq/kg] & [mBq/kg] & [mBq/kg] & [mBq/kg] \\ 
\midrule

\textbf{Sample holders} \\

\multirowcell{2}{UWaterloo homemade \\ sample holder} & \multirow{2}{*}{126.1} & \multicolumn{1}{c}{t: 59.2$\pm$24.2} & \multirow{2}{*}{1.3$\pm$0.5} & \multirow{2}{*}{6.9$\pm$1.6} & \multirow{2}{*}{$<$42384} & \multirow{2}{*}{$<$0.5} & \multirow{2}{*}{$<$1.1} & \multirow{2}{*}{31.0$\pm$13.7} \\
& & b:1.2 $\pm$1.3 \\

\multirow{2}{*}{OQTO sample holder} & \multirow{2}{*}{220.0} & t: 1099.0$\pm$195.4 & \multirow{2}{*}{28.6$\pm$2.6} & \multirow{2}{*}{38.4$\pm$4.7} & \multirow{2}{*}{-} & \multirow{2}{*}{$<$3.0} & \multirow{2}{*}{$<$0.8} & \multirow{2}{*}{60.2$\pm$25.5} \\
& & b: 20.4$\pm$3.3 \\

\midrule
\textbf{Printed-circuit boards} \\

\multirowcell{2}{Gold-plated PCBs \\ (Aspocom/Provexa)} & 8.164 & t: 8911$\pm$1977 & \multirow{2}{*}{79.7$\pm$23.9} & \multirow{2}{*}{2032.0$\pm$117.6} & \multirow{2}{*}{-} & \multirow{2}{*}{$<$72.5} & \multirow{2}{*}{$<$23.7} & \multirow{2}{*}{1840.5$\pm$493.9} \\
& (2 units) & b: 1579.0$\pm$88.9 \\

\multirowcell{2}{Used PCBs RO3010\\ (Rogers Corporation)} & 10.0 & t: 419.3$\pm$94.7 & \multirow{2}{*}{15.2$\pm$2.7} & \multirow{2}{*}{220.5$\pm$18.0} & \multirow{2}{*}{1610.1$\pm$717.7} & \multirow{2}{*}{$<$39.7} & \multirow{2}{*}{9.7$\pm$15.1} & \multirow{2}{*}{3593.4$\pm$686.7} \\
& (5 units) & b: 270.4$\pm$18.2 \\

\multirowcell{2}{New PCBs RO3010\\ (Rogers Corporation)} & 10.7 & t: 459.1$\pm$83.6 & \multirow{2}{*}{11.7$\pm$2.5} & \multirow{2}{*}{208.7$\pm$16.6} & \multirow{2}{*}{$<$984.9} & \multirow{2}{*}{$<$40.5} & \multirow{2}{*}{$<$25.2} & \multirow{2}{*}{1802.2$\pm$517.1} \\
& (5 units) & b: 241.6$\pm$16.4 \\

\multirowcell{2}{Tin-plated PCB \\ (PCBWay)} & 4.806 & t: 9544.0$\pm$688.6 & \multirow{2}{*}{304.2$\pm$16.5} & \multirow{2}{*}{6481.0$\pm$250.1} & \multirow{2}{*}{3163.1$\pm$1665.0} & \multirow{2}{*}{$<$70.6} & \multirow{2}{*}{$<$69.1} & \multirow{2}{*}{7503.0$\pm$1695.0}  \\
& (13 units) & b: 5386.0$\pm$191.3 \\

\multirowcell{2}{PCB, EPIG Plating \\ (Elco BV, Hofstetter PCB)} & 7.835 & t: 6132$\pm$1973 & \multirow{2}{*}{106.5$\pm$27.9} & \multirow{2}{*}{2232.0$\pm$138.9} & \multirow{2}{*}{-} & \multirow{2}{*}{$<$38.0} & \multirow{2}{*}{$<$26.0} & \multirow{2}{*}{1380.1$\pm$567.8} \\
& (2 units) & b: 1499.0$\pm$101.2 \\

\multirowcell{2}{PCB TMM10 \\ (Rogers Corporation)} & \multirow{2}{*}{55.6} & t: 6816.0$\pm$2988.0  & \multirow{2}{*}{871.7$\pm$53.5} & \multirow{2}{*}{4813.0$\pm$202.6} & \multirow{2}{*}{-} & \multirow{2}{*}{4.0$\pm$59.4} & \multirow{2}{*}{35.5$\pm$38.4} & \multirow{2}{*}{15604.0$\pm$1228.0}  \\
& & b: 30350.0$\pm$659.3 \\

\multirowcell{2}{PCB RO4350B \\ (Rogers Corporation)} & \multirow{2}{*}{165.7} & t: 27190.0$\pm$2974.4  & \multirow{2}{*}{913.6$\pm$41.7} & \multirow{2}{*}{18930.0$\pm$499.2} & \multirow{2}{*}{-} & \multirow{2}{*}{17.1$\pm$40.0} & \multirow{2}{*}{27.2$\pm$24.4} & \multirow{2}{*}{15175.0$\pm$949.8} \\
& & b: 16050.0$\pm$348.4 \\

\multirowcell{2}{EPIG-plated 7-layer PCB \\ (Aspocomp)} & 20.56 & t: 3944.0$\pm$432.7 & \multirow{2}{*}{117.3$\pm$15.8} & \multirow{2}{*}{2275.0$\pm$95.4} & \multirow{2}{*}{$<$589.3} & \multirow{2}{*}{$<$12.6} & \multirow{2}{*}{$<$11.2} & \multirow{2}{*}{1390.4$\pm$263.6} \\
& (5 units) & b: 2392.0$\pm$85.1 \\

\multirowcell{2}{AlO$_\text{x}$ tin-plated PCB \\ (CERcuits)} & 0.584 & t: 8104$\pm$1197 & \multirow{2}{*}{189.5$\pm$41.4} & \multirow{2}{*}{6149.0$\pm$407.2} & \multirow{2}{*}{$<$4672} & \multirow{2}{*}{$<$679.1} & \multirow{2}{*}{$<$469.2} & \multirow{2}{*}{14077$\pm$7897}\\
& (2 units) & b: 3337.0$\pm$276.7 \\

\midrule
\multicolumn{9}{l}{\textbf{Components used to manufacture PCBs Al$_\text{2}$O$_\text{3}$ and AlN by CERcuits}} \\

\multirow{2}{*}{PCB Al$_2$O$_3$ - 1} & 1.506 & t: 6719.0$\pm$604.5 &\multirow{2}{*}{215.9$\pm$21.2} & \multirow{2}{*}{5079.0$\pm$256.1} & \multirow{2}{*}{1439$\pm$1707} & \multirow{2}{*}{$<$323.4} & \multirow{2}{*}{$<$153.2} & \multirow{2}{*}{2902$\pm$2820} \\
& (5 units) & b: 3144.0$\pm$161.8 \\

\multirow{2}{*}{PCB Al$_2$O$_3$ - 2} & \multirow{2}{*}{134.3} & t: 15460.0$\pm$732.8 & \multirow{2}{*}{299.9$\pm$8.5 } & \multirow{2}{*}{810.8$\pm$26.7}&\multirow{2}{*}{$<$1688} & \multirow{2}{*}{$<$2.5} & \multirow{2}{*}{$<$4.3}& \multirow{2}{*}{835.2$\pm$76.0} \\
& & b: 1530.0$\pm$36.2 \\

\multirow{2}{*}{PCB AlN - Ceramic strips} & \multirow{2}{*}{38.46} & t: 3233.0$\pm$642.4 & \multirow{2}{*}{50.4$\pm$8.3} & \multirow{2}{*}{137.2$\pm$19.8} & \multirow{2}{*}{-} & \multirow{2}{*}{ $<$11.5} & \multirow{2}{*}{$<$5.0} & \multirow{2}{*}{$<$166.5} \\
& & b: 18.6$\pm$10.1 \\

\multirow{2}{*}{PCB AlN - Ceramic plates} & \multirow{2}{*}{10.645} & t: 3718$\pm$1045 & \multirow{2}{*}{71.2$\pm$16.2} & \multirow{2}{*}{36.6$\pm$31.8} & \multirow{2}{*}{-} & \multirow{2}{*}{$<$30.4} & \multirow{2}{*}{$<$7.3} & \multirow{2}{*}{$<$266.7} \\
& & b: $<$35.0 \\

\multirowcell{2}{PCB AlN - copper \\ backed circuit board plates} & \multirow{2}{*}{45.2} & t: 6641$\pm$489 & \multirow{2}{*}{129.6$\pm$8.5}&\multirow{2}{*}{2296$\pm$75} & \multirow{2}{*}{$<$9088} & \multirow{2}{*}{$<$12.3} &  \multirow{2}{*}{$<$5.9} & \multirow{2}{*}{253.3$\pm$97.6 }\\
& & b: 1439.0$\pm$45.2 \\

\midrule
\textbf{Silicon wafers} \\

\multirow{2}{*}{Si wafer 4"} & \multirow{2}{*}{9.564} & t: 267.1$\pm$305.7 & \multirow{2}{*}{8.0$\pm$4.7} & \multirow{2}{*}{$<$21.5} & \multirow{2}{*}{90958$\pm$75010} & \multirow{2}{*}{$<$10.4} & \multirow{2}{*}{$<$5.9} & \multirow{2}{*}{85.4$\pm$111.4} \\
& & b: $<$4.4 \\

\multirow{2}{*}{Si wafer 2" Type 011} & \multirow{2}{*}{2.726} & t: 308.2$\pm$405.4 & \multirow{2}{*}{7.7$\pm$10.2} & \multirow{2}{*}{$<$45.8} & \multirow{2}{*}{$<$21840} & \multirow{2}{*}{$<$34.4} & \multirow{2}{*}{$<$8.0} & \multirow{2}{*}{$<$290.2} \\
& & b: $<$11.9 \\

\multirow{2}{*}{Si wafer 2" Type 010} & \multirow{2}{*}{2.677} & t: $<$516.8 & \multirow{2}{*}{$<$22.0} & \multirow{2}{*}{$<$28.5} & \multirow{2}{*}{29123$\pm$22150} & \multirow{2}{*}{12.9$\pm$15.9} & \multirow{2}{*}{$<$18.4} & \multirow{2}{*}{$<$336.7} \\
& & b: $<$12.6 \\

\multirow{2}{*}{Si wafer 2" Type 12} & \multirow{2}{*}{2.7} & t: $<$601.7 & \multirow{2}{*}{$<$10.1} & \multirow{2}{*}{$<$20.5} &\multirow{2}{*}{$<$15580}& \multirow{2}{*}{$<$26.5} & \multirow{2}{*}{$<$4.3} & \multirow{2}{*}{166.4$\pm$298.6} \\
& & b: $<$7.10 \\

\multirowcell{2}{UWaterloo Si wafer 4" \\ (University Wafer)} & \multirow{2}{*}{9.515} & t: 125.9$\pm$309.7 & \multirow{2}{*}{9.8$\pm$4.5} & \multirow{2}{*}{$<$25.9} & \multirow{2}{*}{$<$88870}& \multirow{2}{*}{$<$8.9} & \multirow{2}{*}{$<$6.7} & \multirow{2}{*}{$<$121.7} \\
& & b: $<$3.3 \\

\midrule
\textbf{Microwave cables} \\

\multirow{2}{*}{NbTi cable assembly} & \multirow{2}{*}{6.2} & t: 17470.0$\pm$11043.0 & \multirow{2}{*}{352.2$\pm$13.3} & \multirow{2}{*}{509.4$\pm$33.8} & \multirow{2}{*}{11055$\pm$4231} & \multirow{2}{*}{$<$58.4} & \multirow{2}{*}{$<$9.3} & \multirow{2}{*}{137.8$\pm$146.4} \\
& & b: 56.6$\pm$13.8 \\

\multirowcell{2}{Formable non-magnetic cable \\ assembly (EZ Form Cable)} & \multirow{2}{*}{20.2} & t: 787.7$\pm$1024.0 & \multirow{2}{*}{$<$14.2} & \multirow{2}{*}{$<$38.9} & \multirow{2}{*}{-} & \multirow{2}{*}{$<$21.0} & \multirow{2}{*}{$<$2.8} & \multirow{2}{*}{233.6$\pm$289.6} \\
 & & b: $<$37.7 \\

\multirow{2}{*}{CuNi cable assembly} & \multirow{2}{*}{7.2} & t: 322.2$\pm$187.6 & \multirow{2}{*}{5.9$\pm$3.7} & \multirow{2}{*}{$<$21.5} & \multirow{2}{*}{57731$\pm$17900} & \multirow{2}{*}{7.8$\pm$5.1} & \multirow{2}{*}{$<$2.5} & \multirow{2}{*}{$<$75.9} \\
& & b: $<$2.6 \\

\multirow{2}{*}{Nb cable assembly} & \multirow{2}{*}{9.0} & t: $<$1721.0 & \multirow{2}{*}{$<$34.6} & \multirow{2}{*}{290.7$\pm$49.1} & \multirow{2}{*}{$<$26610} & \multirow{2}{*}{$<$23.8} & \multirow{2}{*}{$<$22.7} & \multirow{2}{*}{357.6$\pm$244.1} \\
& & b: 1200.0$\pm$73.8 \\

\multirowcell{2}{Stainless steel \\ cable assembly} & \multirow{2}{*}{39.9} & t: 163.5$\pm$217.8 & \multirow{2}{*}{$<$7.4} & \multirow{2}{*}{103.2$\pm$13.8}  & \multirow{2}{*}{$<$1896} & \multirow{2}{*}{$<$12.8} & \multirow{2}{*}{2.0$\pm$2.8} & \multirow{2}{*}{112.3$\pm$67.7} \\
& & b: $<$10.9 \\

\bottomrule
\multicolumn{9}{r}{continued on next page}
\end{tabular}
}
\end{table}

\begin{table}
\centering

\resizebox{\textwidth}{!}{
\begin{tabular}{ccccccccc} 
\toprule
Component & Mass & $^{238}$U & $^{235}$U & $^{232}$Th & $^{210}$Pb & $^{137}$Cs  & $^{60}$Co & $^{40}$K \\
& [g] & [mBq/kg] & [mBq/kg] & [mBq/kg] & [mBq/kg] & [mBq/kg] & [mBq/kg] & [mBq/kg] \\ 

\midrule
\textbf{Microwave components} \\

\multirow{2}{*}{IR Filter (CNTs) \cite{CNT2019}} & 68.8 & t: 193.3$\pm$37.4 & \multirow{2}{*}{5.4$\pm$0.6} & \multirow{2}{*}{9.6$\pm$2.0} & \multirow{2}{*}{753.2$\pm$876.4} & \multirow{2}{*}{$<$0.9} & \multirow{2}{*}{$<$3.6} & \multirow{2}{*}{57.8$\pm$34.6} \\
& (3 units) & b: 3.5$\pm$1.6 \\

\multirow{2}{*}{IR filter (Bluefors)} & \multirow{2}{*}{12.9568} & t: 1459.0$\pm$216.5 &\multirow{2}{*}{28.9$\pm$3.4} & \multirow{2}{*}{20.6$\pm$9.1} & \multirow{2}{*}{77515$\pm$5479} & \multirow{2}{*}{$<$40.7} & \multirow{2}{*}{$<$21.0}& \multirow{2}{*}{505.5$\pm$297.6} \\
& & b: $<$8.7 \\

\multirow{2}{*}{IR Filter (CliQ10)} & \multirow{2}{*}{25.3} & t: 39.6$\pm$23.3 & \multirow{2}{*}{2.4$\pm$1.2} & \multirow{2}{*}{46.4$\pm$6.1} & \multirow{2}{*}{$<$146.4} & \multirow{2}{*}{$<$15.7} & \multirow{2}{*}{4.7$\pm$9.7} & \multirow{2}{*}{720.3$\pm$246.5} \\
 & & b: 32.5$\pm$6.1 \\

\multirowcell{2}{Amumetal circulator with \\ cryoperm shield (Raditek)} & \multirow{2}{*}{377.9} & t: 135.0$\pm$115.5 & \multirow{2}{*}{10.1$\pm$1.9} & \multirow{2}{*}{22.4$\pm$3.2} & \multirow{2}{*}{-} & \multirow{2}{*}{$<$1.8} & \multirow{2}{*}{2.5$\pm$1.0} & \multirow{2}{*}{45.5$\pm$18.6} \\
& & b: 41.1$\pm$3.3 \\

\multirowcell{2}{Microwave switch \\ R591763600 (Radiall)} & \multirow{2}{*}{95.5} & t: 7554.0$\pm$667.9 & \multirow{2}{*}{133.5$\pm$8.2}  & \multirow{2}{*}{826.0$\pm$32.7} & \multirow{2}{*}{410550$\pm$150300} & \multirow{2}{*}{$<$7.3} & \multirow{2}{*}{2.2$\pm$3.2} & \multirow{2}{*}{806.2$\pm$92.6} \\
& & b: 453.0$\pm$19.8 \\

\multirowcell{2}{Low Pass Filter SLP-1.9+ \\ (Mini-Circuits)} & \multirow{2}{*}{35.5} & t: $<$200.3 & \multirow{2}{*}{$<$8.8} & \multirow{2}{*}{231.4$\pm$17.5} & \multirow{2}{*}{$<$334.6} & \multirow{2}{*}{$<$7.1} & \multirow{2}{*}{11.7$\pm$4.4} & \multirow{2}{*}{366.4$\pm$87.8} \\
& & b: 52.5$\pm$13.8\\

\midrule
\textbf{Connectors and screws} \\

\multirowcell{2}{M2 brass screws \\ (Skruvcenter)} & 5.092 & t: $<$62.2 & \multirow{2}{*}{$<$6.6} & \multirow{2}{*}{ $<$11.8} & \multirow{2}{*}{$<$603.7} & \multirow{2}{*}{$<$79.6} & \multirow{2}{*}{27.7$\pm$28.1} & \multirow{2}{*}{391.5$\pm$525.7} \\
& (16 units) & b: $<$15.0 \\

\multirowcell{2}{Brass washers \\ (Spaenaur)} & \multirow{2}{*}{1.97} & t: $<$166.1 & \multirow{2}{*}{$<$6.9}& \multirow{2}{*}{$<$40.1} & \multirow{2}{*}{$<$994.4} & \multirow{2}{*}{$<$204.9} & \multirow{2}{*}{$<$39.9} & \multirow{2}{*}{4025.6$\pm$1520.0} \\
& & b: $<$65.8 \\

\multirowcell{2}{Brass nuts \\ (Spaenaur)} & \multirow{2}{*}{2.29} & t: $<$168.3 & \multirow{2}{*}{20.4$\pm$7.6} & \multirow{2}{*}{22.5$\pm$30.6} & \multirow{2}{*}{$<$834.7} & \multirow{2}{*}{$<$184.4} & \multirow{2}{*}{$<$52.9} & \multirow{2}{*}{1539.1$\pm$1351.0} \\
& & b: 27.1$\pm$32.7 \\

\multirowcell{2}{SMA connectors \\ (Quantum Microwave)} & 23.237 & t: 96.7$\pm$41.8 & \multirow{2}{*}{5.6$\pm$1.4} & \multirow{2}{*}{1.2$\pm$3.9} & \multirow{2}{*}{6922.0$\pm$407.5} & \multirow{2}{*}{0.6$\pm$11.4} & \multirow{2}{*}{2.0$\pm$5.3} & \multirow{2}{*}{91.7$\pm$99.8} \\
& (10 units) & b: $<$3.8 \\

\multirowcell{2}{SMA pins \\ (Quantum Microwave)} & \multirow{2}{*}{0.048} & t: 34510.0$\pm$10310.0 & \multirow{2}{*}{895.2$\pm$254.6} & \multirow{2}{*}{1688.0$\pm$1006.0} & \multirow{2}{*}{$<$55050.0} & \multirow{2}{*}{$<$8141.0} & \multirow{2}{*}{$<$1505.0} & \multirow{2}{*}{52452.0$\pm$43130.0} \\
& & b: 488.4$\pm$953.0 \\

\multirowcell{2}{SMA PTFE \\ (Quantum Microwave)} & \multirow{2}{*}{0.068} & t: $<$4004.0 & \multirow{2}{*}{50.8$\pm$117.6} & \multirow{2}{*}{$<$933.7} & \multirow{2}{*}{4902.9$\pm$32940.0} & \multirow{2}{*}{$<$5595.0} & \multirow{2}{*}{$<$902.2} & \multirow{2}{*}{26711.0$\pm$25970.0} \\
& & b: $<$476.4 \\

\multirowcell{2}{SMA cap \\ (Amphenol RF)} & 12.67 & t: $<$33.2 & \multirow{2}{*}{$<$3.2} & \multirow{2}{*}{8.9$\pm$5.6} & \multirow{2}{*}{2224.3$\pm$203.7} & \multirow{2}{*}{$<$33.8} & \multirow{2}{*}{$<$36.4} & \multirow{2}{*}{$<$398.4} \\
& (5 units) & b: 12.0$\pm$6.2 \\

\midrule
\textbf{Magnetic shields} \\

\multirow{2}{*}{Amumetal shield (Bluefors)} & \multirow{2}{*}{49.6} & t: 852.6$\pm$133.5 & \multirow{2}{*}{12.5$\pm$1.9} & \multirow{2}{*}{9.9$\pm$4.4} & \multirow{2}{*}{27642$\pm$16580} & \multirow{2}{*}{$<$2.4} & \multirow{2}{*}{$<$1.9} & \multirow{2}{*}{45.0$\pm$34.6} \\
& & b: $<$1.2 \\

\multirow{2}{*}{Al shield} & \multirow{2}{*}{593.3} & t: 6446.0$\pm$288.8 & \multirow{2}{*}{116.2$\pm$2.9} & \multirow{2}{*}{125.8$\pm$4.6} & \multirow{2}{*}{7346.1$\pm$840.5}  & \multirow{2}{*}{$<$1.6} & \multirow{2}{*}{0.2$\pm$0.3} & \multirow{2}{*}{12.2$\pm$6.5} \\
& & b: 6.2$\pm$1.2 \\

\multirow{2}{*}{Amumetal plate} & \multirow{2}{*}{15.6} & t: 527.5$\pm$544.6 & \multirow{2}{*}{$<$15.1} & \multirow{2}{*}{$<$36.9} & \multirow{2}{*}{-} & \multirow{2}{*}{$<$13.7} & \multirow{2}{*}{$<$4.4} & \multirow{2}{*}{1017.9$\pm$305.1}  \\
& & b: $<$20.0 \\

\multirowcell{2}{Magnetic shield plate CP-EXP-1184 \\ (Ad-Vance Magnetics Inc)} & \multirow{2}{*}{89.9} & t: $<$171.4 & \multirow{2}{*}{$<$14.6} & \multirow{2}{*}{134.1$\pm$21.4} & \multirow{2}{*}{$<$3703} & \multirow{2}{*}{$<$13.4} & \multirow{2}{*}{7.1$\pm$5.4} & \multirow{2}{*}{$<$126.6}  \\
& & b: 3.7$\pm$19.8 \\

\midrule
\textbf{Aluminum} \\

\multirowcell{2}{Used aluminum (99.999\%) \\ (Kurt J Lesker)} & \multirow{2}{*}{139.063} & t: $<$85.7 & \multirow{2}{*}{$<$2.8} & \multirow{2}{*}{9.5$\pm$3.6} & \multirow{2}{*}{-} & \multirow{2}{*}{$<$1.7} & \multirow{2}{*}{$<$0.7} & \multirow{2}{*}{$<$26.9} \\
& & b: $<$3.9 \\

\multirowcell{2}{Used aluminum (99.999\%) \\ (Kurt J Lesker)} & \multirow{2}{*}{24.851} & t: $<$403.5 & \multirow{2}{*}{$<$4.8} & \multirow{2}{*}{83.0$\pm$12.1} & \multirow{2}{*}{-} & \multirow{2}{*}{$<$5.4} & \multirow{2}{*}{$<$3.4} & \multirow{2}{*}{97.5$\pm$86.5} \\
& & b: $<$3.6 \\

\multirowcell{2}{Al pellets (99.999\%) \\ (Kurt-J Lesker)} & \multirow{2}{*}{4.246} & t: $<$133.0 & \multirow{2}{*}{$<$6.7} & \multirow{2}{*}{16.2$\pm$13.4} & \multirow{2}{*}{$<$967.1} & \multirow{2}{*}{$<$90.3} & \multirow{2}{*}{$<$12.6} & \multirow{2}{*}{$<$754.9} \\
& & b: $<$13.6 \\

\multirowcell{2}{Al slugs (99.999\%) \\ (Alfa Aesar, Puratronic)} & \multirow{2}{*}{1.97} & t: $<$131.9 & \multirow{2}{*}{$<$7.8} & \multirow{2}{*}{$<$18.9} & \multirow{2}{*}{$<$1762} & \multirow{2}{*}{$<$209.0} & \multirow{2}{*}{$<$38.1} & \multirow{2}{*}{1780.1$\pm$962.0} \\
& & b: $<$17.3 \\

\multirowcell{2}{Aluminum alloy semi-disc \\ (Bluefors)} & \multirow{2}{*}{33.207} & t: 7385$\pm$1143 & \multirow{2}{*}{166.1$\pm$15.8} & \multirow{2}{*}{339.4$\pm$33.4} & \multirow{2}{*}{-} & \multirow{2}{*}{$<$29.5} & \multirow{2}{*}{$<$3.8} & \multirow{2}{*}{$<$193.2} \\
& & b: 16.2$\pm$14.4 \\

\multirow{2}{*}{Aluminum alloy discs} & 18.036 & t: 465.7$\pm$335.2 & \multirow{2}{*}{$<$11.0} & \multirow{2}{*}{273.1$\pm$21.0} & \multirow{2}{*}{-} & \multirow{2}{*}{$<$11.1} & \multirow{2}{*}{$<$2.0} & \multirow{2}{*}{$<$106.9} \\
& (4 units) & b: $<$9.6 \\

\midrule
\textbf{Other raw materials} \\

\multirow{2}{*}{Copper sample (Bluefors)} & \multirow{2}{*}{94.08} & t: $<$130.8 & \multirow{2}{*}{1.2$\pm$1.6} & \multirow{2}{*}{$<$4.9} & \multirow{2}{*}{-} & \multirow{2}{*}{$<$5.9} & \multirow{2}{*}{$<$0.8} & \multirow{2}{*}{$<$65.1} \\
& & b: $<$2.4 \\

\multirowcell{2}{Ti pellets (99.999\%) \\ (Kurt-J Lesker)} & \multirow{2}{*}{2.66} & t: $<$154.4 & \multirow{2}{*}{5.0$\pm$4.3} & \multirow{2}{*}{$<$23.1} & \multirow{2}{*}{$<$2802} & \multirow{2}{*}{$<$129.8} & \multirow{2}{*}{$<$72.5} & \multirow{2}{*}{$<$1538} \\
& & b: 22.0$\pm$20.1 \\

\multirowcell{2}{BF-6 glue \\ (Ukrvet Biopharm)} & \multirow{2}{*}{94.0} & t: $<$44.7 & \multirow{2}{*}{$<$1.6} & \multirow{2}{*}{26.6$\pm$4.5} & \multirow{2}{*}{$<$130} & \multirow{2}{*}{$<$3.7} & \multirow{2}{*}{$<$0.8} & \multirow{2}{*}{42.9$\pm$24.4} \\
& & b: $<$1.3 \\

\multirowcell{2}{Copper powder (99\%) \\ (thermo scientific)} & \multirow{2}{*}{24.5} & t: $<$385.6 & \multirow{2}{*}{$<$9.9} & \multirow{2}{*}{$<$50.5} & \multirow{2}{*}{$<$1718} & \multirow{2}{*}{12.3$\pm$8.8} & \multirow{2}{*}{$<$9.7} & \multirow{2}{*}{195.2$\pm$88.8} \\
& & b: $<$16.2 \\

\multirowcell{2}{Glass beads \\ (thermo scientific)} & \multirow{2}{*}{28.9} & t: 14190$\pm$1195 & \multirow{2}{*}{409.7$\pm$36.4} & \multirow{2}{*}{5785$\pm$233.4} & \multirow{2}{*}{$<$934.6} & \multirow{2}{*}{$<$80.3} & \multirow{2}{*}{$<$59.6} & \multirow{2}{*}{101610$\pm$5762} \\
 & & b: 8855$\pm$270.7 \\

\multirowcell{2}{Carbon black (99.9+\%) \\ (thermo scientific)} & \multirow{2}{*}{1.45} & t: $<$4130 & \multirow{2}{*}{$<$89} & \multirow{2}{*}{1199$\pm$259} & \multirow{2}{*}{$<$13270} & \multirow{2}{*}{$<$128.1} & \multirow{2}{*}{$<$98.2} & \multirow{2}{*}{4492.1$\pm$1484.0} \\
& & b: $<$71.6 \\

\multirowcell{2}{NbTi superconducting wire \\ (SUPERCON)} & \multirow{2}{*}{1.8} & t: $<$202 & \multirow{2}{*}{13.1$\pm$15.8} & \multirow{2}{*}{$<$115.8} & \multirow{2}{*}{$<$518} & \multirow{2}{*}{$<$225} & \multirow{2}{*}{$<$188.5} & \multirow{2}{*}{2310.4$\pm$2834.0} \\
& & b: $<$106.2 \\

\multirowcell{2}{Unfluxed desolder braid \\ (Chemtronics)} & \multirow{2}{*}{8.745} & t: $<$50.5 & \multirow{2}{*}{3.0$\pm$2.1} & \multirow{2}{*}{$<$17.8} & \multirow{2}{*}{$<$160.9} & \multirow{2}{*}{$<$48.7} & \multirow{2}{*}{$<$23.8} & \multirow{2}{*}{1211.4$\pm$426.5} \\
& & b: $<$13.5 \\

\multirow{2}{*}{RuO2 thermometer} & \multirow{2}{*}{10.3} & t: 222.7$\pm$75.0 & \multirow{2}{*}{7.9$\pm$3.2} & \multirow{2}{*}{203.0$\pm$18.3} & \multirow{2}{*}{1188.9$\pm$291.1} & \multirow{2}{*}{$<$42.1} & \multirow{2}{*}{$<$26.0} & \multirow{2}{*}{1150.7$\pm$532.8} \\
& & b: 168.6$\pm$17.9 \\

\multirowcell{2}{OFHC copper \\ (McMaster-Carr)} & \multirow{2}{*}{8.49} & t: $<$84.3 & \multirow{2}{*}{2.8$\pm$2.3} & \multirow{2}{*}{$<$14.5} & \multirow{2}{*}{$<$214.4} & \multirow{2}{*}{$<$58.8} & \multirow{2}{*}{$<$11.6} & \multirow{2}{*}{831.5$\pm$466.2} \\
& & b: $<$16.9 \\

\multirowcell{2}{Conductive Copper Tape \\ (McMaster-Carr)} & \multirow{2}{*}{9.52} & t: $<$49.1 & \multirow{2}{*}{$<$2.5} & \multirow{2}{*}{0.13$\pm$0.51} & \multirow{2}{*}{$<$191.7} & \multirow{2}{*}{$<$48.6} & \multirow{2}{*}{$<$12.0} & \multirow{2}{*}{868.5$\pm$331.2} \\
& & b: $<$5.1 \\

\bottomrule

\end{tabular}
}
\end{table}

\section{Background contribution of potential components}
\label{sec:potential_components}

In this section, we compare some components which were not included in the background projections discussed in section~\ref{sec:bulk_contamination}.

We have assayed several PCBs to select which one is best suited for our purpose.
The PCB with EPIG Plating (Elco BV, Hofstetter PCB) which was provided by Chalmers University was chosen relatively early because its contributing background rate of 0.163\,mHz in the Si chip was significantly lower compared to others, e.g.\ the PCB TMM10 (Rogers Corporation) with 0.945\,mHz or PCB RO4350B (Rogers Corporation) with 1.199 mHz.
Because the PCB turned out to dominate the background rate together with the Si chip itself and the OQTO holder, the University of Waterloo explored other PCB options.
A more recent assay of an EPIG-plated 7-layer PCB (Aspocomp) shows promising results of 0.159\,mHz (without $^{210}$Pb) which is comparable to the original PCB choice and will be considered for future QUTEbits runs.

In order to mitigate ambient IR, we considered two potential IR absorbers: one as a layer inside the CM shield which is in line of sight with the Al cavities, and another one at the OQTO holder surfaces in direct line of sight with the Si chip.
Table~\ref{tab:background_rates_IR_absorber} summarizes the background rates from relevant IR absorber components.
We do not list Stycast 1266 because its background contribution is negligible compared to the other components.
Carbon black and the glass beads dominate.
The former could be replaced by copper powder, which decreases the induced background rate by more than a factor of ten.
The plots shown in figure~\ref{fig:bulk_rates_Si} and \ref{fig:bulk_rates_AlCav} include the potential contribution of the IR absorbers consisting of a copper sheet, glass beads and copper powder. 
We are currently exploring glass beads from various vendors to find a version that suits our low-radioactivity requirements.
Future assay results will be published on \href{https://www.radiopurity.org/}{radiopurity.org} \cite{RadiopurityWeb} as a service to the community. 

\begin{table}[ht!]
\centering
\caption{Simulated background rates in one Si chip for potential IR absorber components and locations. The results do not take into account the $^{210}$Pb bulk assays.}
\label{tab:background_rates_IR_absorber}
\begin{tabular}{c|cc}
\toprule
                 & IR absorber at CM shield & IR absorber inside OQTO holder \\
Recipe component & background rate [mHz]    & background rate [mHz] \\
\midrule
Copper sheet     & $1.44\cdot10^{-4}$       & $2.70\cdot10^{-2}$ \\
Glass beads      & $7.49\cdot10^{-3}$       & $9.34\cdot10^{-1}$ \\
Carbon black     & $8.48\cdot10^{-3}$       & $7.94\cdot10^{-1}$ \\
Copper powder    & $4.07\cdot10^{-4}$       & $5.19\cdot10^{-2}$ \\
\bottomrule
\end{tabular}
\end{table}

\section{Uncertainties in background simulations}
\label{sec:uncertainties}

Table~\ref{tab:uncertainties_Sichip} and table~\ref{tab:uncertainties_Alcavity} summarize how the simulated background rates are composed of measurements and limits, because some assay results are reported as a contamination measurement with an uncertainty, while others were only upper limits because the radioactivity was below the detector's sensitivity.
These differences have been propagated to the determined event rates in the following way. 
For the case that a simulation results in no detector hit, a 90\% C.L. upper limit is calculated, i.e.\ some assay results with uncertainties are accounted for as limits if the simulation statistics did not produce a statistically significant result.
This only occurred for components which are very far away from the Si chips, e.g.\ the outermost lead shield and components outside of it, such as the water tank of CUTE, and the volume attributed to the interstitial air in the shield.
All other simulations have small statistical uncertainties, which are usually well below the systematic uncertainties of the assay results or other known uncertainties such as the radon concentration.

In general, the bulk background rates are roughly equally composed of components reported as measurements or limits, respectively.
The same is true for the interstitial air, which takes into account the radon concentration measured inside SNOLAB and the purged air in the CUTE lead shield.
The surface contamination contains only a tiny portion of upper limits for the Si chips which arise solely from simulations with no detector hits for the presently simulated statistics.

\begin{table}[ht!]
\centering
\caption{Uncertainties and composition of simulated background rates for a Si chip.}
\label{tab:uncertainties_Sichip}
\resizebox{\textwidth}{!}{
\begin{tabular}{c|ccc}
\toprule
Background category & Total rate [mHz] & Value $\pm$ uncertainties [mHz] & Limit [mHz] \\
\midrule
Bulk                & $5.51\cdot 10^{-1}$ 
                    & $(3.39 \pm 0.02 (\text{stat.}) \pm 0.48 (\text{syst.})) \cdot 10^{-1}$ 
                    & $<2.12\cdot 10^{-1}$ \\  
Interstitial air    & $1.39\cdot 10^{-3}$
                    & $(9.57 \pm 5.53 (\text{stat.}) \pm 0.48 (\text{syst.})) \cdot 10^{-4}$
                    & $<4.29\cdot 10^{-4}$ \\
Surface $^{210}$Pb  & $1.37\cdot 10^{-1}$ 
                    & $(1.37 \pm 0.002 (\text{stat.}) \pm 0.76 (\text{syst.})) \cdot 10^{-1}$
                    & $<1.90\cdot 10^{-6}$ \\
\bottomrule
\end{tabular}
}
\end{table}

\begin{table}[ht!]
\centering
\caption{Uncertainties and composition of simulated background rates for an Al cavity.}
\label{tab:uncertainties_Alcavity}
\resizebox{\textwidth}{!}{
\begin{tabular}{c|ccc}
\toprule
Background category & Total rate [mHz] & Value $\pm$ uncertainties [mHz] & Limit [mHz] \\
\midrule
Bulk                & $7.78\cdot 10^{1}$ 
                    & $(1.24 \pm 0.004 (\text{stat.}) \pm 0.23 (\text{syst.})) \cdot 10^{1}$ 
                    & $<6.54\cdot 10^{1}$ \\  
Interstitial air    & $4.55\cdot 10^{-1}$
                    & $(2.60 \pm 0.09 (\text{stat.}) \pm 0.13 (\text{syst.})) \cdot 10^{-1}$
                    & $<1.95\cdot 10^{-1}$ \\
Surface $^{210}$Pb  & $1.02\cdot 10^{1}$
                    & $(1.02 \pm 0.001 (\text{stat.}) \pm 0.57 (\text{syst.})) \cdot 10^{1}$
                    & - \\
\bottomrule
\end{tabular}
}
\end{table}

The split time used to separate and combine hits into events (see section \ref{sec:hit_processing}) does affect the spectral shape in certain cases, which has been described in section \ref{sec:spectral_analysis}.
For the present analysis, we apply a split time of $\Delta t = 15\,\unit{\mu s}$, which has been informed by simulations performed with G4CMP (see section \ref{sec:G4CMP_collection}).
In order to cross check how the split time affects the estimated background rate, the analysis was run with a split time of $\Delta t = 1\,\unit{ms}$.
The largest impact was observed for the bottom part of the $^{238}$U decay chain (e.g.\ when contaminating the Si substrate and calculating the rate in the Si chip), which decreased the $^{238}$U rate by 10\%.
The same level of rate decrease was found for the $^{238}$U contamination in the Al cavities.
For $^{235}$U, the observed decrease is only about 2-3\%, while for $^{232}$Th the effect is less than 0.1\%.
In combination, a larger split time leads to a relative decrease of the total background rate in the Si chip by 0.2\% and in the Al cavity by 0.6\%.

For the surface $^{210}$Pb simulations, the systematic uncertainty is composed of the radon concentration measured on Earth's surface and inside SNOLAB.
The former has a large uncertainty, which is propagated into the estimated $^{210}$Pb emission rates from the materials' surfaces.
Summing up all components, the $^{210}$Pb emission rate is $178.1 \pm 142.5\,\unit{mBq}$ due to radon exposure on Earth's surface and $78.2 \pm 3.9\,\unit{mBq}$ from underground exposure, leading to a total emission rate of $256.3 \pm 142.6\,\unit{mBq}$ for the medium exposure scenario (see table~\ref{tab:Rn_exposure_scenarious}).
The uncertainties were added in quadrature as they are assumed to be uncorrelated.

As mentioned in section~\ref{sec:surface_contamination}, the Jacobi model was used to determine the ratio of adsorbed vs.\ implanted $^{210}$Pb on surfaces and it has been found that 55.8\% of the $^{210}$Pb nuclei are implanted.
In order to calculate this ratio, an assumption of the initial fraction of $^{218}$Po vs.\ $^{214}$Po adhered on surfaces needs to be made.
The volume of the room, its ventilation and recirculation rate, and whether radon is filtered out determine the plate-out height.
Together with the half-lives of the $^{222}$Rn progeny, the fraction of $^{218}$Po vs.\ $^{214}$Po adhered on surfaces was determined to be about 80\% vs.\ 20\%.
When adjusting these parameters into extreme ranges, one may end up with a fraction of 100\% or 50\% for $^{218}$Po.
These three scenarios introduce an absolute systematic uncertainty of 2.6\% on the ratio of adsorbed vs.\ implanted $^{210}$Pb, which is an order of magnitude below the systematic uncertainties of the radon concentration, and thus has been neglected.

The simulations to determine the ratio of adsorbed vs.\ implanted $^{210}$Pb were run for different materials -- including Al, Cu, mu-metal and Si -- to confirm that the ratio is independent of the material properties.
The final ratio of 55.8\% implanted $^{210}$Pb was calculated as an average of these simulations with a standard deviation of 0.9\%.
The statistical uncertainties in these simulations are well below 0.1\%.

\section{G4CMP configuration and runtime optimization}
\label{sec:G4CMP_config}

The configuration of G4CMP for our showcase study using a Si chip with attached superconducting Al films, in particular the configuration of the surface properties of the interface between Si and Al, follows the procedure described in the appendix of Ref.~\cite{Yelton2024qpmodeling}.
For a general overview of the capabilities of G4CMP and its physics models see Ref.~\cite{Kelsey2023G4CMP}.

An excerpt of the most relevant G4CMP configuration parameters for our study is shown in table~\ref{tab:G4CMP_parameters}.
The majority of the material properties are inherited from the default values shipped with the library.
Moreover, for the thickness of our Al mask of $d_\text{Al} = 300$\,nm, the phonon absorption by Al can be modeled with an energy independent probability $p_\text{abs}$:
\begin{equation}
p_\text{abs} = p_\text{trans} (1-p_\text{esc}) \approx p_\text{trans} \text{ with } p_\text{esc} \rightarrow 0.
\label{eq:phonon_absorption}
\end{equation}

Equation~\eqref{eq:phonon_absorption} includes a transmission term $p_\text{trans}$ for the Si-to-Al boundary taken from \cite{Yelton2024qpmodeling} and an escape probability $p_\text{esc}$ for phonons that traverse the Al film thickness and return to the substrate without breaking a Cooper pair.
The latter can be expressed as \cite{Yelton2024qpmodeling}:
\begin{equation}
    p_\text{esc} = \exp\left( -\frac{4d_\text{Al}}{\lambda(E_\text{ph})} \right).
\end{equation}

For superconducting Al, the phonon mean free path $\lambda(E_\text{ph})$ can be expressed as the ratio of the isotropic speed of sound in Al, denoted as $v_s$, and the Cooper-pair-breaking rate $\Gamma^b_\text{ph}$:
\begin{equation}
    \lambda(E_\text{ph}) = v_s / \Gamma^b_\text{ph}
\end{equation}
with
\begin{equation}
\Gamma^b_\text{ph} = \frac{1}{\tau_0^\text{ph}} \left\lbrace 1 + 0.29\left[ \left( \frac{E_\text{ph}}{\Delta_\text{Al}} \right) -2 \right] \right\rbrace.
\label{eq:cooper_rate}
\end{equation}
Table~\ref{tab:G4CMP_parameters} demonstrates the outcome of eq.~\eqref{eq:phonon_absorption} -- \eqref{eq:cooper_rate} assuming a phonon energy of \mbox{$E_\text{ph} = 4\,$meV} which is representative for our case study.
For a thinner Al film, e.g.\ $d_\text{Al} = 30$\,nm, the phonon escape probability would result in $p_\text{esc} = 38.6\%$ for the same inputs in comparison.

\begin{table}[ht!]
\centering
\caption{Excerpt of relevant G4CMP simulation parameters. The majority of the material parameters for Al and Si are taken from Ref.~\cite{Kelsey2023G4CMP} and \cite{Yelton2024qpmodeling}. The calculated quantities were evaluated for a phonon energy of $E_\text{ph} = 4\,$meV.}
\label{tab:G4CMP_parameters}
\begin{tabular}{lccc}
\toprule
Parameter & Symbol & Value & Reference\\
\midrule
Al mask thickness & $d_\text{Al}$ & 0.3 $\mu$m & input \\
Phonon energy & $E_\text{ph}$ & 4.0 meV & input \\
\midrule
Phonon absorption probability in Al &$p_\text{abs}$ & 0.795 & calculated \\
Phonon escape probability in Al &$p_\text{esc}$ & $\approx$ 0 & calculated \\
Phonon transmission probability for Si-Al &$p_\text{trans}$ & 0.795 & \cite{Yelton2024qpmodeling} \\
Pair-breaking rate in Al &$\Gamma^b_\text{ph}$ & 29.33 1/ns & calculated \\
Phonon mean free path in Al &$\lambda(E_\text{ph})$ & 0.11 $\mu$m & calculated \\
\midrule
Isotropic speed of sound in Al &$v_s$ & 3.26 $\mu$m/ns & \cite{Kelsey2023G4CMP} (Al) \\
Phonon lifetime in Al &$\tau_0^\text{ph}$ & 0.242 ns & \cite{Kelsey2023G4CMP} (Al) \\
Al superconducting bandgap &$\Delta_\text{Al}$ & 0.174 meV & \cite{Kelsey2023G4CMP} (Al) \\
\midrule
Si bandgap & $\varepsilon_g$ & 1.17 eV & \cite{Kelsey2023G4CMP} (Si) \\
Average energy to create e$^-$h$^+$ pair in Si & $\varepsilon_\text{eh}$ & 3.81 eV & \cite{Kelsey2023G4CMP} (Si) \\
Si Fano factor & $F$ & 0.15 & \cite{Kelsey2023G4CMP} (Si) \\
Debye frequency / energy in Si  & $\omega_D$ & 15 THz / 62 meV & \cite{Kelsey2023G4CMP} (Si) \\
Longitudinal speed of sound in Si  & $v_s$ & 9000 m/s & \cite{Kelsey2023G4CMP} (Si) \\
\bottomrule
\end{tabular}
\end{table}

In addition to the basic configuration of the material properties and phonon interfaces, we optimized the runtime of our G4CMP-based simulations by making use of several available energy downsampling features (see Ref.~\cite{Kelsey2023G4CMP} for a detailed description).
The first measure is to terminate all phonon tracks that are not able to break Cooper pairs in Al because they do not carry a sufficient energy of $E_\text{ph} \geq 2\Delta_\text{Al}$ with $\Delta_\text{Al}$ being the superconducting gap of Al (see table~\ref{tab:G4CMP_parameters}).
This greatly reduces the number of low-energy phonons to be tracked in the simulation per event.

For the simulation of the high-energy particle hits by $\beta$ electrons, which can spread out over the entire Si chip volume for sufficiently energetic electrons, we make use of combining hits that are closer together in space than 2\,mm.
This measure reduces the overall tracking time significantly for high-energy electrons in particular because each \textsc{Geant4} particle hit would otherwise be processed separately in G4CMP.
Additionally, each energy deposit is downsampled to an equivalent energy deposit of 200\,eV (see section~\ref{sec:G4CMP_multiplicity}) to reduce the number of charge carrier pairs and primary phonons per event to a representative ensemble with appropriate track weights (see \cite{Kelsey2023G4CMP} for details).
For an ER-like energy deposit of 200\,eV in Si an average of 52.5 e$^-$h$^+$ pairs will be created.
For an NR-like energy deposit of the same energy but $Y=0.1$, on average 5 e$^-$h$^+$ pairs and about 2800 prompt phonons will be created.
Hits grouped together by the hit-merging algorithm are downsampled together according to the set energy partition value.
Finally, we restrict the maximum number of phonons created by lattice scattering (referred to as ``Luke'' phonons) and the maximum number of phonon reflections after which tracks get terminated. 
A summary is presented in table~\ref{tab:downsampling}.

\begin{table}[ht!]
\caption{Summary of G4CMP downsampling parameters. The options are roughly presented in order of effectiveness with respect to the runtime improvement per event. The last two restrictions barely had an impact on any of the simulated cases discussed in section~\ref{sec:G4CMP_studies} when combined with the previous downsampling methods.}
\label{tab:downsampling}
\centering
\begin{tabular}{ll}
\toprule
Description & G4CMP command \\
\midrule
Do not track phonons with $E_\text{ph} < 2\Delta_\text{Al}$ & \texttt{/g4cmp/minEPhonons 348e-6 eV} \\
Do not record below-minimum energy tracks & \texttt{/g4cmp/recordMinETracks false} \\
Combine hits below step length of 2\,mm & \texttt{/g4cmp/combiningStepLength 2 mm} \\
Downsample energy deposits to 200\,eV & \texttt{/g4cmp/samplingEnergy 200 eV} \\
Set maximum of Luke phonons per event to $10^5$ & \texttt{/g4cmp/maxLukePhonons 100000} \\
Set maximum phonon reflections to $10^4$ & \texttt{/g4cmp/phononBounces 10000} \\
\bottomrule 
\end{tabular}
\end{table}

\section{Computing resources}
\label{sec:computing_resources}

The \textsc{Geant4} background simulations consumed in total 32.8 CPU years.
The simulations are composed of 482 combinations of components with their corresponding isotopes in the bulk, two special configurations for the interstitial air, and 20 components with $^{210}$Pb adsorbed and implanted on surfaces.
The cavern $\gamma$ ray simulations have been broken down into their contributions from $^{232}$Th, $^{238}$U and $^{40}$K.
The cavern neutrons have been simulated separately.
In total, $3.63\cdot10^{12}$ primary decays were generated for the background simulations, which comprises $6.1\cdot10^{11}$ events for the bulk, $2.1\cdot10^{10}$ events for the surface and $3.0\cdot10^{12}$ for the cavern simulations.

The calibration source simulations consumed about 4.9 CPU years for the $^{133}$Ba source studies and 0.5 CPU years for $^{252}$Cf.
For each of the twelve possible payload rotations, $10^{10}$ primary $^{133}$Ba decays were simulated.
The $^{252}$Cf induced ER and NR rates were investigated at two different positions, with $10^{9}$ primary decays each.

The G4CMP simulations were optimized for runtime as discussed in appendix~\ref{sec:G4CMP_config} before running the final production batch with high statistics (up to $10^{7}$ events per configuration) in a multithreaded fashion.
The optimization consumed about 1.1 CPU years and the final production batch comprising simulations of a bulk contaminant source and a point source at the center of the Si chip consumed about 0.4 CPU years.
The multithreaded simulations achieved on average a CPU efficiency of 80\% with eight worker threads.

In total, the simulations performed for this work consumed about 39.7 CPU years on the Fir cluster of the Digital Research Alliance of Canada.

\end{document}